\documentclass[aps,prd,a4paper,twocolumn,amsmath,showpacs,superscriptaddress,nofootinbib,preprintnumbers]{revtex4-1}
\pdfoutput=1
\usepackage{amssymb,amsmath,latexsym,mathrsfs}
\usepackage[sort&compress]{natbib}
\usepackage{graphicx,subfigure}
\usepackage{epsfig}
\usepackage{varioref,xr-hyper}
\usepackage{color}
\usepackage{multirow}
\usepackage{array}
\usepackage{hyperref}
\usepackage{wasysym}
\usepackage{color}
\usepackage{float}
\usepackage{xcolor}
\usepackage[utf8]{inputenc}
\usepackage[T1]{fontenc}
\hypersetup{colorlinks,linkcolor={blue},citecolor={red},urlcolor={blue}}

\begin{document}

\title{IWDM: The fate of an interacting non-cold dark matter \textemdash{} vacuum  scenario}

\author{Supriya Pan}
\email{supriya.maths@presiuniv.ac.in}
\affiliation{Department of Mathematics, Presidency University, 86/1 College Street,  Kolkata 700073, India}
\affiliation{Institute of Systems Science, Durban University of Technology, PO Box 1334, Durban 4000, Republic of South Africa}

\author{Weiqiang Yang}
\email{d11102004@163.com}
\affiliation{Department of Physics, Liaoning Normal University, Dalian, 116029, P. R. China}

\author{Eleonora Di Valentino}
\email{e.divalentino@sheffield.ac.uk}
\affiliation{School of Mathematics and Statistics, University of Sheffield, Hounsfield Road, Sheffield S3 7RH, United Kingdom}

\author{David F. Mota}
\email{d.f.mota@astro.uio.no}
\affiliation{Institute of Theoretical Astrophysics, University of Oslo, P.O. Box 1029 Blindern, N-0315 Oslo, Norway}

\author{Joseph Silk}
\email{silk@iap.fr}
\affiliation{Institut d'Astrophysique de Paris (UMR7095: CNRS \& UPMC- Sorbonne Universities), F-75014, Paris, France}
\affiliation{Department of Physics and Astronomy, The Johns Hopkins University Homewood Campus, Baltimore, MD 21218, USA}
\affiliation{BIPAC, Department of Physics, University of Oxford, Keble Road, Oxford OX1 3RH, UK}


\begin{abstract}
In most cosmological models, the equation of state of the dark matter is assumed to be zero,
which means that the dark matter is pressure-less or cold. While this hypothesis is based on the abundance of cold dark matter in the universe, however, there is no compelling 
reason to assume that the equation of state of dark matter is exactly zero. A more general approach would be to allow for a range of values for the dark matter equation of state and use the observational data to determine which values are most likely. With the increasing accuracy of experimental data, we have chosen to explore the possibility of interacting non-cold dark matter $-$ 
vacuum scenario, where the equation of state of the dark matter is constant but can take different values 
within a specific range. Using the Cosmic Microwave Background (CMB) anisotropies and the CMB lensing reconstruction from the Planck legacy release, plus other non-CMB measurements, namely, the baryon acoustic oscillations distance measurements, and the Pantheon catalogue from Type Ia Supernovae, we have analyzed this scenario and found that a non-zero value for the dark matter equation of state is preferred with a confidence level of over 68\%. While this is not significant by itself, however, it does suggest that investigating the possibility of non-cold dark matter in the universe is worth exploring further to gain a better understanding of the nature of dark matter.
\end{abstract}

\maketitle
\section{Introduction}

Observational evidence suggests that our universe is dominated by two dark fluids: one is dark matter (DM) and the other is dark energy (DE) (in the context of Einstein's General Relativity)  or  geometrical DE (in the context of modified gravity theories)~\cite{Planck:2018vyg}. However, the nature of these dark fluids has remained mysterious in recent years. Although the standard $\Lambda$-Cold Dark Matter scenario (henceforth $\Lambda$CDM) has been very successful in fitting a series of astronomical data sets,  several theoretical and observational signatures argue that $\Lambda$CDM cosmology needs to be revised. The tensions in cosmological parameters have been quite serious in recent years~\cite{DiValentino:2021izs,Perivolaropoulos:2021jda,Schoneberg:2021qvd,Abdalla:2022yfr}. This has motivated our scientific community to build new cosmological scenarios and test them against observational evidences, and, as a result, this has led to a vast literature of DE and geometrical DE models, see for instance Refs.~\cite{Peebles:2002gy,Copeland:2006wr,Nojiri:2006ri,Sahni:2006pa,Padmanabhan:2007xy,Capozziello:2007ec,Sotiriou:2008rp,Frieman:2008sn,Caldwell:2009ix,Silvestri:2009hh,DeFelice:2010aj,Nojiri:2010wj,Clifton:2011jh,Capozziello:2011et,Bamba:2012cp,Li:2012dt,Cai:2015emx,Nojiri:2017ncd,Bahamonde:2021gfp}. If one carefully examines the existing literature, one discovers that a common practice in building cosmological models is to assume the DM equation of state equal to zero, or, in other words, that the universe is filled with cold DM. While this assumption is motivated by structure formation, and is indeed based on the enormous success of the $\Lambda$CDM cosmology, however, let us recall that the nature of DE and DM is not yet truly known. So instead of setting the DM equation of state  to zero by hand, as the sensitivity in the experimental data grows over time, we could test this assumption by allowing a free DM equation of state and let the observational data come to a conclusion. 
This approach has motivated several researchers to examine whether observational data indicate any non-cold nature in DM~\cite{Muller:2004yb,Kumar:2012gr,Armendariz-Picon:2013jej,Kopp:2018zxp,Ilic:2020onu,Naidoo:2022rda}.

In the present article we actually focus on an interacting cosmological theory between DM and DE, where DM has an equation of state that varies freely over a certain range, a well justified assumption in an interacting scenario, and DE represents the vacuum energy. The interacting DM-DE models, widely known as the Interacting Dark Energy (IDE) models/Coupled dark Energy (CDE)  models, are very rich both theoretically and observationally. In recent years, IDE models have been extensively investigated by many researchers, which has led to a number of interesting possibilities, including a possible alleviation  of the cosmic coincidence problem~\cite{Amendola:1999er,Huey:2004qv,Cai:2004dk,Pavon:2005yx,delCampo:2006vv,delCampo:2008sr,delCampo:2008jx}, phantom crossing~\cite{Wang:2005jx,Das:2005yj,Sadjadi:2006qb,Pan:2014afa} and recently the alleviation of the cosmological tensions~\cite{DiValentino:2017iww,Kumar:2017dnp,Yang:2018euj,Pan:2019jqh,Yang:2019uog,Pan:2019gop,Pan:2020bur,DiValentino:2019ffd,DiValentino:2019jae,Lucca:2020zjb,Anchordoqui:2021gji,Gao:2021xnk,Pourtsidou:2016ico,An:2017crg,Lucca:2021dxo,Gariazzo:2021qtg,Yang:2021hxg,Yao:2022kub}. We recall that the origin of IDE theory was motivated by a previous proposal by Wetterich in which the author argued that the cosmological constant problem might be solved through an interaction in the cosmic sector~\cite{Wetterich:1994bg}. The dynamics of the IDE models is primarily controlled by an interaction function that modifies the expansion history of the universe at the background and perturbation levels. 
We refer to an incomplete list of works on the IDE models studied over the years~\cite{Amendola:1999er,Huey:2004qv,Cai:2004dk,Pavon:2005yx,Amendola:2006dg,Barrow:2006hia,Berger:2006db,delCampo:2006vv,Valiviita:2008iv,delCampo:2008sr,delCampo:2008jx,Gavela:2009cy,Majerotto:2009np,Gavela:2010tm,Clemson:2011an,Salvatelli:2014zta,Wang:2014xca,Yang:2014gza,yang:2014vza,Yang:2014hea,Li:2015vla,vandeBruck:2016jgg,Feng:2016djj,vandeBruck:2016hpz,Sharov:2017iue,DiValentino:2017iww,Mifsud:2017fsy,Kumar:2017dnp,VanDeBruck:2017mua,Yang:2018euj,Yang:2018pej,Li:2018jiu,Pan:2019jqh,Martinelli:2019dau,Mifsud:2019fut,Oikonomou:2019boy,Yang:2019uog, Pan:2019gop,Cheng:2019bkh,Pan:2020bur,DiValentino:2019ffd,DiValentino:2019jae,Yao:2020hkw,Yao:2020pji,Jesus:2020tby,Sa:2020fvn,Sa:2021eft,Yang:2021oxc,Kang:2021osc,Bonilla:2021dql,Harko:2022unn,Yengejeh:2022tpa,Chatzidakis:2022mpf,Yang:2022csz}. In this article, we consider a very well known and most used interaction function and constrain the scenario using various observational data sets, namely, the Cosmic Microwave Background (CMB) anisotropies, the CMB Lensing reconstruction from the Planck legacy release, Baryon Acoustic Oscillations (BAO) distance measurements from several astronomical missions and finally the Pantheon catalogue from Type Ia Supernovae, with the aim of understanding whether a non-zero equation-of-state of DM is suggested by current observational data sets.

The manuscript is organized as follows. In \autoref{sec-2} we describe the key equations of the interacting non-cold dark matter $-$ vacuum scenario. Then in  \autoref{sec-data sets } we describe observational data sets and statistical methodology. In \autoref{sec-results} we describe the observational constraints and implications of the interacting non-cold DM $-$ vacuum scenario. Finally, we summarize the results of the manuscript in \autoref{sec-summary}.

\section{Interacting non-cold dark matter and vacuum energy}
\label{sec-2}

We consider an interacting scenario between vacuum energy and a non-cold dark matter fluid with equation of state $W$ (labeled as \textbf{``IWDM''}, read as Interacting DM with equation-of-state $W$), 
in a homogeneous and isotropic universe  characterized by the spatially flat Friedmann-Lema\^{i}tre-Robertson-Walker (FLRW) universe.  The vacuum energy is characterized by its equation of state $w_{de} = -1$ and the equation-of-state of the non-cold DM, $W$, could be either constant in time or dynamical. Here we explore the simplest scenario where $W$ is constant (labeled as $w_{dm}$ for simplicity and to match with the DM abbreviation), which can vary freely over a non-negative interval\footnote{We limit ourselves to non-negative values of the DM equation-of-state to avoid any exotic touch on the DM since the negative values of $w_{dm}$ could raise some unphysical issues, for example, if $w_{dm}$ becomes negative, then it might be difficult to distinguish between the DM and DE sectors from their evolution equations. }. 
In presence of an interaction between the non-cold DM and the vacuum sector, the continuity equations read 
\begin{eqnarray}
\label{interaction}
\nabla_{\mu}T_{j}^{\mu \nu }=Q_{j}^{\nu}\,, \quad \sum\limits_{\mathrm{j}}{%
Q_{j}^{\mu }}=0~,
\end{eqnarray}
where $j$ runs for the non-cold DM and the vacuum sector.  The four-vector $Q_{j}^{\mu}$ controls the energy exchange between the dark sectors. We assume that $Q_{j}^{\mu}$ is given by~\cite{Valiviita:2008iv,Wang:2014xca}
\begin{eqnarray}
Q_{j}^{\mu}=(Q_{j}+\delta Q_{j})u^{\mu}+ f^{\mu}_{j}, 
\end{eqnarray}
in which  $u^{\mu}$ denotes the four-vector velocity; $Q_j$ denotes the  background energy transfer; $f_j$ stands for momentum transfer. Note that from now on we will use the more commonly used notation $Q_j \equiv Q$.  In the FLRW background, from \autoref{interaction}, the conservation equations of the interacting fluids can be written in their well known forms and are,
\begin{eqnarray}
\dot{\rho}_{de}&=& Q (t), \label{cons-de}\\
\dot{\rho}_{dm}+3H(1+w_{dm}) \rho_{dm} &=& - Q (t),\label{cons-dm}
\end{eqnarray}
where an overhead dot represents the derivative with respect to the cosmic time; $\rho_{de}$, $\rho_{dm}$ are respectively the energy density of the vacuum sector and DM; $H = \dot{a}/a$ (`$a$' denotes the scale factor of the FLRW universe) is the Hubble rate of the FLRW universe which provides an additional constraint on the total energy density of the universe as $3H^2 = 8 \pi G (\rho_{de}+ \rho_{dm}+ \rho_b +\rho_r)$, where the remaining two terms, namely, $\rho_{b}$ and $\rho_{r}$ denote the energy density of baryons and radiation, respectively; $G$ is the Newton's gravitational constant; the
remaining $Q (t)$ denotes the  interaction function that determines the energy flux between the vacuum energy and the non-cold DM. For  $Q (t) > 0$, the energy flows from the non-cold DM to the vacuum while for $Q (t) <0$, the energy flows in the opposite direction, i.e. from the vacuum  to the non-cold DM. Here, we consider the best known interaction function~\cite{Gavela:2009cy,DiValentino:2017iww,Yang:2019uog}:
\begin{eqnarray}\label{model-Q}
Q (t) = 3 H \xi\rho_{de}
\end{eqnarray}
where $\xi$ is the coupling parameter of the interaction function. Following the sign convention in $Q (t)$, one infers that $\xi > 0$ denotes the energy transfer from the non-cold DM to the vacuum sector while $\xi < 0$ denotes the energy flow in the opposite direction (i.e. from the vacuum sector to the non-cold DM).  We note that the choice of the interaction function in \autoref{model-Q} is not phenomenological, on the contrary, this particular choice of the interaction function can be derived from an action formalism~\cite{Pan:2020zza}. Furthermore, as we will show below, for this interaction function, the evolution of the non-cold DM and the vacuum sector in terms of their energy densities can be found analytically. 
Now in presence of the above interaction function,  the conservation equations, i.e.~\autoref{cons-de} and~\autoref{cons-dm} can also be rewritten as 
\begin{eqnarray}
\dot{\rho}_{de} + 3 H (1+ w^{\rm eff}_{de}) \rho_{de} = 0,\\
\dot{\rho}_{dm} + 3 H (1+ w^{\rm eff}_{dm}) \rho_{dm} = 0,
\end{eqnarray}
where $w^{\rm eff}_{de}$ and $w^{\rm eff}_{dm}$ termed as the effective equation-of-state parameters of the vacuum and the non-cold DM, are given by 

\begin{eqnarray}
&&w^{\rm eff}_{de} = -1 - \frac{Q (t)}{3H \rho_{de}}, \label{eff-eos-vac}\\
&&w^{\rm eff}_{dm} =  w_{dm} + \frac{Q (t)}{3H \rho_{dm}}. \label{eff-eos-dm}
\end{eqnarray}
Notice from ~\autoref{eff-eos-vac} that for $Q (t) > 0$, $w^{\rm eff}_{de}$ goes beyond the cosmological constant boundary (i.e. $w^{\rm eff}_{de} < -1$) and this is justified because $Q (t) >0$ implies an energy transfer from DM to the vacuum sector. While for $Q (t) < 0$, which represents an energy transfer from the vacuum sector to non-cold DM, the effective equation of state of the vacuum energy, i.e. $w^{\rm eff}_{de}$ may behave like a quintessence DE ($w^{\rm eff}_{de} > -1$). In principle, the exact evolution of the effective equation-of-state parameters can be traced once the evolution of the energy densities of the dark components has been found analytically and this certainly depends on the interaction function, $Q (t)$ itself.  While we note that it is always possible to find out the numerical evolution of all the quantities involved regardless of the choice of the interaction function, $Q (t)$. Interestingly, for the present interaction model of~\autoref{model-Q}, the energy densities of vacuum and non-cold DM can be expressed analytically:

\begin{align}
\rho_{de} &= \rho_{de,0}\;a^{3\xi}, \label{rho-de}\\
\rho_{dm} &= \rho_{dm,0}\; a^{-3(1+w_{dm})} + \frac{\xi \rho_{de,0}}{\delta} \left[a^{-3(1+w_{dm})} -a^{3\xi}\right],\label{rho-dm}
\end{align}
where $\rho_{dm,0}$, $\rho_{de,0}$ are respectively the present day values of $\rho_{dm}$, $\rho_{de}$ and $\delta = 1+\xi +w_{dm}$. 
From \autoref{rho-de} and \autoref{rho-dm} one can quickly see that for $\xi =0$ (i.e. in absence of any interaction), one recovers the usual evolution equations for DE and DM. 
It is essential to mention that the parameter $\delta$ should not vanish because for $\delta = 0$, the energy density of the DM sector will diverge at a finite time and this is unphysical.  Consequently, the effective equation-of-state of the vacuum sector, i.e.~\autoref{eff-eos-vac} becomes trivial leading to $w^{\rm eff}_{de} = -1 -\xi$ and  the effective equation-of-state of the cold DM sector of~\autoref{eff-eos-dm} becomes 

\begin{eqnarray}
w^{\rm eff}_{dm} =  w_{dm} + \frac{\xi \delta a^{3\delta }}{\xi \left(1 - a^{3 \delta} \right) + r_0 \delta},
\end{eqnarray}
where $r_0 = \rho_{dm,0}/\rho_{de,0}$ and $\delta $ is defined above. The previous expression for  $w^{\rm eff}_{dm}$ offers an effective evolution of the DM sector in presence of the interaction function of~\autoref{model-Q}. Note that such effective prescription is subject to the choice of the interaction function.

As already noted, the evolution of the interacting scenario at the background level is influenced by the equation-of-state of DM, the new ingredient of the interacting scenario. Similarly, the equation-of-state of DM, could equally affect the evolution of the interacting scenario at the perturbation level and, consequently, the resulting perturbation equations become different than the perturbation equations for $w_{dm} = 0$.  In the following we describe the evolution of the interacting scenario at the level of perturbations.

In a general gauge, the scalar perturbations of the flat FLRW metric are given by  \cite{Mukhanov:1990me,Ma:1995ey,Malik:2008im}
\begin{eqnarray}
ds^2=a^2(\tau)\Bigg[ -(1+2\phi)d\tau^2+2\partial_iBd\tau dx^i \nonumber\\ + \bigg\{(1-2\psi)\delta_{ij}+2\partial_i\partial_jE \bigg\} dx^i dx^j \Bigg],
\label{eq:perturbed-metric}
\end{eqnarray}
where $\tau$ is the conformal time and $\phi$, $B$, $\psi$, $E$ are the gauge-dependent scalar perturbations quantities. 
The four-velocity of the fluid $A= \{de, dm\}$ represents either DE or DM, is given by~\cite{Valiviita:2008iv,Yang:2014gza}:

\begin{eqnarray}
u^{\mu}_A=a^{-1}(1-\phi,\partial^i v_A),      
\end{eqnarray}
where $v_A$ is the peculiar velocity potential of the fluid $A$ whose relation with the volume expansion  in the Fourier space $k$ is given by $\theta_A = -k^2(v_A + B)$.   In the interacting background characterized by the evolution equations in (\ref{interaction}), a general energy-momentum transfer can be divided relative to the total four-velocity as~\cite{Valiviita:2008iv,Yang:2014gza}
\begin{eqnarray}
Q^{\mu}_A=\widetilde{Q}_Au^{\mu}+F^{\mu}_A,
\label{eq:Q-vector}
\end{eqnarray}
where $\widetilde{Q}_A=Q_A+\delta Q_A$ and 
$F^{\mu}_A=a^{-1}(0,\partial^if_A)$; here $Q_A$ refers to the background term of the interaction rate, and $f_A$ is a momentum transfer potential. The perturbed energy-momentum transfer four-vector can be decomposed as~\cite{Valiviita:2008iv,Yang:2014gza}
\begin{eqnarray}
&& Q^A_0=-a[Q_A(1+\phi)+\delta Q_A], \\ 
&& Q^A_i=a\partial_i[Q_A(v+B)+f_A],
\label{eq:Q-component}
\end{eqnarray}
The perturbed energy and momentum balance equations for each fluid $A$ can be derived easily as follows~\cite{Valiviita:2008iv,Yang:2014gza}
\begin{eqnarray}
\delta \rho'_A+ 3\mathcal{H}(\delta \rho_A+\delta p_A)-3(\rho_A+p_A)\psi' \nonumber\\ -k^2(\rho_A+p_A)(v_A+E')
=aQ_A\phi+a\delta Q_A, \label{eq:delta-rhoA}
\end{eqnarray}
\begin{eqnarray}
\delta p_A+[(\rho_A+p_A)(v_A+B)]'+4\mathcal{H}(\rho_A+p_A)(v_A+B) \nonumber\\ +(\rho_A+p_A)\phi
=aQ_A(v+B)+af_A,
\label{eq:delta-PA}
\end{eqnarray}
where the prime denotes the derivative with respect to the conformal time $\tau$, and $\mathcal{H}$ is the conformal Hubble rate. Note that to derive the above equations we have neglected the presence of the anisotropic stress in the energy-momentum tensor \cite{Valiviita:2008iv}. While one can consider the presence of the anisotropic stress in the energy-momentum tensor and as a result, the resulting equation (\ref{eq:delta-PA}) will include an extra term due to the anisotropic stress, see Ref. \cite{Valiviita:2008iv}. 
Now introducing the density contrast as $\delta_A=\delta\rho_A/\rho_A$,  one can find the evolution equations for the density perturbations and velocity perturbations of the fluid $A$ \cite{Valiviita:2008iv,Yang:2014gza}

\begin{eqnarray}
\delta'_A+3\mathcal{H}(c^2_{sA}-w_A)\delta_A
+9\mathcal{H}^2(1+w_A)(c^2_{sA}-c^2_{aA})\frac{\theta_A}{k^2}
\nonumber \\ +(1+w_A)\theta_A
-3(1+w_A)\psi'+(1+w_A)k^2(B-E')
\nonumber \\
=\frac{a}{\rho_A}(-Q_A\delta_A+\delta Q_A)
+\frac{aQ_A}{\rho_A}\Bigg[\phi +3\mathcal{H}(c^2_{sA}-c^2_{aA})\frac{\theta_A}{k^2}\Bigg],
\label{eq:general-deltaA} \\
\theta'_A+\mathcal{H}(1-3c^2_{sA})\theta_A-\frac{c^2_{sA}}{(1+w_A)}k^2\delta_A
-k^2\phi 
\nonumber \\
=\frac{a}{(1+w_A)\rho_A}\Bigg[(Q_A\theta-k^2f_A)-(1+c^2_{sA})Q_A\theta_A \Bigg],
\label{eq:general-thetaA}
\end{eqnarray}
where $c^2_{sA}$ is the physical sound speed of the fluid $A$ in the rest frame ($\mbox{rf}$), defined as $c^2_{sA}=(\delta p_A/\delta\rho_A)|_{\mbox{rf}}$ and $c^2_{aA}$ is the adiabatic sound speed of the fluid $A$ defined as  $c^2_{aA}=p'_A/\rho'_A=w_A+w'_A/(\rho'_A/\rho_A)$.  
These are the general equations of the interacting scenario between DE and DM at the level of perturbations where $w_A$ denotes the equation-of-state parameter of the fluid $A$. However, when DE is represented by the vacuum energy sector characterized by the equation-of-state $-1$, then the evolution equations of the interacting vacuum model at the level of perturbations will be much simplified that we shall describe below. Now,  focusing on the general equations, one can notice that  for a barotropic fluid, $c^2_{sA} =  c^2_{aA}$, and in addition, if $w_A$ is constant, then $c^2_{sA} =  c^2_{aA} = w_A$. Now if we consider that DE is an adiabatic fluid, then   $c^2_{s,de} =  c^2_{a,de} = w_{de} < 0$, and hence $c_{s,de}$ becomes imaginary and as a result we encounter with the instabilities in the DE sector.  Therefore, in order to fix this issue, one needs to impose $c^2_{s,de} > 0$ by hand  \cite{Gordon:2004ez} and set a positive value for $c^2_{s,de}$. The natural choice for  $c^2_{s,de} $ is $1$ as obtained in the scalar field models \cite{Valiviita:2008iv}.  This is what is done in the cosmological packages, e.g. CAMB \footnote{\href{see here}{https://camb.info/}} and others. 
In our case we have a simple cosmological scenario in which vacuum energy interacts with the DM sector, hence the resulting perturbations equations will be different from the non-interacting cosmological scenarios involving DE and DM as described in \cite{Ma:1995ey} for both the synchronous and conformal Newtonian gauges.  Here, we work in the synchronous gauge, that means, $\phi =B=0$, $\psi =\eta $, and $k^{2}E=-h/2-3\eta$ ($h$, $\eta$ denote the metric perturbations). Within this present interaction scheme, one can see that the evolution equations at the perturbations level take simple forms. We adopt the methodology already described in Refs. ~\cite{Wands:2012vg,Wang:2014xca,Yang:2019uog}. We consider an energy flow parallel to the 4-velocity of the non-cold DM: $Q^{\mu}_{dm} = - Q u^{\mu}_{dm}$ \cite{Wands:2012vg,Wang:2014xca,Yang:2019uog}. In this case, the DM follows the geodesics~\cite{Wands:2012vg,Wang:2013qy} which means that in the non-cold DM comoving frame, the vacuum energy perturbations vanish ~\cite{Wands:2012vg,Wands:2012vg,Wang:2014xca,Yang:2019uog}. Subsequently, for the non-cold DM, the evolution equations for the density perturbations ($\delta_{dm}$) and the velocity perturbations ($\theta_{dm}$) for the interaction function of \autoref{model-Q}  can explicitly be recast as 

\begin{eqnarray}
\delta _{dm}^{\prime } =-(1+w_{dm})\left( \theta _{dm}+\frac{h^{\prime }}{2}%
\right) \nonumber\\ -3\mathcal{H}(c_{s,dm}^{2}-w_{dm})\Bigg[ \delta _{dm} +3\mathcal{H}%
(1+w_{dm})\frac{\theta _{dm}}{k^{2}}\Bigg] 
\nonumber\\ - 3\mathcal{H}\xi\frac{\rho _{de}}{\rho _{dm}}\Bigg[ \delta _{dm} - 3\mathcal{H}%
(c_{s,dm}^{2}-w_{dm})\frac{\theta _{dm}}{k^{2}}\Bigg], \\
\theta_{dm}^{\prime} =-\mathcal{H}(1-3c_{s,dm}^{2})\theta _{dm}+\frac{%
c_{s,dm}^{2}}{(1+w_{dm})}k^{2}\delta _{dm} \nonumber\\+3\mathcal{H}\xi\frac{\rho _{de}}{\rho _{dm}}\left[ \frac{%
\theta _{dm}-(1+c_{s,dm}^{2})\theta _{dm}}{1+w_{dm}}\right]
\end{eqnarray}
where $c^2_{s,dm}$ is the physical sound speed of non-cold DM in the rest frame. As the DM sector is responsible for the structure formation of the universe, so the sound speed of non-cold DM is  assumed to be $c^2_{s,dm} = 0$. Thus, having the evolution equations of the dark components at the level of background and perturbations,  one can now proceed to constrain the interacting scenario using the available cosmological probes. 

\begin{table}
\begin{center}
\renewcommand{\arraystretch}{1.4}
\begin{tabular}{|c@{\hspace{1 cm}}|@{\hspace{1 cm}} c|}
\hline
\textbf{Parameter}           & \textbf{Prior}\\
\hline\hline
$\Omega_{b} h^2$             & $[0.005,0.1]$ \\
$\Omega_{dm} h^2$            & $[0.01,0.99]$ \\
$\tau$                       & $[0.01,0.8]$ \\
$n_s$                        & $[0.5, 1.5]$ \\
$\log[10^{10}A_{s}]$         & $[2.4,4]$ \\
$100\theta_{MC}$             & $[0.5,10]$ \\ 
$w_{dm}$                     & $[0, 1]$ \\
$\xi$                        & $[-1, 1]$ \\

\hline
\end{tabular}
\end{center}
\caption{We show the flat prior used on the cosmological parameters varied independently during the statistical analysis. }
\label{tab:priors}
\end{table}

\section{Observational data sets}
\label{sec-data sets } 

Here we describe the observational data sets that we used to constrain the interacting scenario.

\begin{enumerate}

\item {\bf Cosmic Microwave Background Radiation}: Cosmic Microwave Background (CMB) data from Planck 2018~\cite{Planck:2018vyg,Aghanim:2019ame} have been used. In particular, we have used the CMB temperature and polarization angular power spectra {\it plikTTTEEE+lowl+lowE}.

\item {\bf Lensing}: CMB Lensing reconstruction likelihood from Planck 2018 team~\cite{Aghanim:2018oex} has been considered. 
    
\item {\bf Baryon acoustic oscillations}:  Baryon acoustic oscillations (BAO) distance measurements from various astronomical surveys, such as  6dFGS~\cite{Beutler:2011hx}, SDSS-MGS~\cite{Ross:2014qpa}, and BOSS DR12~\cite{Alam:2016hwk} have been considered in the analysis. 
    
\item {\bf Pantheon}: We also include the Pantheon catalogue of the Type Ia Supernovae~\cite{Scolnic:2017caz}. 

\end{enumerate}

To constrain the interacting scenario, we make use of the 
modified \texttt{CosmoMC} package~\cite{Lewis:2002ah}. This package is freely available and supports Planck 2018 likelihood~\cite{Aghanim:2019ame}. Additionally, \texttt{CosmoMC} package is equipped with convergence diagnostic by Gelman and Rubin~\cite{Gelman:1992zz}. In \autoref{tab:priors} we display the priors on the free parameters of this model, that are, the baryon density $\Omega_b h^2$, the dark matter density assumed to be non-cold instead of cold $\Omega_{dm} h^2$, the optical depth $\tau$, the spectral index and the amplitude of the primordial scalar perturbations $n_s$ and $A_s$, the ratio of the sound horizon to the angular diameter distance $\theta_{MC}$, the equation of state of the non-cold DM $w_{dm}$, and the coupling parameter of the interaction function $Q (t)$ between the non-cold DM and the vacuum~$\xi$.

\begingroup                                                                                                                     
\begin{center}                                                                                                                  
\begin{table*}
\scalebox{0.65}
{                                                               
\begin{tabular}{ccccccccc}                                      
\hline\hline                                                                                                                    
Parameters & CMB & CMB+Lensing & CMB+BAO & CMB+Pantheon & CMB+Lensing+BAO+Pantheon\\ \hline

$\Omega_b h^2$ & $    0.02248_{-    0.00017-    0.00033}^{+    0.00017+    0.00035}$ & $    0.02249_{-    0.00016-    0.00032}^{+    0.00016+    0.00032}$  & $    0.02250_{-    0.00016-    0.00031}^{+    0.00016+    0.00032}$  & $    0.02247_{-    0.00017-    0.00032}^{+    0.00016+    0.00033}$ & $    0.02252_{-    0.00015-    0.00030}^{+    0.00015+    0.00031}$  \\

$\Omega_{dm} h^2$ & $    0.077_{- 0.056}^{+ 0.033} < 0.14$ & $    0.082_{-    0.036}^{+    0.055} < 0.14$  & $    0.104_{-    0.015-    0.040}^{+    0.023+    0.036}$  & $    0.109_{-    0.011-    0.024}^{+    0.013+    0.022}$  & $    0.112_{-    0.009-    0.020}^{+    0.011+    0.020}$  \\

$100\theta_{MC}$ & $    1.0436_{-    0.0036-    0.0046}^{+    0.0024+    0.0049}$ & $    1.0433_{-    0.0037-    0.0043}^{+    0.0020+    0.0051}$ & $    1.0419_{-    0.0014-    0.0021}^{+    0.0008+    0.0024}$  & $    1.04147_{-    0.00076-    0.0014}^{+    0.00065+    0.0015}$ & $    1.04137_{-    0.00066-    0.0012}^{+    0.00054+    0.0012}$ \\

$\tau$ & $    0.0534_{-    0.0074-    0.015}^{+    0.0075+    0.015}$ & $    0.0526_{-    0.0072-    0.014}^{+    0.0072+    0.015}$  & $    0.0541_{-    0.0075-    0.015}^{+    0.0077+    0.016}$   & $    0.0533_{-    0.0074-    0.015}^{+    0.0075+    0.015}$  & $    0.0539_{-    0.0071-    0.014}^{+    0.0070+    0.015}$ \\

$n_s$ & $    0.9628_{-    0.0046-    0.0088}^{+    0.0046+    0.0093}$ & $    0.9636_{-    0.0042-    0.0082}^{+    0.0042+    0.0082}$  & $    0.9637_{-    0.0042-    0.0083}^{+    0.0042+    0.0083}$   & $    0.9623_{-    0.0044-    0.0089}^{+    0.0044+    0.0088}$   & $    0.9646_{-    0.0039-    0.0078}^{+    0.0039+    0.0078}$   \\

${\rm{ln}}(10^{10} A_s)$ & $    3.046_{-    0.016-    0.030}^{+    0.015+    0.030}$ & $    3.043_{-    0.014-    0.028}^{+    0.014+    0.029}$ & $    3.046_{-    0.016-    0.031}^{+    0.016+    0.032}$   & $    3.046_{-    0.015-    0.030}^{+    0.015+    0.031}$  & $    3.044_{-    0.014-    0.027}^{+    0.014+    0.028}$  \\

$w_{dm}$ & $ 0.00122_{-    0.00097}^{+    0.00053} <0.0025$ & $    0.00112_{- 0.00091}^{+0.00047} < 0.0023$ & $    0.00115_{-    0.00098}^{+    0.00045} < 0.0025$  & $    0.00123_{-    0.00099}^{+    0.00052} < 0.0031$  & $    0.00108_{-    0.00096}^{+    0.00040} < 0.0023$    \\

$\xi$ & $    0.11_{-    0.07-    0.20}^{+    0.14+    0.17}$ & $    0.10_{-    0.08-    0.19}^{+    0.14+    0.17}$  & $    0.048_{-    0.064-    0.11}^{+    0.053+    0.12}$ & $    0.036_{-    0.036-    0.072}^{+    0.036+    0.072}$ & $    0.025_{-    0.032-    0.063}^{+    0.032+    0.063}$ \\

$\Omega_{m}$ & $    0.21_{-    0.14-    0.16}^{+    0.07+    0.19}$ & $    0.22_{-    0.14-    0.17}^{+    0.08+    0.18}$ & $    0.270_{-    0.046-    0.11}^{+    0.057+    0.10}$  & $    0.285_{-    0.032-    0.066}^{+    0.033+    0.066}$  & $    0.289_{-    0.028-    0.055}^{+    0.028+    0.055}$  \\

$\sigma_8$ & $    1.40_{-    0.80-    0.9}^{+    0.25+    1.4}$ & $    1.32_{-    0.73-    0.9}^{+    0.18+    1.5}$ & $    0.95_{-    0.20-    0.28}^{+    0.08+    0.36}$  & $    0.898_{-    0.097-    0.16}^{+    0.062+    0.17}$ & $    0.869_{-    0.082-    0.13}^{+  0.050+    0.14}$  \\

$H_0$ [Km/s/Mpc] & $   70.6_{-    2.4-    6.0}^{+    4.3+    5.2}$ & $   70.3_{-    2.8-    5.6}^{+    4.3+    5.4}$ & $   68.8_{-    1.6-    2.7}^{+    1.3+    3.0}$ & $   68.2_{-    1.0-    2.1}^{+    1.0+    2.1}$  & $   68.33_{-    0.81-    1.5}^{+    0.82+    1.6}$  \\

$S_8$ & $    1.01_{-    0.23-    0.27}^{+    0.08+    0.41}$ & $    0.98_{-    0.21-    0.24}^{+    0.06+    0.40}$ & $    0.880_{-    0.071-    0.11}^{+    0.032+    0.13}$  & $    0.869_{-    0.038-    0.066}^{+    0.028+    0.070}$ & $    0.849_{-    0.031-    0.052}^{+    0.022+    0.056}$ \\

$r_{\rm{drag}}$ [Mpc] & $  146.80_{-    0.33-    0.65}^{+    0.33+    0.64}$ & $  146.91_{-    0.29-    0.57}^{+    0.29+    0.56}$ & $  146.90_{-    0.31-    0.63}^{+    0.31+    0.61}$ & $  146.78_{-    0.33-    0.65}^{+    0.33+    0.64}$  & $  146.98_{-    0.27-    0.53}^{+    0.27+    0.53}$ \\

$\Omega_{m} h^2$ & $    0.100_{-    0.053-    0.068}^{+    0.037+    0.071}$ & $    0.105_{-    0.036-    0.070}^{+    0.055+    0.067}$ & $    0.127_{-    0.015-    0.040}^{+    0.023+    0.037}$  & $    0.132_{-    0.011-    0.024}^{+    0.013+    0.022}$  & $    0.135_{-    0.0092-    0.020}^{+    0.011+    0.020}$ \\

\hline\hline                                                                                                                    
\end{tabular} }                                                                
\caption{68\% and 95\% CL constraints on the \textbf{IWDM} scenario are presented for CMB, CMB+Lensing, CMB+BAO, CMB+Pantheon and CMB+Lensing+BAO+Pantheon data sets.  }
\label{tab:IVS}                   
\end{table*}                                     
\end{center}                                                                                                                    
\endgroup                                                                                                                       
\begin{figure*}
\includegraphics[width=0.9\textwidth]{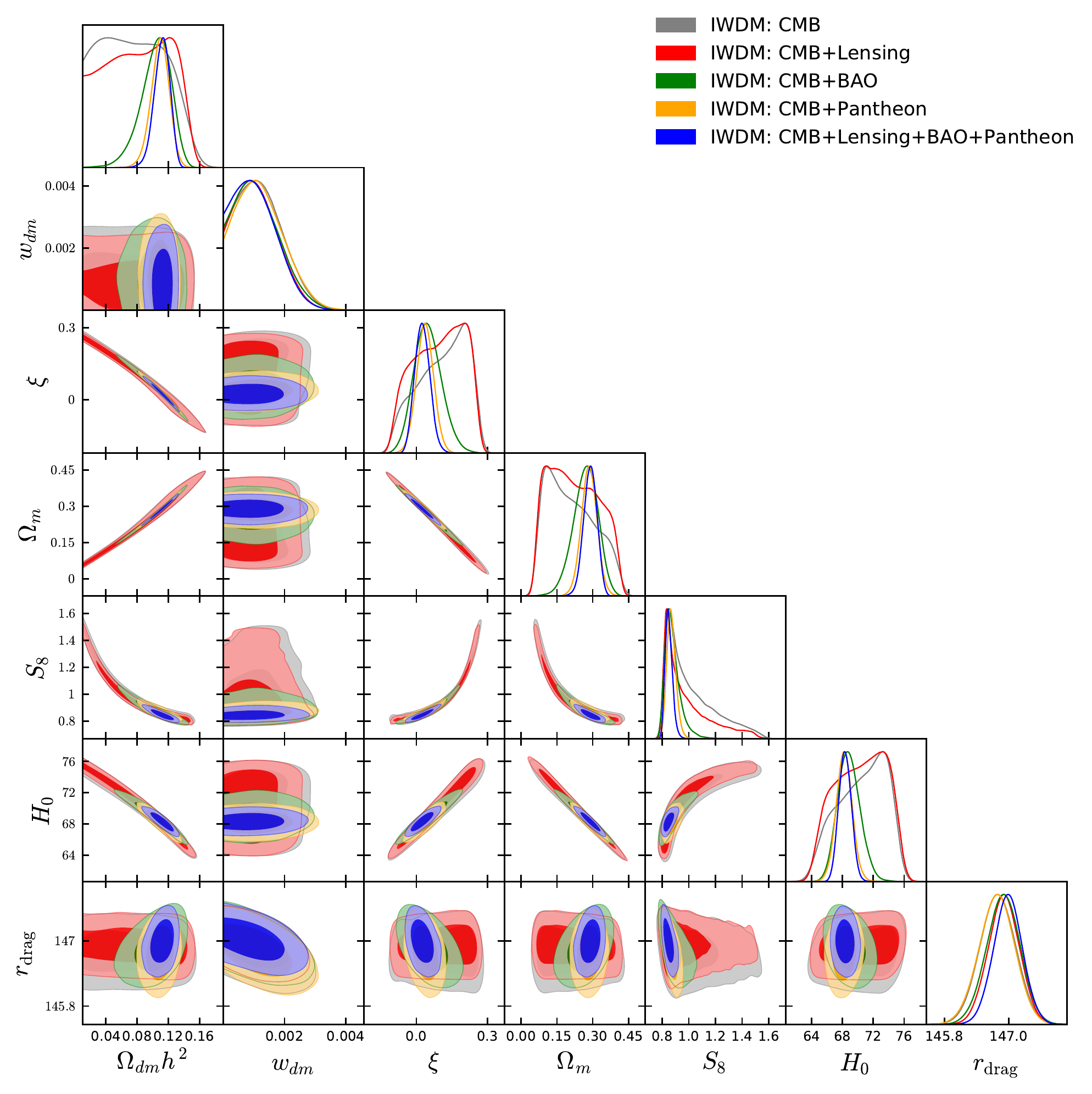}
\caption{We show the one dimensional posterior distributions and two dimensional joint contours for the most relevant parameters of the \textbf{IWDM} scenario using several observational data sets, namely, CMB, CMB+Lensing, CMB+BAO, CMB+Pantheon and CMB+Lensing+BAO+Pantheon.   }
\label{fig:interacting-vacuum-DM}
\end{figure*}
\begin{figure*}
     \includegraphics[width=0.48\textwidth]{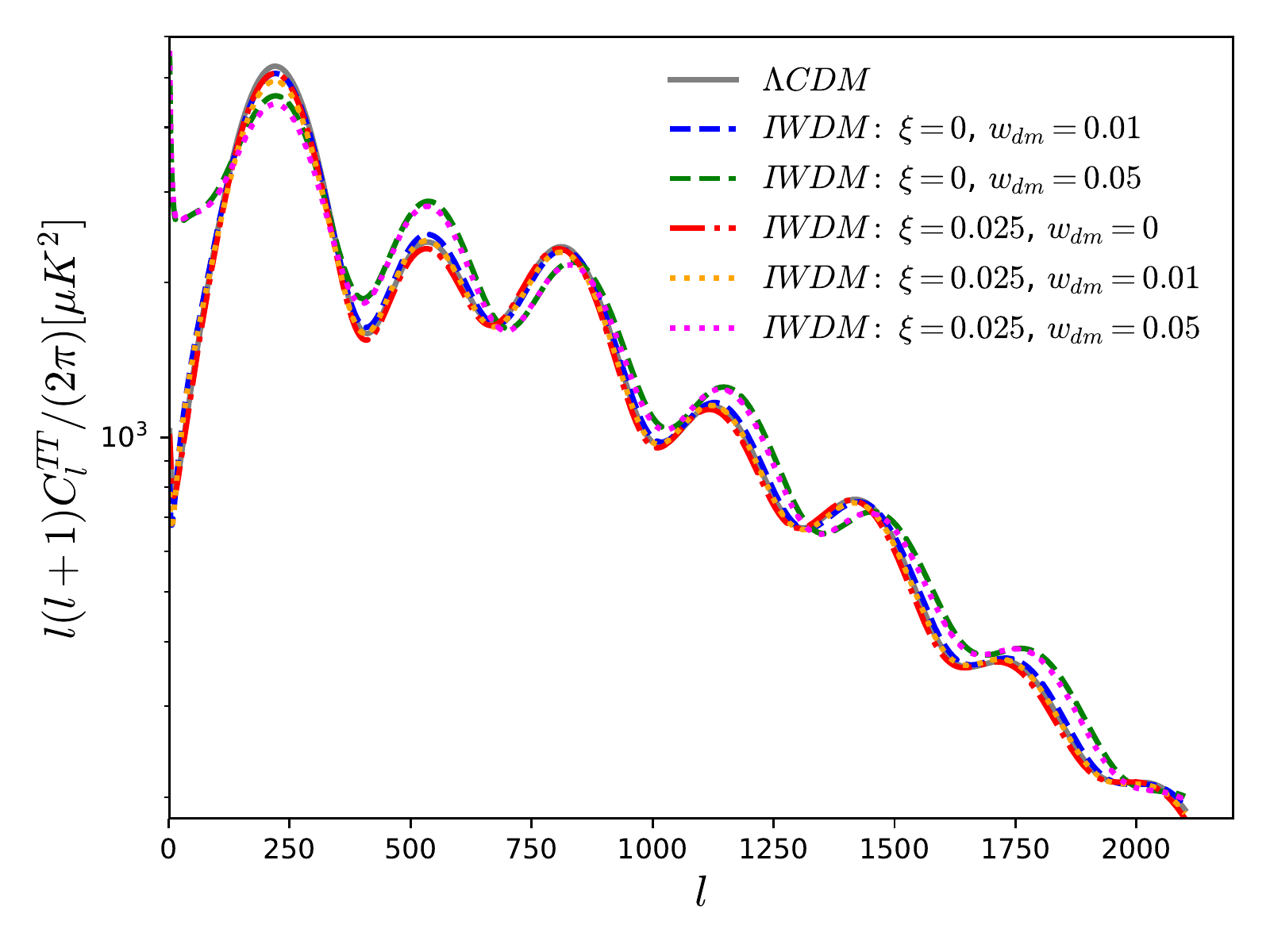}
    \includegraphics[width=0.48\textwidth]{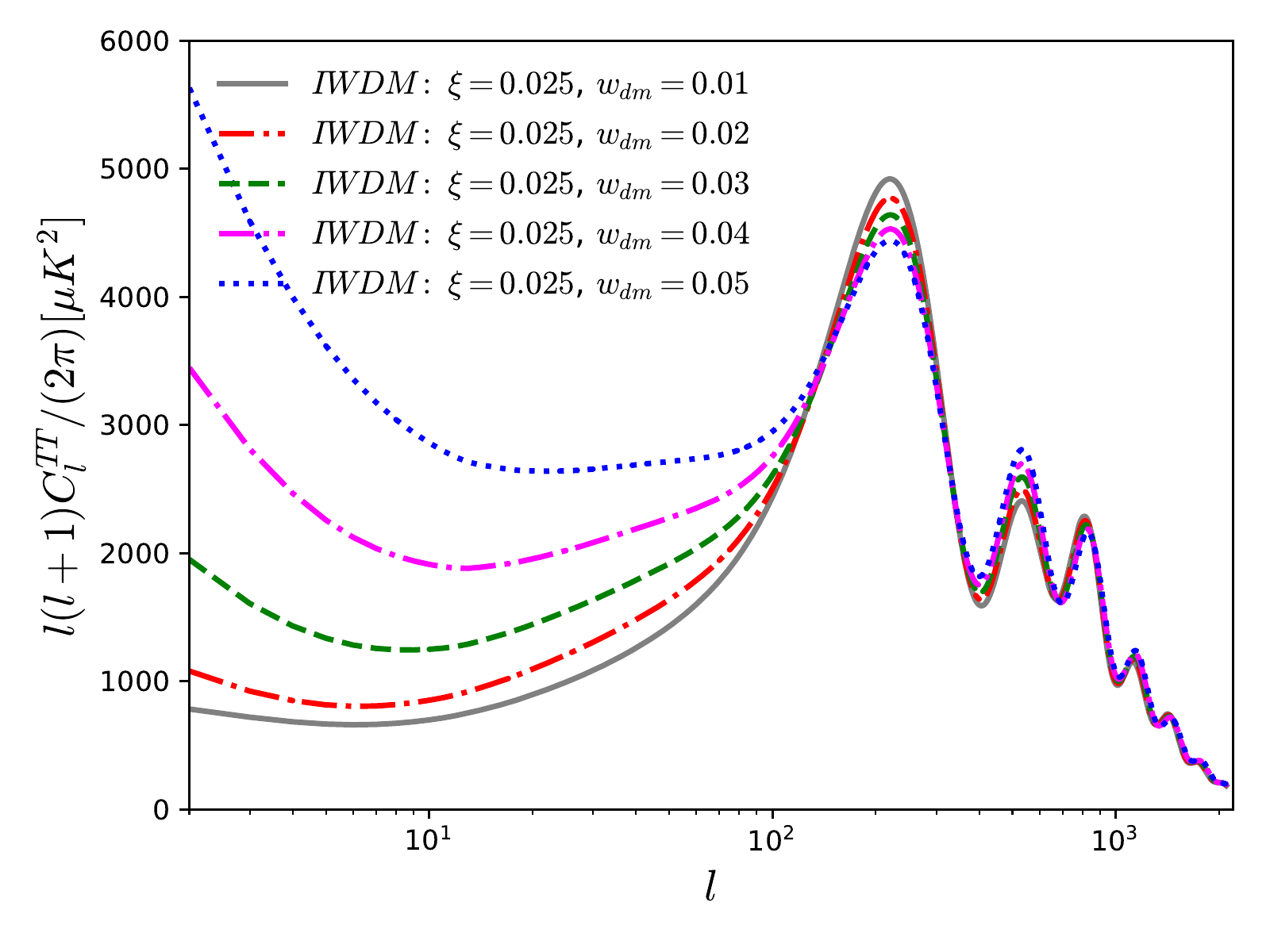}
    \includegraphics[width=0.48\textwidth]{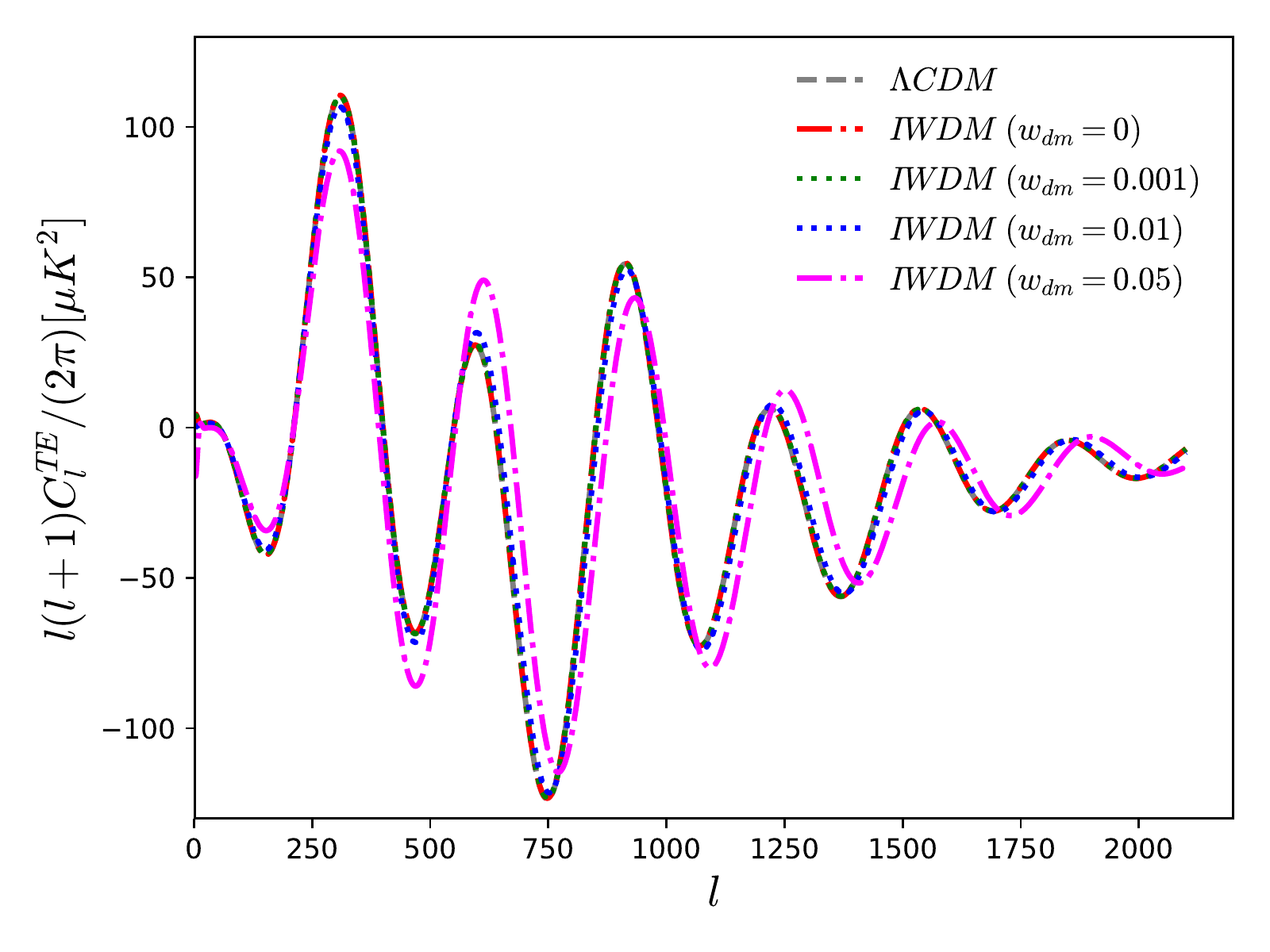}
    \includegraphics[width=0.48\textwidth]{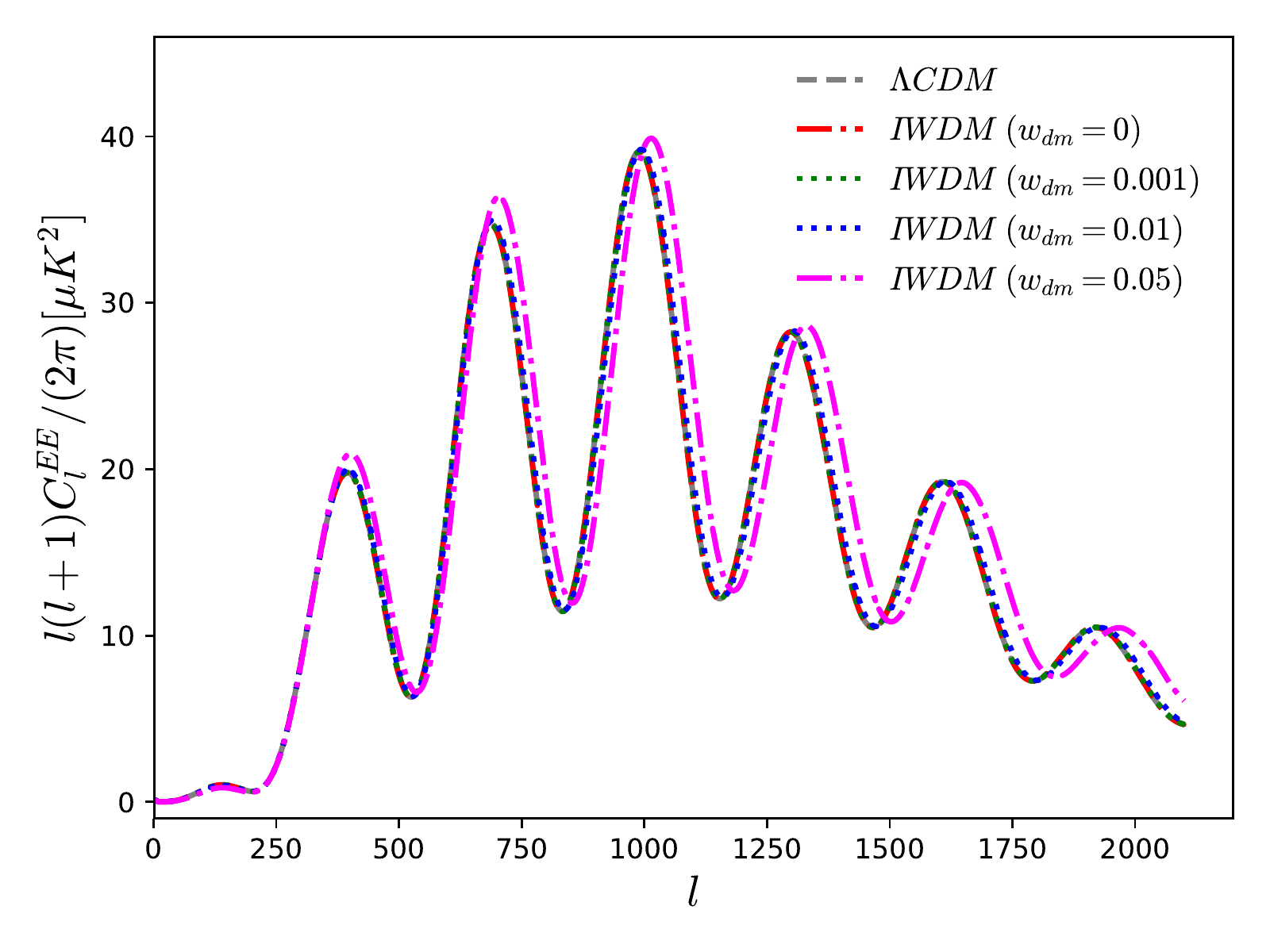}
    \caption{We present the CMB TT, TE, and EE spectra for the \textbf{IWDM} scenario for various values of the DM equation-of-state parameter $w_{dm}$ in comparison with the $\Lambda$CDM scenario. Regarding the CMB TT spectra, for the upper right panel, we have used the logarithmic scale to more accurately show the effects in the low multipole regime, that are not so clear in the upper left panel, where we instead prefer a linear scale.  }
    \label{fig:CMB-spectra}
\end{figure*}
\begin{figure}
    \centering
    \includegraphics[width=0.45\textwidth]{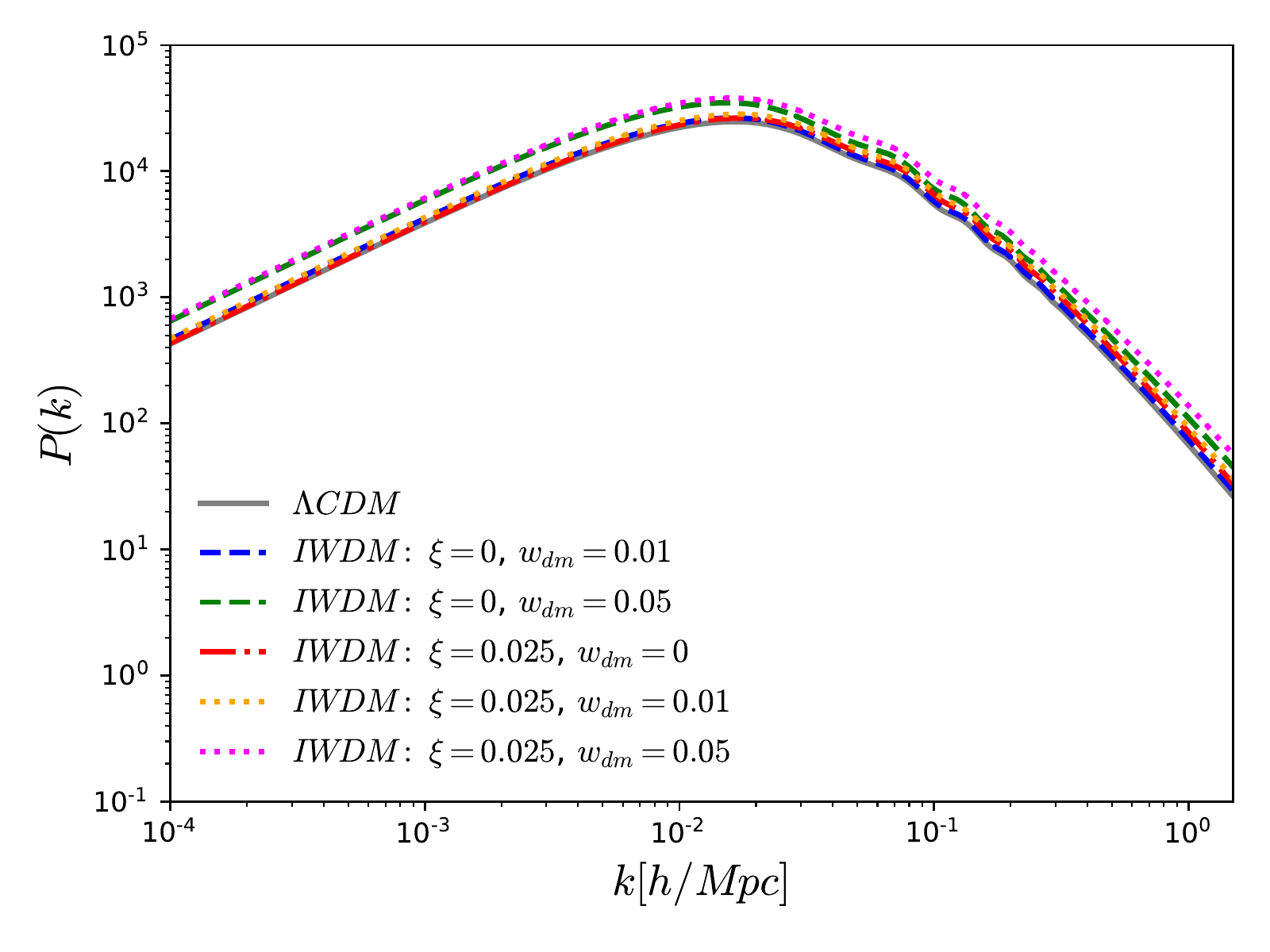}
    \caption{We show the matter power spectra for the \textbf{IWDM} scenario for various values of the DM equation-of-state parameter $w_{dm}$ in comparison with the $\Lambda$CDM scenario. }
    \label{fig:matter-power-spectra}
\end{figure}

\begingroup                                                                                                                     
\squeezetable                                                                                                                   
\begin{center}                                                                                                                  
\begin{table*}                                                                                                                   
\begin{tabular}{cccccccc}                                                                                                            
\hline\hline                                                                                                                    
Parameters & CMB & CMB+BAO & CMB+Pantheon \\ \hline

$\Omega_b h^2$ & $    0.02233_{-    0.00015-    0.00029}^{+    0.00015+    0.00030}$ & $    0.02233_{-    0.00014-    0.00029}^{+    0.00014+    0.00028}$  & $    0.02233_{-    0.00015-    0.00029}^{+    0.00015+    0.00029}$ \\

$\Omega_{dm} h^2$ & $    0.1169_{-    0.0020-    0.0065}^{+    0.0037+    0.0054}$ &  $    0.1186_{-    0.0011-    0.0022}^{+    0.0011+    0.0022}$  & $    0.1178_{-    0.0017-    0.0042}^{+    0.0022+    0.0038}$  \\

$100\theta_{MC}$ & $    1.04111_{-    0.00039-    0.00068}^{+    0.00035+    0.00075}$ &  $    1.04101_{-    0.00029-    0.00058}^{+    0.00029+    0.00059}$  & $    1.04106_{-    0.00032-    0.00063}^{+    0.00032+    0.00062}$ \\

$\tau$ & $    0.0551_{-    0.0075-    0.015}^{+    0.0076+    0.016}$ & $    0.0546_{-    0.0081-    0.015}^{+    0.0073+    0.016}$  & $    0.0547_{-    0.0076-    0.015}^{+    0.0076+    0.016}$  \\

$n_s$ & $    0.9674_{-    0.0048-    0.0092}^{+    0.0048+    0.0098}$ & $    0.9658_{-    0.0039-    0.0075}^{+    0.0038+    0.0076}$   & $    0.9665_{-    0.0042-    0.0083}^{+    0.0042+    0.0081}$  \\

${\rm{ln}}(10^{10} A_s)$ & $    3.048_{-    0.016-    0.031}^{+    0.016+    0.033}$ & $    3.046_{-    0.016-    0.031}^{+    0.016+    0.032}$ & $    3.046_{-    0.016-    0.032}^{+    0.016+    0.032}$ \\

$w_{dm}$ &  $ < 0.0013 < 0.0030 $ &  $ < 0.00072 < 0.0012  $  & $ < 0.0010 < 0.0020 $ \\

$\Omega_m$ & $    0.290_{-    0.014-    0.047}^{+    0.027+    0.038}$ & $    0.3030_{-    0.0075-    0.016}^{+    0.0082+    0.015}$  & $    0.297_{-    0.012-    0.030}^{+    0.017+    0.027}$ \\

$\sigma_8$ & $    0.835_{-    0.026-    0.035}^{+    0.012+    0.045}$ & $    0.824_{-    0.014-    0.022}^{+    0.010+    0.024}$ & $    0.829_{-    0.018-    0.028}^{+    0.012+    0.032}$  \\

$H_0$ & $   69.5_{-    2.4-    3.2}^{+    1.0+    4.3}$ & $   68.36_{-    0.70-    1.2}^{+    0.58+    1.3}$   & $   68.88_{-    1.48-    2.2}^{+    0.90+    2.6}$  \\

$S_8$ & $    0.821_{-    0.018-    0.039}^{+    0.020+    0.036}$ & $    0.828_{-    0.013-    0.025}^{+    0.013+    0.026}$ & $    0.825_{-    0.016-    0.031}^{+    0.016+    0.031}$  \\

$r_{\rm{drag}}$ & $  147.17_{-    0.31-    0.61}^{+    0.31+    0.62}$ & $  147.10_{-    0.26-    0.51}^{+    0.26+    0.49}$ & $  147.13_{-    0.28-    0.56}^{+    0.28+    0.56}$ \\

$\Omega_m h^2$ & $    0.1398_{-    0.0019-    0.0065}^{+    0.0037+    0.0053}$ & $    0.1416_{-    0.0011-    0.0022}^{+    0.0011+    0.0021}$  & $    0.1408_{-    0.0016-    0.0041}^{+    0.0022+    0.0037}$ \\


\hline\hline                                                                                    
\end{tabular}                                                                 
\caption{Summary of the observational constraints at 68\% and 95\% CL on the first simplest extension of the $\Lambda$CDM cosmological model (Non-interacting scenario 1) where the dark energy equation of state, $w_{de}  = -1$, and the CDM part has been replaced by the non-cold DM characterized by $w_{dm}$ using three important data sets, namely, CMB, CMB+BAO, and CMB+Pantheon.  }
\label{tab:IA}                                                                           
\end{table*}                                                                               
\end{center}                                           
\endgroup
\begin{figure*}
     \includegraphics[width=0.85\textwidth]{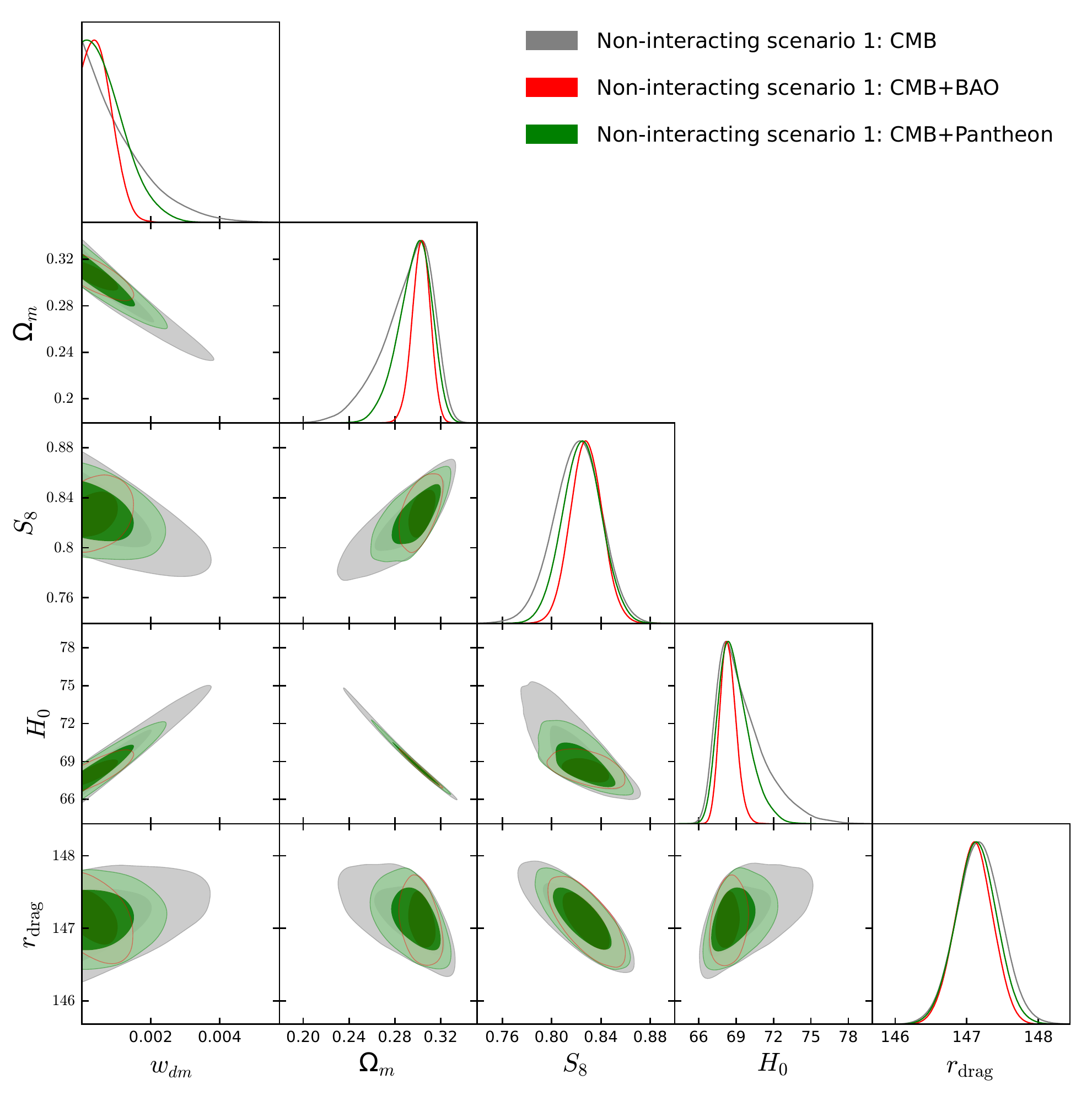}
    \caption{One dimensional posterior distributions and two dimensional joint contours for the most relevant parameters of the Non-interacting scenario 1 using three important data sets,  namely, CMB, CMB+BAO, CMB+Pantheon. }
    \label{fig:non-interacting-1}
\end{figure*}
\begingroup                                                                                                                     
\squeezetable                                                                                                                   
\begin{center}                                                                                                                  
\begin{table*}                                                                                                                   
\begin{tabular}{cccccccc}                                                                                                            
\hline\hline                                                                                                                    
Parameters & CMB & CMB+BAO & CMB+Pantheon  \\ \hline

$\Omega_b h^2$ & $    0.02236_{-    0.00015-    0.00030}^{+    0.00015+    0.00030}$ & $    0.02232_{-    0.00015-    0.00030}^{+    0.00015+    0.00029}$  & $    0.02233_{-    0.00015-    0.00029}^{+    0.00015+    0.00029}$ \\

$\Omega_{dm} h^2$ & $    0.1166_{-    0.0020-    0.0066}^{+    0.0037+    0.0055}$ & $    0.1176_{-    0.0016-    0.0045}^{+    0.0025+    0.0040}$  & $    0.1176_{-    0.0018-    0.0057}^{+    0.0032+    0.0048}$   \\

$100\theta_{MC}$ & $    1.04115_{-    0.00039-    0.00069}^{+    0.00035+    0.00075}$ &  $    1.04108_{-    0.00032-    0.00064}^{+    0.00032+    0.00064}$  & $    1.04107_{-    0.00033-    0.00065}^{+    0.00034+    0.00070}$  \\

$\tau$ & $    0.0546_{-    0.0076-    0.015}^{+    0.0077+    0.016}$ &  $    0.0548_{-    0.0075-    0.015}^{+    0.0075+    0.016}$  & $    0.0548_{-    0.0078-    0.015}^{+    0.0073+    0.016}$ \\

$n_s$ & $    0.9679_{-    0.0048-    0.0092}^{+    0.0048+    0.0097}$ & $    0.9666_{-    0.0043-    0.0083}^{+    0.0043+    0.0083}$ & $    0.9665_{-    0.0046-    0.0090}^{+    0.0046+    0.0093}$ \\

${\rm{ln}}(10^{10} A_s)$ & $    3.046_{-    0.016-    0.032}^{+    0.016+    0.033}$ &  $    3.047_{-    0.016-    0.032}^{+    0.016+    0.032}$  & $    3.047_{-    0.016-    0.031}^{+    0.016+    0.032}$  \\

$w_{dm}$ & $  < 0.0013 < 0.0030 $ & $  < 0.0011 < 0.0022 $  & $ < 0.0011 < 0.0025 $  \\

$w_{de}$ & $   -1.52_{-    0.30-    0.43}^{+    0.18+    0.48}$ & $   -0.966_{-    0.072-    0.14}^{+    0.073+    0.15}$  & $   -1.004_{-    0.046-    0.081}^{+    0.040+    0.088}$  \\

$\Omega_m$ & $    0.186_{-    0.050-    0.060}^{+    0.016+    0.089}$ & $    0.305_{-    0.011-    0.023}^{+    0.012+    0.022}$  & $    0.294_{-    0.012-    0.031}^{+    0.017+    0.028}$  \\

$\sigma_8$ & $    0.990_{-    0.042-    0.15}^{+    0.092+    0.12}$ & $    0.820_{-    0.019-    0.037}^{+    0.019+    0.038}$ & $    0.832_{-    0.018-    0.032}^{+    0.015+    0.034}$ \\

$H_0$ & $   88_{-    5-   17}^{+   12+   14}$ & $   67.9_{-    1.6-    2.9}^{+    1.4+    2.9}$ & $   69.1_{-    1.4-    2.5}^{+    1.1+    2.7}$ \\

$S_8$ & $    0.769_{-    0.034-    0.050}^{+    0.021+    0.058}$ & $    0.827_{-    0.013-    0.026}^{+    0.014+    0.027}$  & $    0.824_{-    0.017-    0.034}^{+    0.017+    0.032}$ \\

$r_{\rm{drag}}$ & $  147.20_{-    0.31-    0.60}^{+    0.32+    0.63}$ & $  147.14_{-    0.29-    0.56}^{+    0.29+    0.55}$  & $  147.13_{-    0.31-    0.61}^{+    0.30+    0.61}$ \\

$\Omega_m h^2$ & $    0.1396_{-    0.0019-    0.0066}^{+    0.0037+    0.0054}$ & $    0.1405_{-    0.0016-    0.0045}^{+    0.0025+    0.0040}$ & $    0.1405_{-    0.0017-    0.0057}^{+    0.0032+    0.0047}$  \\


\hline\hline                                                                                                                    
\end{tabular}                                                                                                                   
\caption{Summary of the observational constraints at 68\% and 95\% CL on the second extension of the $\Lambda$CDM cosmological model (Non-interacting scenario 2) where the dark energy equation of state, $w_{de}$, has been kept free and the CDM part has been replaced by the non-cold DM characterized by $w_{dm}$ using three important data sets, namely, CMB, CMB+BAO, and CMB+Pantheon.}
\label{tab:IB}                            
\end{table*}                                                                                   
\end{center}                                             
\endgroup                                          
\begin{figure*}
     \includegraphics[width=0.85\textwidth]{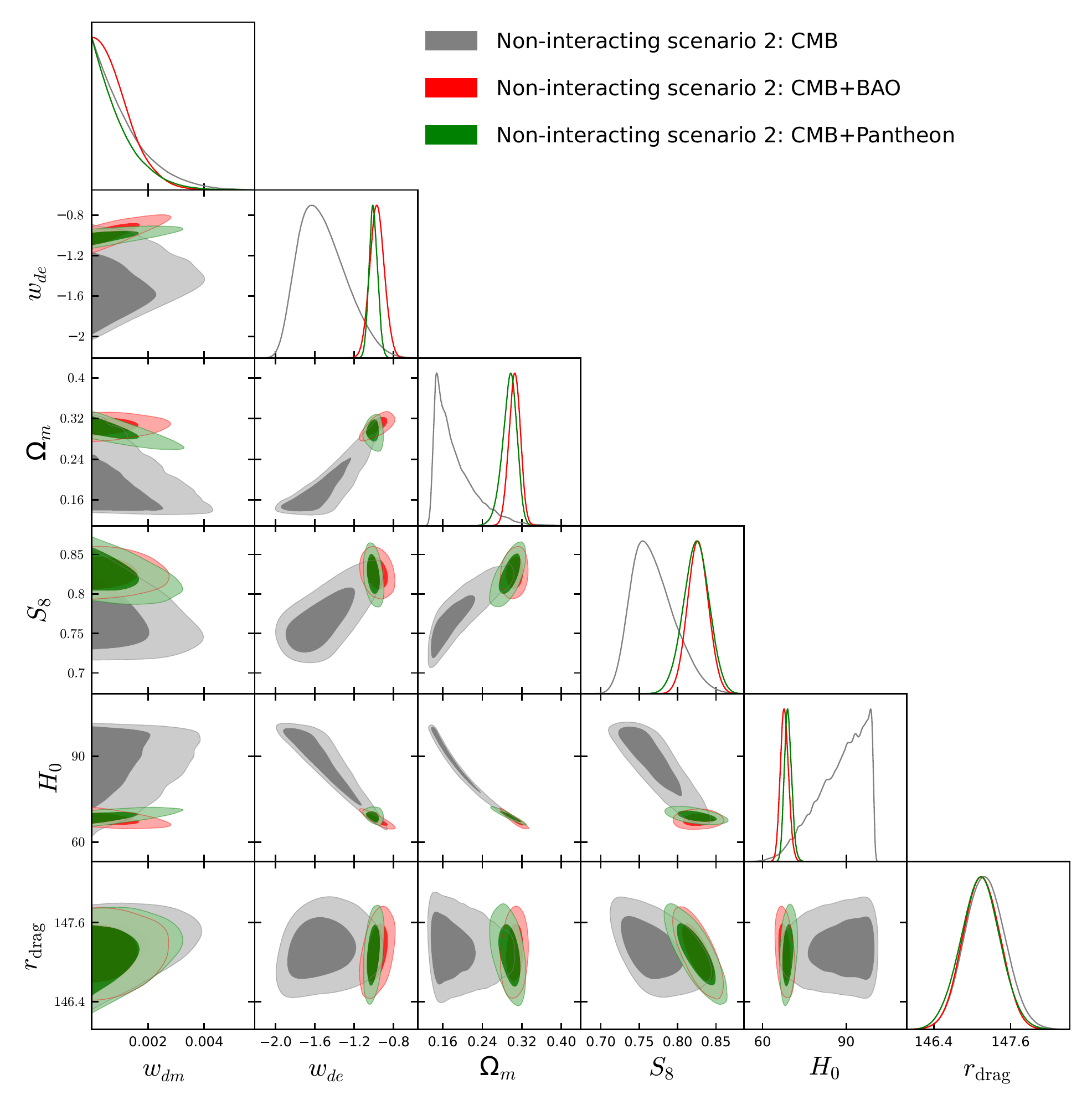}
    \caption{One dimensional posterior distributions and two dimensional joint contours for the most relevant parameters of the Non-interacting scenario 2 using three important data sets, namely, CMB, CMB+BAO, CMB+Pantheon. }
    \label{fig:non-interacting-2}
\end{figure*}

\section{Results}
\label{sec-results}

In \autoref{tab:IVS} and \autoref{fig:interacting-vacuum-DM} we summarize the constraints on the interacting scenario for various cosmological probes, namely, CMB, CMB+Lensing, CMB+BAO, CMB+Lensing+BAO+Pantheon.  The two key free parameters of this scenario are the coupling parameter $\xi$ and the eos of DM, $w_{dm}$.

We begin by discussing the constraints on the interacting scenario for CMB alone and then gradually investigate the effects of other cosmological probes as they are added to CMB. For CMB alone, we find an indication for a non-cold DM at more than 68\% CL ($w_{dm} = 0.00122_{-    0.00097}^{+    0.00053}$ at 68\% CL). This supports the decay of the non-cold DM in vacuum (i.e. an energy transfer from non-cold DM to vacuum) because we also notice an indication for a non-zero coupling at more than 68\% CL ($\xi = 0.11_{-    0.07}^{+  0.14}$ at 68\% CL). Due to the transfer of energy from the non-cold DM to the vacuum, we detect a smaller amount of DM which is clearly reflected by the estimated value of the matter density parameter $\Omega_{m} =  0.21_{-    0.14}^{+    0.07}$ at 68\% CL (for CMB alone) compared to the $\Lambda$CDM value obtained from Planck~\cite{Planck:2018vyg}. It is worth noting that given
only an upper limit on $\Omega_{dm} h^2$ in the CMB only case due to the interaction, and given the strong correlation with many parameters of the model (see \autoref{fig:interacting-vacuum-DM}), this produces highly non-Gaussian posteriors for all of them.  Furthermore, this results in a higher value of the Hubble constant ($H_0 = 70.6_{-    2.4}^{+  4.3}$~km/s/Mpc at 68\% CL for CMB alone) than the $H_0$ value obtained by Planck alone (within the $\Lambda$CDM paradigm)~\cite{Planck:2018vyg} due to the anti-correlation existing between $H_0$ and $\Omega_{m}$, 
see \autoref{fig:interacting-vacuum-DM}.  Actually, the positivity of the coupling parameter (i.e. $\xi >0$) implies an energy flow from the non-cold DM to the vacuum sector which produces as effect a lower value of $\Omega_m$ and, consequently, a higher value of $H_0$ is returned.  We also notice a very large error bars on $H_0$ due to an increase in the volume of the parameter space.  
As a result, the $H_0$ constraint is significantly relaxed due to such large error bars, and as a result,
the $5\sigma$ tension on $H_0$ between Planck (within the $\Lambda$CDM paradigm)~\cite{Planck:2018vyg} and the SH0ES (Supernovae and $H_0$ for the Equation of State of dark energy) collaboration ($H_0= 73.04 \pm 1.04$ km/s/Mpc)~\cite{Riess:2021jrx} is reduced to $1 \sigma$ for this case. It should be noted that this alleviation of the $H_0$ tension depends on the shift in the mean value of $H_0$ and 
 mainly on the increased error bars, caused by the increase in the volume of the parameter space.  The equation-of-state of the non-cold DM, $w_{dm}$, plays no role in alleviating the $H_0$ tension, it is the interaction between the dark components. As seen from \autoref{fig:interacting-vacuum-DM}, no correlation (anti-correlation)  is observed between $H_0$ and $w_{dm}$ unlike the existing correlation (anti-correlation) between $H_0$ and $\xi$ ($\Omega_m$).

The inclusion of CMB Lensing to the CMB temperature and polarization data (i.e. CMB+Lensing) offers results almost similar to those of the CMB alone. For example, also here we obtain an upper limit on $\Omega_{dm} h^2$ because of the interaction, which is reflected on highly non-Gaussian posteriors for the parameters correlated with it.
This analysis also indicates an existence of non-cold dark matter at more than 68\% CL ($w_{dm} = 0.00112_{-    0.00091}^{+    0.00047}$ at 68\% CL) and of a non-zero coupling in the dark sector at more than 68\% CL ($\xi = 0.10_{- 0.08}^{+ 0.14}$ at 68\% CL), alleviating the $H_0$ tension within $1\sigma$. Similar to the CMB alone case as described above, the high value of $H_0$ is due to the interaction between the non-cold DM and the vacuum sector but not to the equation-of-state in the non-cold DM, $w_{dm}$.  Again we note that the alleviation on the $H_0$ tension is driven mainly by its very high error bars caused by an increase in the volume of the parameter space rather than the actual shift in the mean value of $H_0$. Analogous to the CMB alone case, here too, the equation-of-state of the non-cold DM, $w_{dm}$, plays no role in alleviating the $H_0$ tension, but this is due to the interaction between the dark components.

When BAO data are added to CMB, we observe some changes in the constraints. In particular, we observe that the mean values of the cosmological parameters are shifted towards the $\Lambda$CDM values, while an indication for a non-cold DM is still retained at more than 68\% CL ($w_{dm} = 0.00115_{-    0.00098}^{+    0.00045}$ at 68\% CL for CMB+BAO). Unlike the previous two cases, here the coupling parameter is back in agreement with $\xi = 0$ within $1\sigma$ and the mean value of the coupling parameter is decreased to $\xi = 0.048_{- 0.064}^{+    0.053}$ at 68\% CL. Hence, we find no strong evidence of $\xi \neq 0$ for this dataset since both positive and negative values are equally allowed in 1 standard deviation. Due to the positive correlation between $\xi$ and $H_0$, the reduction in the mean value in $\xi$ implies a reduction in the mean value of $H_0 = 68.8_{-    1.6}^{+    1.3}$ km/s/Mpc at 68\% CL and consequently we find a higher value of $\Omega_{m} = 0.270_{-    0.046 }^{+    0.057}$ than the previous two constraints obtained in CMB only and CMB+Lensing cases. We note a shift in the mean value of $H_0$ away from the SH0ES measure~\cite{Riess:2021jrx}. Finally, we note that the $H_0$ tension between Planck (within the $\Lambda$CDM paradigm)~\cite{Planck:2018vyg} and the SH0ES  collaboration~\cite{Riess:2021jrx} is however reduced to $2.6\sigma$  and this reduction is mainly influenced by the large error bars in $H_0$, not due to the 
presence of $w_{dm}$.

The inclusion of Pantheon to the CMB shifts the mean values of the parameters to the $\Lambda$CDM values but the indication for a non-cold DM ($w_{dm} \neq 0$ at more than 68\% CL) is kept: $w_{dm} = 0.00123_{-    0.00099}^{+    0.00052}$ at 68\% CL. Similar to the CMB+BAO analysis, we find no clear evidence of a non-zero coupling. The correlation between $\xi$ and $H_0$ remains the same as observed in the previous cases and also the matter density parameter increases. Finally, we note that 
the $H_0$ tension between Planck (within the $\Lambda$CDM paradigm)~\cite{Planck:2018vyg} and the SH0ES collaboration~\cite{Riess:2021jrx} is reduced to 
$3.4\sigma$ for this dataset combination. The reduction in the $H_0$ tension in this case is not comparable to the CMB and CMB+Lensing analyses since we  notice (i) a shift in the mean value of $H_0$ away from the SH0ES estimate~\cite{Riess:2021jrx}, and (ii) the error bars on $H_0$ are stronger than both CMB and CMB+Lensing analyses. Similar to the earlier cases, we confirm that $w_{dm}$ does not affect $H_0$.

For the most complete dataset, i.e.  CMB+Lensing+BAO+Pantheon, we get the most stringent constraints on the interacting scenario. 
We still find a non-zero value of the DM eos at more than 68\% CL ($w_{dm} = 0.00108_{-    0.00096}^{+    0.00040}$ at 68\% CL). Similarly to the previous case (i.e. CMB+Pantheon), we find no evidence for $\xi \neq 0$ and the mean value of $\xi$ is further reduced with respect to CMB+Pantheon. The Hubble constant turns out to be $H_0 =  68.33_{-    0.81}^{+    0.82}$ km/s/Mpc at 68\% CL, which in any case reduces the $5\sigma$ tension between Planck (within the  $\Lambda$CDM paradigm)~\cite{Planck:2018vyg} and the SH0ES  collaboration~\cite{Riess:2021jrx} down to $3.6 \sigma$.  We note that the slight increase in the mean value of $H_0$ with respect to the Planck's estimate within the $\Lambda$CDM paradigm~\cite{Planck:2018vyg} is solely responsible for the energy flow from the non-cold DM sector to the vacuum  sector (i.e. $\xi > 0$) since, due to the existing anti-correlation between $H_0$ and $\Omega_m$ as explained previously, we have a slight increase in the mean value of $H_0$. However, the dark matter equation-of-state, $w_{dm}$, does not affect $H_0$ in any way.

We also analyze the impacts of the \textbf{IWDM} scenario on the CMB  power spectra  and matter power spectra  for  different values of $w_{dm}$ as well as the coupling parameter $\xi$. In \autoref{fig:CMB-spectra} we plot the CMB TT, TE, and EE spectra by setting the cosmological parameters to the best fit values of the parameters obtained from CMB+Lensing+BAO+Pantheon data analysis, with a fixed value of $\xi = 0.025$\footnote{Note that \textbf{IWDM} with $\xi = 0$ and $w_{dm} = 0.01$ or $0.05$ as kept in the upper left plot of \autoref{fig:CMB-spectra} refers to the non-interacting scenario. }. We then consider different values $w_{dm}$, starting from $w_{dm} = 0$, i.e. the usual interacting cold DM scenario, up to $0.05$ in different combinations, together with the spectra representing the non-interacting $\Lambda$CDM cosmological model as the reference scenario. 
Focusing on the CMB TT spectra, we see that the value of the equation-of-state parameter of the non-cold DM, i.e. $w_{dm}$, affects the entire CMB spectrum and this effect increases for enhancement of $w_{dm}$. In particular, as $w_{dm}$ increases, the amplitude of the first acoustic peak, and in general, the amplitude of the odd peaks, in the CMB TT and TE spectra is suppressed with respect to the non-interacting standard $\Lambda$CDM model (solid grey curve) and the interacting cold DM (red dashed curve). However, the amplitude of the second acoustic peak, and in general, the amplitude of the even peaks, increases for increasing values of $w_{dm}$. In other words, we see changes in the ratio of odd and even numbered peak heights of the TT (and TE) power spectrum. This effect resembles the one obtained by decreasing the baryon density, because both of them are weakening the gravity: the asymmetry of the oscillations is altered, and a reduction of the compression peaks (odd) compared to the rarefaction peaks (even) is observed.  
We also see that the amplitude of the ISW plateau in the low multipole regime ($\ell \leq 100$) of the CMB TT spectrum increases significantly by enhancing $w_{dm}$ with respect to the $\Lambda$CDM and interacting cold DM scenarios, indicating a dominance of the DE component.
Finally, in \autoref{fig:matter-power-spectra} we show the increase of the amplitude of the matter power spectrum as the $w_{dm}$ value increases.

\subsection{\textbf{IWDM} and the non-interacting scenarios}

After presenting the results on the \textbf{IWDM} scenario, we now proceed to investigate the physics of the non-interacting scenarios, that means we aim to investigate the dark sectors physics when $Q$ is absent in the picture. We consider two very well known extensions of the standard $\Lambda$CDM cosmological scenario, namely, (i) the cosmological scenario with $ w_{de} = -1$ being the dark energy equation of state but the DM equation-of-state $w_{dm}$  is constant (labeled as Non-interacting scenario 1), and (ii) the cosmological scenario where the dark energy equation-of-state $w_{de}$ and the DM equation-of-state $w_{dm}$ are constants  (labeled as Non-interacting scenario 2). We note that Non-interacting scenario 1 has ``seven free parameters'' \{$\Omega_{b} h^2$,  $\Omega_{dm} h^2$,  $\tau$,  $n_s$ $\log[10^{10}A_{s}]$, $100\theta_{MC}$,  $w_{dm}$\} and the Non-interacting scenario 2 has ``eight  free parameters'' \{$\Omega_{b} h^2$,  $\Omega_{dm} h^2$,  $\tau$,  $n_s$ $\log[10^{10}A_{s}]$, $100\theta_{MC}$,  $w_{dm}$, $w_{de}$\}. We constrained both the scenarios considering three effective data sets, namely CMB, CMB+BAO and CMB+Pantheon\footnote{While it is possible to perform various combinations as presented in Table~\ref{tab:IVS}, however the intrinsic nature of the cosmological scenarios does not change significantly, which does not indicate anything particularly appealing. } considering the flat priors on the parameters \{$\Omega_{b} h^2$,  $\Omega_{dm} h^2$,  $\tau$,  $n_s$ $\log[10^{10}A_{s}]$, $100\theta_{MC}$,  $w_{dm}$\} as in Table~\ref{tab:priors} while we use $[-3, 0]$ as the flat prior for $w_{de}$. We have considered the well known perturbations equations as described in Ref.~\cite{Ma:1995ey}.

In \autoref{tab:IA} and \autoref{tab:IB}, we present the observational constraints on these two scenarios for CMB, CMB+BAO and CMB+Pantheon data sets.  And in \autoref{fig:non-interacting-1} and \autoref{fig:non-interacting-2} we show the corresponding graphical variations. 
From \autoref{tab:IA}, we notice that for all three data sets, $w_{dm}$ does not exhibit any non-null behaviour in the Non-interacting scenario 1 and it is actually consistent with $w_{dm} = 0$ but interestingly we observe from \autoref{fig:non-interacting-1} that $w_{dm}$ is correlated with most of the parameters. However, we find that the Hubble constant takes a relatively higher value compared to what we obtain in the $\Lambda$CDM paradigm by Planck 2018 collaboration~\cite{Planck:2018vyg}. This is mainly due to the existing correlation with $w_{dm}$ and $H_0$ as seen in \autoref{fig:non-interacting-1}.  
For Non-interacting scenario 2, we also do not find any evidence of $w_{dm} \neq 0$, and all three data sets indicate that $w_{dm}$ is consistent to zero. On the other hand, the dark energy equation of state crosses the phantom divide line $ w_{de}  = -1.52_{-    0.30}^{+    0.18}$ at 68\% CL in agreement with the usual $w$CDM scenario~\cite{Yang:2021flj}. When the external probes, namely, BAO and Pantheon are independently added to CMB, the cosmological constant is restored.  This effectively means that this extended model has similar features as in the usual $w$CDM model.  Contrary to this, if we allow for an interaction between the dark sectors, we observe only a very weak evidence (at more than 68\% CL) of non-cold DM, as indicated by all the observational data sets (see Table~\ref{tab:IVS}). This observation is noteworthy because, based on the current observational data, the presence of an interaction between DE and non-cold DM suggests that the non-cold DM sector may not be entirely excluded, even though the evidence for such a sector remains weak. 
\begin{table}
\begin{center}
\renewcommand{\arraystretch}{1.4}
\begin{tabular}{|c@{\hspace{1 cm}}|@{\hspace{1 cm}} c|}
\hline
\textbf{Dataset}           & \textbf{$\ln B_{ij}$}\\
\hline\hline

CMB &  $-2.5$\\

CMB+Lensing &      $-3.3$\\

CMB+BAO &      $-4.2$\\

CMB+Pantheon  &   $-4.9$ \\

CMB+Lensing+BAO+Pantheon &      $-5.1$  \\ 

\hline
\end{tabular}
\end{center}
\caption{Summary of the $\ln B_{ij}$ values ($i$ refers to \textbf{IWDM} and $j$ refers to the reference model $\Lambda$CDM) for all the data sets.  The negative value in $\ln B_{ij}$ indicates that the data set prefers $\Lambda$CDM over the \textbf{IWDM} scenario. }
\label{tab:BE}
\end{table}

\subsection{Bayesian evidence}
The Bayesian evidence plays a very crucial role to understand the fitness of a cosmological scenario with respect to a given set of cosmological probes and a given reference model. As the \textbf{IWDM} is a new cosmological scenario, therefore, it is essential to understand the soundness of the model. Concerning the reference model, the $\Lambda$CDM model is undoubtedly the best choice for this purpose even though one can set a different reference model. In this work we have considered the $\Lambda$CDM model as the reference model and performed the Bayesian evidence analysis (see~\cite{Yang:2018qmz} for more details on the methodology).  We have used the MCEvidence package~\cite{Heavens:2017hkr,Heavens:2017afc} to compute the Bayes factor $\ln B_{ij}$ of the \textbf{IWDM} scenario compared to the $\Lambda$CDM scenario for all the data sets  (see Table~\ref{tab:BE}) where the negative value of $\ln B_{ij}$ indicates that data prefer the $\Lambda$CDM model over the \textbf{IWDM} model. From the Table~\ref{tab:BE}, we find that $\Lambda$CDM remains preferred over the interacting scenario proposed in this work.  This is due to the extra free parameters in the \textbf{IWDM} scenario  compared to the $\Lambda$CDM model, which are complicating the model and are therefore disfavored by Occam's razor principle. Note in fact that \textbf{IWDM} has two extra free parameters than the 6-parameter $\Lambda$CDM model.

\section{Summary and Conclusions}
\label{sec-summary}

Cosmological scenarios in which  DM and DE can  exchange energies with each other have been found very appealing both from a theoretical and observational perspective. Such scenarios are generalized versions of non-interacting cosmologies and offer a rich phenomenology of the dark sector. In most of the interacting scenarios between DM and DE (even in the non-interacting DE-DM scenarios), the equation of state of DM is assumed to be zero, which means that the DM sector is assumed to be cold or pressure-less. While the assumption of a pressure-less DM is a very natural choice that is motivated by the formation of structure of our universe on the large scale and the marvellous success of the standard $\Lambda$CDM cosmology, nevertheless, let us recall that the intrinsic nature of DM remained mysterious after a series of astronomical tests. Furthermore, since the interaction between DE and DM involves an energy transfer between them, it cpuld therefore be argued that the abundances of the pressure-less DM in the universe sector could be the result of a decaying DM with a non-vanishing equation-of-state in DE during the evolution of the universe.  The statement $-$ ``whether one should consider a pressure-less DM or a non-cold DM'' (alternatively, whether there is a non-cold DM species in the universe sector) can continue to be debated, however, with the increasing sensitivity in experimental data, probably the best approach is to consider a free-to-vary equation-of-state of the DM  and let observational data decide the most preferable scenario.

Following this, in this article, we have considered an interacting (non-cold) DM scenario (labeled as ``\textbf{IWDM}'') with the vacuum sector through a very well known interaction function $Q (t) = 3 H \xi \rho_{de}$. The vacuum sector has the equation-of-state $w_{de} =-1$ and the (non-cold) DM has a constant equation-of-state $w_{dm}$ which is free-to-vary in $[0, 1]$. Considering the evolution of the \textbf{IWDM} scenario at both background and perturbation levels, we constrained the scenario using a set of cosmological probes, such as CMB from Planck 2018, CMB+Lensing, CMB+BAO, CMB+Pantheon and CMB+Lensing+BAO+Pantheon (see \autoref{sec-data sets } for the details of the data sets). The  results are summarized in \autoref{tab:IVS} and \autoref{fig:interacting-vacuum-DM}. We find that $w_{dm}$ remains non-zero at more than 68\% CL for all observational data sets. This means  that the existence of a non-cold DM (albeit slight) in the universe sector cannot be discarded. The coupling parameter $\xi$ remains non-zero at more than 68\% CL for CMB and CMB+Lensing which signals an energy transfer from DM to DE. However, for the remaining data sets, i.e. for CMB+BAO, CMB+Pantheon and CMB+Lensing+BAO+Pantheon, we find no strong evidence for  $\xi \neq 0$. We also find that the existing $5\sigma$ tension on the Hubble constant between Planck (within the $\Lambda$CDM paradigm)~\cite{Planck:2018vyg} and the SH0ES  collaboration~\cite{Riess:2021jrx} is reduced by many standard deviations for the data sets used. We found that in this interacting scenario, this tension on $H_0$ is reduced down to $0.6\sigma $ (for CMB and CMB+Lensing), $2.5\sigma$ (CMB+BAO), $3.3\sigma$ (CMB+Pantheon) and $3.6\sigma$ (CMB+Lensing+BAO+Pantheon).  And this reduction is mainly due to: (i) the energy flow from the non-cold DM to the vacuum sector, and (ii) the large error bars on $H_0$ as a result of the increase in the volume of the parameter space.  The inclusion of the dark matter equation-of-state does not affect $H_0$ and most of the constraints, see \autoref{fig:interacting-vacuum-DM}.

Additionally, we plot the CMB TT, TE, and EE spectra (\autoref{fig:CMB-spectra}) and matter power spectra (\autoref{fig:matter-power-spectra}) for the \textbf{IWDM} scenario with the goal to understand how differently the values of $w_{dm}$ affect them. We find significant evidence in CMB TT spectra showing that with the increase of $w_{dm}$, the amplitude of the odd acoustic peaks in the CMB TT and TE spectra is suppressed and the amplitude of the even acoustic peaks increases, relative to the non-interacting $\Lambda$CDM  and the interacting cold DM scenarios. Differences are also observed in the low multipole regime ($\ell \leq 100$) of the CMB TT spectrum for increasing values of $w_{dm}$, affecting the ISW plateau. In the matter power spectra, the effects of the non-cold DM are also captured where we see that the amplitude of the matter power spectrum increases as $w_{dm}$ increases.

On another side, we have compared the \textbf{IWDM} scenario with two non-interacting scenarios which are the minimal extensions of the $\Lambda$CDM cosmological model 
aiming to distinguish between the interacting and non-interacting scenarios in the light of the non-cold DM component. In Tables~\ref{tab:IA} and \ref{tab:IB} we have summarized the constraints on these non-interacting scenarios considering three important data sets , namely CMB, CMB+BAO and CMB+Pantheon. However, our results indicate that there is no evidence of $w_{dm} \neq 0$ for any of the data sets. This is in contrast to the \textbf{IWDM} scenario, where there is still a suggestion of $w_{dm} \neq 0$, albeit very weak, based on the data.  
According to the Bayesian evidence analysis (Table~\ref{tab:BE}), the $\Lambda$CDM model still remains the favored cosmological scenario. However, it is important to note that the \textbf{IWDM} scenario introduces two additional free parameters compared to the $\Lambda$CDM model, so it is disfavored by the Occam's razor principle. This is one of the reasons why the $\Lambda$CDM model is consistently preferred over \textbf{IWDM} in our comparisons.

Based on the current outcomes, it could be argued that setting the equation-of-state of DM to zero could make the underlying cosmological scenario simple but at the expense of some important information loss regarding the dark sector. The cosmology with non-cold DM may be more exciting as we note that such scenarios have not received much attention without compelling reasons.  We also add that the evolution of this interacting scenario beyond the linear perturbation regime could be an interesting investigation. 
We trust that with the emergence of potential cosmological probes in the near future more exciting news on the non-cold DM cosmologies are yet to appear. We anticipate that this could herald the beginning of a new era in cosmology.

\section{Acknowledgments}
The authors sincerely thank the referee for his/her important comments that helped us to improve the article considerably. SP acknowledges the financial support from  the Department of Science and Technology (DST), Govt. of India under the Scheme  ``Fund for Improvement of S\&T Infrastructure (FIST)'' [File No. SR/FST/MS-I/2019/41]. 
WY was supported by the National Natural Science Foundation of China under Grants No. 12175096 and No. 11705079. EDV is supported by a Royal Society Dorothy Hodgkin Research Fellowship. This article is based upon work from COST Action CA21136 Addressing observational tensions in cosmology with systematics and fundamental physics (CosmoVerse) supported by COST (European Cooperation in Science and Technology). D.F.M. acknowledges support from the Research Council of Norway and UNINETT Sigma2 -- the National Infrastructure for High Performance Computing and Data Storage in Norway.

\bibliography{biblio}

\begin{thebibliography}{118}%
\makeatletter
\providecommand \@ifxundefined [1]{%
 \@ifx{#1\undefined}
}%
\providecommand \@ifnum [1]{%
 \ifnum #1\expandafter \@firstoftwo
 \else \expandafter \@secondoftwo
 \fi
}%
\providecommand \@ifx [1]{%
 \ifx #1\expandafter \@firstoftwo
 \else \expandafter \@secondoftwo
 \fi
}%
\providecommand \natexlab [1]{#1}%
\providecommand \enquote  [1]{``#1''}%
\providecommand \bibnamefont  [1]{#1}%
\providecommand \bibfnamefont [1]{#1}%
\providecommand \citenamefont [1]{#1}%
\providecommand \href@noop [0]{\@secondoftwo}%
\providecommand \href [0]{\begingroup \@sanitize@url \@href}%
\providecommand \@href[1]{\@@startlink{#1}\@@href}%
\providecommand \@@href[1]{\endgroup#1\@@endlink}%
\providecommand \@sanitize@url [0]{\catcode `\\12\catcode `\$12\catcode
  `\&12\catcode `\#12\catcode `\^12\catcode `\_12\catcode `\%12\relax}%
\providecommand \@@startlink[1]{}%
\providecommand \@@endlink[0]{}%
\providecommand \url  [0]{\begingroup\@sanitize@url \@url }%
\providecommand \@url [1]{\endgroup\@href {#1}{\urlprefix }}%
\providecommand \urlprefix  [0]{URL }%
\providecommand \Eprint [0]{\href }%
\providecommand \doibase [0]{http://dx.doi.org/}%
\providecommand \selectlanguage [0]{\@gobble}%
\providecommand \bibinfo  [0]{\@secondoftwo}%
\providecommand \bibfield  [0]{\@secondoftwo}%
\providecommand \translation [1]{[#1]}%
\providecommand \BibitemOpen [0]{}%
\providecommand \bibitemStop [0]{}%
\providecommand \bibitemNoStop [0]{.\EOS\space}%
\providecommand \EOS [0]{\spacefactor3000\relax}%
\providecommand \BibitemShut  [1]{\csname bibitem#1\endcsname}%
\let\auto@bib@innerbib\@empty
\bibitem [{\citenamefont {Aghanim}\ \emph {et~al.}(2020)\citenamefont {Aghanim}
  \emph {et~al.}}]{Planck:2018vyg}%
  \BibitemOpen
  \bibfield  {author} {\bibinfo {author} {\bibfnamefont {N.}~\bibnamefont
  {Aghanim}} \emph {et~al.} (\bibinfo {collaboration} {Planck}),\ }\href
  {\doibase 10.1051/0004-6361/201833910} {\bibfield  {journal} {\bibinfo
  {journal} {Astron. Astrophys.}\ }\textbf {\bibinfo {volume} {641}},\ \bibinfo
  {pages} {A6} (\bibinfo {year} {2020})},\ \bibinfo {note} {[Erratum:
  Astron.Astrophys. 652, C4 (2021)]},\ \Eprint
  {http://arxiv.org/abs/1807.06209} {arXiv:1807.06209 [astro-ph.CO]}
  \BibitemShut {NoStop}%
\bibitem [{\citenamefont {Di~Valentino}\ \emph {et~al.}(2021)\citenamefont
  {Di~Valentino}, \citenamefont {Mena}, \citenamefont {Pan}, \citenamefont
  {Visinelli}, \citenamefont {Yang}, \citenamefont {Melchiorri}, \citenamefont
  {Mota}, \citenamefont {Riess},\ and\ \citenamefont
  {Silk}}]{DiValentino:2021izs}%
  \BibitemOpen
  \bibfield  {author} {\bibinfo {author} {\bibfnamefont {E.}~\bibnamefont
  {Di~Valentino}}, \bibinfo {author} {\bibfnamefont {O.}~\bibnamefont {Mena}},
  \bibinfo {author} {\bibfnamefont {S.}~\bibnamefont {Pan}}, \bibinfo {author}
  {\bibfnamefont {L.}~\bibnamefont {Visinelli}}, \bibinfo {author}
  {\bibfnamefont {W.}~\bibnamefont {Yang}}, \bibinfo {author} {\bibfnamefont
  {A.}~\bibnamefont {Melchiorri}}, \bibinfo {author} {\bibfnamefont {D.~F.}\
  \bibnamefont {Mota}}, \bibinfo {author} {\bibfnamefont {A.~G.}\ \bibnamefont
  {Riess}}, \ and\ \bibinfo {author} {\bibfnamefont {J.}~\bibnamefont {Silk}},\
  }\href {\doibase 10.1088/1361-6382/ac086d} {\bibfield  {journal} {\bibinfo
  {journal} {Class. Quant. Grav.}\ }\textbf {\bibinfo {volume} {38}},\ \bibinfo
  {pages} {153001} (\bibinfo {year} {2021})},\ \Eprint
  {http://arxiv.org/abs/2103.01183} {arXiv:2103.01183 [astro-ph.CO]}
  \BibitemShut {NoStop}%
\bibitem [{\citenamefont {Perivolaropoulos}\ and\ \citenamefont
  {Skara}(2022)}]{Perivolaropoulos:2021jda}%
  \BibitemOpen
  \bibfield  {author} {\bibinfo {author} {\bibfnamefont {L.}~\bibnamefont
  {Perivolaropoulos}}\ and\ \bibinfo {author} {\bibfnamefont {F.}~\bibnamefont
  {Skara}},\ }\href {\doibase 10.1016/j.newar.2022.101659} {\bibfield
  {journal} {\bibinfo  {journal} {New Astron. Rev.}\ }\textbf {\bibinfo
  {volume} {95}},\ \bibinfo {pages} {101659} (\bibinfo {year} {2022})},\
  \Eprint {http://arxiv.org/abs/2105.05208} {arXiv:2105.05208 [astro-ph.CO]}
  \BibitemShut {NoStop}%
\bibitem [{\citenamefont {Sch\"oneberg}\ \emph {et~al.}(2022)\citenamefont
  {Sch\"oneberg}, \citenamefont {Franco~Abell\'an}, \citenamefont
  {P\'erez~S\'anchez}, \citenamefont {Witte}, \citenamefont {Poulin},\ and\
  \citenamefont {Lesgourgues}}]{Schoneberg:2021qvd}%
  \BibitemOpen
  \bibfield  {author} {\bibinfo {author} {\bibfnamefont {N.}~\bibnamefont
  {Sch\"oneberg}}, \bibinfo {author} {\bibfnamefont {G.}~\bibnamefont
  {Franco~Abell\'an}}, \bibinfo {author} {\bibfnamefont {A.}~\bibnamefont
  {P\'erez~S\'anchez}}, \bibinfo {author} {\bibfnamefont {S.~J.}\ \bibnamefont
  {Witte}}, \bibinfo {author} {\bibfnamefont {V.}~\bibnamefont {Poulin}}, \
  and\ \bibinfo {author} {\bibfnamefont {J.}~\bibnamefont {Lesgourgues}},\
  }\href {\doibase 10.1016/j.physrep.2022.07.001} {\bibfield  {journal}
  {\bibinfo  {journal} {Phys. Rept.}\ }\textbf {\bibinfo {volume} {984}},\
  \bibinfo {pages} {1} (\bibinfo {year} {2022})},\ \Eprint
  {http://arxiv.org/abs/2107.10291} {arXiv:2107.10291 [astro-ph.CO]}
  \BibitemShut {NoStop}%
\bibitem [{\citenamefont {Abdalla}\ \emph {et~al.}(2022)\citenamefont {Abdalla}
  \emph {et~al.}}]{Abdalla:2022yfr}%
  \BibitemOpen
  \bibfield  {author} {\bibinfo {author} {\bibfnamefont {E.}~\bibnamefont
  {Abdalla}} \emph {et~al.},\ }\href {\doibase 10.1016/j.jheap.2022.04.002}
  {\bibfield  {journal} {\bibinfo  {journal} {JHEAp}\ }\textbf {\bibinfo
  {volume} {34}},\ \bibinfo {pages} {49} (\bibinfo {year} {2022})},\ \Eprint
  {http://arxiv.org/abs/2203.06142} {arXiv:2203.06142 [astro-ph.CO]}
  \BibitemShut {NoStop}%
\bibitem [{\citenamefont {Peebles}\ and\ \citenamefont
  {Ratra}(2003)}]{Peebles:2002gy}%
  \BibitemOpen
  \bibfield  {author} {\bibinfo {author} {\bibfnamefont {P.~J.~E.}\
  \bibnamefont {Peebles}}\ and\ \bibinfo {author} {\bibfnamefont
  {B.}~\bibnamefont {Ratra}},\ }\href {\doibase 10.1103/RevModPhys.75.559}
  {\bibfield  {journal} {\bibinfo  {journal} {Rev. Mod. Phys.}\ }\textbf
  {\bibinfo {volume} {75}},\ \bibinfo {pages} {559} (\bibinfo {year} {2003})},\
  \Eprint {http://arxiv.org/abs/astro-ph/0207347} {arXiv:astro-ph/0207347}
  \BibitemShut {NoStop}%
\bibitem [{\citenamefont {Copeland}\ \emph {et~al.}(2006)\citenamefont
  {Copeland}, \citenamefont {Sami},\ and\ \citenamefont
  {Tsujikawa}}]{Copeland:2006wr}%
  \BibitemOpen
  \bibfield  {author} {\bibinfo {author} {\bibfnamefont {E.~J.}\ \bibnamefont
  {Copeland}}, \bibinfo {author} {\bibfnamefont {M.}~\bibnamefont {Sami}}, \
  and\ \bibinfo {author} {\bibfnamefont {S.}~\bibnamefont {Tsujikawa}},\ }\href
  {\doibase 10.1142/S021827180600942X} {\bibfield  {journal} {\bibinfo
  {journal} {Int. J. Mod. Phys. D}\ }\textbf {\bibinfo {volume} {15}},\
  \bibinfo {pages} {1753} (\bibinfo {year} {2006})},\ \Eprint
  {http://arxiv.org/abs/hep-th/0603057} {arXiv:hep-th/0603057} \BibitemShut
  {NoStop}%
\bibitem [{\citenamefont {Nojiri}\ and\ \citenamefont
  {Odintsov}(2006)}]{Nojiri:2006ri}%
  \BibitemOpen
  \bibfield  {author} {\bibinfo {author} {\bibfnamefont {S.}~\bibnamefont
  {Nojiri}}\ and\ \bibinfo {author} {\bibfnamefont {S.~D.}\ \bibnamefont
  {Odintsov}},\ }\href {\doibase 10.1142/S0219887807001928} {\bibfield
  {journal} {\bibinfo  {journal} {eConf}\ }\textbf {\bibinfo {volume}
  {C0602061}},\ \bibinfo {pages} {06} (\bibinfo {year} {2006})},\ \Eprint
  {http://arxiv.org/abs/hep-th/0601213} {arXiv:hep-th/0601213} \BibitemShut
  {NoStop}%
\bibitem [{\citenamefont {Sahni}\ and\ \citenamefont
  {Starobinsky}(2006)}]{Sahni:2006pa}%
  \BibitemOpen
  \bibfield  {author} {\bibinfo {author} {\bibfnamefont {V.}~\bibnamefont
  {Sahni}}\ and\ \bibinfo {author} {\bibfnamefont {A.}~\bibnamefont
  {Starobinsky}},\ }\href {\doibase 10.1142/S0218271806009704} {\bibfield
  {journal} {\bibinfo  {journal} {Int. J. Mod. Phys. D}\ }\textbf {\bibinfo
  {volume} {15}},\ \bibinfo {pages} {2105} (\bibinfo {year} {2006})},\ \Eprint
  {http://arxiv.org/abs/astro-ph/0610026} {arXiv:astro-ph/0610026} \BibitemShut
  {NoStop}%
\bibitem [{\citenamefont {Padmanabhan}(2008)}]{Padmanabhan:2007xy}%
  \BibitemOpen
  \bibfield  {author} {\bibinfo {author} {\bibfnamefont {T.}~\bibnamefont
  {Padmanabhan}},\ }\href {\doibase 10.1007/s10714-007-0555-7} {\bibfield
  {journal} {\bibinfo  {journal} {Gen. Rel. Grav.}\ }\textbf {\bibinfo {volume}
  {40}},\ \bibinfo {pages} {529} (\bibinfo {year} {2008})},\ \Eprint
  {http://arxiv.org/abs/0705.2533} {arXiv:0705.2533 [gr-qc]} \BibitemShut
  {NoStop}%
\bibitem [{\citenamefont {Capozziello}\ and\ \citenamefont
  {Francaviglia}(2008)}]{Capozziello:2007ec}%
  \BibitemOpen
  \bibfield  {author} {\bibinfo {author} {\bibfnamefont {S.}~\bibnamefont
  {Capozziello}}\ and\ \bibinfo {author} {\bibfnamefont {M.}~\bibnamefont
  {Francaviglia}},\ }\href {\doibase 10.1007/s10714-007-0551-y} {\bibfield
  {journal} {\bibinfo  {journal} {Gen. Rel. Grav.}\ }\textbf {\bibinfo {volume}
  {40}},\ \bibinfo {pages} {357} (\bibinfo {year} {2008})},\ \Eprint
  {http://arxiv.org/abs/0706.1146} {arXiv:0706.1146 [astro-ph]} \BibitemShut
  {NoStop}%
\bibitem [{\citenamefont {Sotiriou}\ and\ \citenamefont
  {Faraoni}(2010)}]{Sotiriou:2008rp}%
  \BibitemOpen
  \bibfield  {author} {\bibinfo {author} {\bibfnamefont {T.~P.}\ \bibnamefont
  {Sotiriou}}\ and\ \bibinfo {author} {\bibfnamefont {V.}~\bibnamefont
  {Faraoni}},\ }\href {\doibase 10.1103/RevModPhys.82.451} {\bibfield
  {journal} {\bibinfo  {journal} {Rev. Mod. Phys.}\ }\textbf {\bibinfo {volume}
  {82}},\ \bibinfo {pages} {451} (\bibinfo {year} {2010})},\ \Eprint
  {http://arxiv.org/abs/0805.1726} {arXiv:0805.1726 [gr-qc]} \BibitemShut
  {NoStop}%
\bibitem [{\citenamefont {Frieman}\ \emph {et~al.}(2008)\citenamefont
  {Frieman}, \citenamefont {Turner},\ and\ \citenamefont
  {Huterer}}]{Frieman:2008sn}%
  \BibitemOpen
  \bibfield  {author} {\bibinfo {author} {\bibfnamefont {J.}~\bibnamefont
  {Frieman}}, \bibinfo {author} {\bibfnamefont {M.}~\bibnamefont {Turner}}, \
  and\ \bibinfo {author} {\bibfnamefont {D.}~\bibnamefont {Huterer}},\ }\href
  {\doibase 10.1146/annurev.astro.46.060407.145243} {\bibfield  {journal}
  {\bibinfo  {journal} {Ann. Rev. Astron. Astrophys.}\ }\textbf {\bibinfo
  {volume} {46}},\ \bibinfo {pages} {385} (\bibinfo {year} {2008})},\ \Eprint
  {http://arxiv.org/abs/0803.0982} {arXiv:0803.0982 [astro-ph]} \BibitemShut
  {NoStop}%
\bibitem [{\citenamefont {Caldwell}\ and\ \citenamefont
  {Kamionkowski}(2009)}]{Caldwell:2009ix}%
  \BibitemOpen
  \bibfield  {author} {\bibinfo {author} {\bibfnamefont {R.~R.}\ \bibnamefont
  {Caldwell}}\ and\ \bibinfo {author} {\bibfnamefont {M.}~\bibnamefont
  {Kamionkowski}},\ }\href {\doibase 10.1146/annurev-nucl-010709-151330}
  {\bibfield  {journal} {\bibinfo  {journal} {Ann. Rev. Nucl. Part. Sci.}\
  }\textbf {\bibinfo {volume} {59}},\ \bibinfo {pages} {397} (\bibinfo {year}
  {2009})},\ \Eprint {http://arxiv.org/abs/0903.0866} {arXiv:0903.0866
  [astro-ph.CO]} \BibitemShut {NoStop}%
\bibitem [{\citenamefont {Silvestri}\ and\ \citenamefont
  {Trodden}(2009)}]{Silvestri:2009hh}%
  \BibitemOpen
  \bibfield  {author} {\bibinfo {author} {\bibfnamefont {A.}~\bibnamefont
  {Silvestri}}\ and\ \bibinfo {author} {\bibfnamefont {M.}~\bibnamefont
  {Trodden}},\ }\href {\doibase 10.1088/0034-4885/72/9/096901} {\bibfield
  {journal} {\bibinfo  {journal} {Rept. Prog. Phys.}\ }\textbf {\bibinfo
  {volume} {72}},\ \bibinfo {pages} {096901} (\bibinfo {year} {2009})},\
  \Eprint {http://arxiv.org/abs/0904.0024} {arXiv:0904.0024 [astro-ph.CO]}
  \BibitemShut {NoStop}%
\bibitem [{\citenamefont {De~Felice}\ and\ \citenamefont
  {Tsujikawa}(2010)}]{DeFelice:2010aj}%
  \BibitemOpen
  \bibfield  {author} {\bibinfo {author} {\bibfnamefont {A.}~\bibnamefont
  {De~Felice}}\ and\ \bibinfo {author} {\bibfnamefont {S.}~\bibnamefont
  {Tsujikawa}},\ }\href {\doibase 10.12942/lrr-2010-3} {\bibfield  {journal}
  {\bibinfo  {journal} {Living Rev. Rel.}\ }\textbf {\bibinfo {volume} {13}},\
  \bibinfo {pages} {3} (\bibinfo {year} {2010})},\ \Eprint
  {http://arxiv.org/abs/1002.4928} {arXiv:1002.4928 [gr-qc]} \BibitemShut
  {NoStop}%
\bibitem [{\citenamefont {Nojiri}\ and\ \citenamefont
  {Odintsov}(2011)}]{Nojiri:2010wj}%
  \BibitemOpen
  \bibfield  {author} {\bibinfo {author} {\bibfnamefont {S.}~\bibnamefont
  {Nojiri}}\ and\ \bibinfo {author} {\bibfnamefont {S.~D.}\ \bibnamefont
  {Odintsov}},\ }\href {\doibase 10.1016/j.physrep.2011.04.001} {\bibfield
  {journal} {\bibinfo  {journal} {Phys. Rept.}\ }\textbf {\bibinfo {volume}
  {505}},\ \bibinfo {pages} {59} (\bibinfo {year} {2011})},\ \Eprint
  {http://arxiv.org/abs/1011.0544} {arXiv:1011.0544 [gr-qc]} \BibitemShut
  {NoStop}%
\bibitem [{\citenamefont {Clifton}\ \emph {et~al.}(2012)\citenamefont
  {Clifton}, \citenamefont {Ferreira}, \citenamefont {Padilla},\ and\
  \citenamefont {Skordis}}]{Clifton:2011jh}%
  \BibitemOpen
  \bibfield  {author} {\bibinfo {author} {\bibfnamefont {T.}~\bibnamefont
  {Clifton}}, \bibinfo {author} {\bibfnamefont {P.~G.}\ \bibnamefont
  {Ferreira}}, \bibinfo {author} {\bibfnamefont {A.}~\bibnamefont {Padilla}}, \
  and\ \bibinfo {author} {\bibfnamefont {C.}~\bibnamefont {Skordis}},\ }\href
  {\doibase 10.1016/j.physrep.2012.01.001} {\bibfield  {journal} {\bibinfo
  {journal} {Phys. Rept.}\ }\textbf {\bibinfo {volume} {513}},\ \bibinfo
  {pages} {1} (\bibinfo {year} {2012})},\ \Eprint
  {http://arxiv.org/abs/1106.2476} {arXiv:1106.2476 [astro-ph.CO]} \BibitemShut
  {NoStop}%
\bibitem [{\citenamefont {Capozziello}\ and\ \citenamefont
  {De~Laurentis}(2011)}]{Capozziello:2011et}%
  \BibitemOpen
  \bibfield  {author} {\bibinfo {author} {\bibfnamefont {S.}~\bibnamefont
  {Capozziello}}\ and\ \bibinfo {author} {\bibfnamefont {M.}~\bibnamefont
  {De~Laurentis}},\ }\href {\doibase 10.1016/j.physrep.2011.09.003} {\bibfield
  {journal} {\bibinfo  {journal} {Phys. Rept.}\ }\textbf {\bibinfo {volume}
  {509}},\ \bibinfo {pages} {167} (\bibinfo {year} {2011})},\ \Eprint
  {http://arxiv.org/abs/1108.6266} {arXiv:1108.6266 [gr-qc]} \BibitemShut
  {NoStop}%
\bibitem [{\citenamefont {Bamba}\ \emph {et~al.}(2012)\citenamefont {Bamba},
  \citenamefont {Capozziello}, \citenamefont {Nojiri},\ and\ \citenamefont
  {Odintsov}}]{Bamba:2012cp}%
  \BibitemOpen
  \bibfield  {author} {\bibinfo {author} {\bibfnamefont {K.}~\bibnamefont
  {Bamba}}, \bibinfo {author} {\bibfnamefont {S.}~\bibnamefont {Capozziello}},
  \bibinfo {author} {\bibfnamefont {S.}~\bibnamefont {Nojiri}}, \ and\ \bibinfo
  {author} {\bibfnamefont {S.~D.}\ \bibnamefont {Odintsov}},\ }\href {\doibase
  10.1007/s10509-012-1181-8} {\bibfield  {journal} {\bibinfo  {journal}
  {Astrophys. Space Sci.}\ }\textbf {\bibinfo {volume} {342}},\ \bibinfo
  {pages} {155} (\bibinfo {year} {2012})},\ \Eprint
  {http://arxiv.org/abs/1205.3421} {arXiv:1205.3421 [gr-qc]} \BibitemShut
  {NoStop}%
\bibitem [{\citenamefont {Li}\ \emph {et~al.}(2013)\citenamefont {Li},
  \citenamefont {Li}, \citenamefont {Wang},\ and\ \citenamefont
  {Wang}}]{Li:2012dt}%
  \BibitemOpen
  \bibfield  {author} {\bibinfo {author} {\bibfnamefont {M.}~\bibnamefont
  {Li}}, \bibinfo {author} {\bibfnamefont {X.-D.}\ \bibnamefont {Li}}, \bibinfo
  {author} {\bibfnamefont {S.}~\bibnamefont {Wang}}, \ and\ \bibinfo {author}
  {\bibfnamefont {Y.}~\bibnamefont {Wang}},\ }\href {\doibase
  10.1007/s11467-013-0300-5} {\bibfield  {journal} {\bibinfo  {journal} {Front.
  Phys. (Beijing)}\ }\textbf {\bibinfo {volume} {8}},\ \bibinfo {pages} {828}
  (\bibinfo {year} {2013})},\ \Eprint {http://arxiv.org/abs/1209.0922}
  {arXiv:1209.0922 [astro-ph.CO]} \BibitemShut {NoStop}%
\bibitem [{\citenamefont {Cai}\ \emph {et~al.}(2016)\citenamefont {Cai},
  \citenamefont {Capozziello}, \citenamefont {De~Laurentis},\ and\
  \citenamefont {Saridakis}}]{Cai:2015emx}%
  \BibitemOpen
  \bibfield  {author} {\bibinfo {author} {\bibfnamefont {Y.-F.}\ \bibnamefont
  {Cai}}, \bibinfo {author} {\bibfnamefont {S.}~\bibnamefont {Capozziello}},
  \bibinfo {author} {\bibfnamefont {M.}~\bibnamefont {De~Laurentis}}, \ and\
  \bibinfo {author} {\bibfnamefont {E.~N.}\ \bibnamefont {Saridakis}},\ }\href
  {\doibase 10.1088/0034-4885/79/10/106901} {\bibfield  {journal} {\bibinfo
  {journal} {Rept. Prog. Phys.}\ }\textbf {\bibinfo {volume} {79}},\ \bibinfo
  {pages} {106901} (\bibinfo {year} {2016})},\ \Eprint
  {http://arxiv.org/abs/1511.07586} {arXiv:1511.07586 [gr-qc]} \BibitemShut
  {NoStop}%
\bibitem [{\citenamefont {Nojiri}\ \emph {et~al.}(2017)\citenamefont {Nojiri},
  \citenamefont {Odintsov},\ and\ \citenamefont {Oikonomou}}]{Nojiri:2017ncd}%
  \BibitemOpen
  \bibfield  {author} {\bibinfo {author} {\bibfnamefont {S.}~\bibnamefont
  {Nojiri}}, \bibinfo {author} {\bibfnamefont {S.~D.}\ \bibnamefont
  {Odintsov}}, \ and\ \bibinfo {author} {\bibfnamefont {V.~K.}\ \bibnamefont
  {Oikonomou}},\ }\href {\doibase 10.1016/j.physrep.2017.06.001} {\bibfield
  {journal} {\bibinfo  {journal} {Phys. Rept.}\ }\textbf {\bibinfo {volume}
  {692}},\ \bibinfo {pages} {1} (\bibinfo {year} {2017})},\ \Eprint
  {http://arxiv.org/abs/1705.11098} {arXiv:1705.11098 [gr-qc]} \BibitemShut
  {NoStop}%
\bibitem [{\citenamefont {Bahamonde}\ \emph {et~al.}(2023)\citenamefont
  {Bahamonde}, \citenamefont {Dialektopoulos}, \citenamefont
  {Escamilla-Rivera}, \citenamefont {Farrugia}, \citenamefont {Gakis},
  \citenamefont {Hendry}, \citenamefont {Hohmann}, \citenamefont {Levi~Said},
  \citenamefont {Mifsud},\ and\ \citenamefont
  {Di~Valentino}}]{Bahamonde:2021gfp}%
  \BibitemOpen
  \bibfield  {author} {\bibinfo {author} {\bibfnamefont {S.}~\bibnamefont
  {Bahamonde}}, \bibinfo {author} {\bibfnamefont {K.~F.}\ \bibnamefont
  {Dialektopoulos}}, \bibinfo {author} {\bibfnamefont {C.}~\bibnamefont
  {Escamilla-Rivera}}, \bibinfo {author} {\bibfnamefont {G.}~\bibnamefont
  {Farrugia}}, \bibinfo {author} {\bibfnamefont {V.}~\bibnamefont {Gakis}},
  \bibinfo {author} {\bibfnamefont {M.}~\bibnamefont {Hendry}}, \bibinfo
  {author} {\bibfnamefont {M.}~\bibnamefont {Hohmann}}, \bibinfo {author}
  {\bibfnamefont {J.}~\bibnamefont {Levi~Said}}, \bibinfo {author}
  {\bibfnamefont {J.}~\bibnamefont {Mifsud}}, \ and\ \bibinfo {author}
  {\bibfnamefont {E.}~\bibnamefont {Di~Valentino}},\ }\href {\doibase
  10.1088/1361-6633/ac9cef} {\bibfield  {journal} {\bibinfo  {journal} {Rept.
  Prog. Phys.}\ }\textbf {\bibinfo {volume} {86}},\ \bibinfo {pages} {026901}
  (\bibinfo {year} {2023})},\ \Eprint {http://arxiv.org/abs/2106.13793}
  {arXiv:2106.13793 [gr-qc]} \BibitemShut {NoStop}%
\bibitem [{\citenamefont {Muller}(2005)}]{Muller:2004yb}%
  \BibitemOpen
  \bibfield  {author} {\bibinfo {author} {\bibfnamefont {C.~M.}\ \bibnamefont
  {Muller}},\ }\href {\doibase 10.1103/PhysRevD.71.047302} {\bibfield
  {journal} {\bibinfo  {journal} {Phys. Rev. D}\ }\textbf {\bibinfo {volume}
  {71}},\ \bibinfo {pages} {047302} (\bibinfo {year} {2005})},\ \Eprint
  {http://arxiv.org/abs/astro-ph/0410621} {arXiv:astro-ph/0410621} \BibitemShut
  {NoStop}%
\bibitem [{\citenamefont {Kumar}\ and\ \citenamefont
  {Xu}(2014)}]{Kumar:2012gr}%
  \BibitemOpen
  \bibfield  {author} {\bibinfo {author} {\bibfnamefont {S.}~\bibnamefont
  {Kumar}}\ and\ \bibinfo {author} {\bibfnamefont {L.}~\bibnamefont {Xu}},\
  }\href {\doibase 10.1016/j.physletb.2014.08.059} {\bibfield  {journal}
  {\bibinfo  {journal} {Phys. Lett. B}\ }\textbf {\bibinfo {volume} {737}},\
  \bibinfo {pages} {244} (\bibinfo {year} {2014})},\ \Eprint
  {http://arxiv.org/abs/1207.5582} {arXiv:1207.5582 [gr-qc]} \BibitemShut
  {NoStop}%
\bibitem [{\citenamefont {Armendariz-Picon}\ and\ \citenamefont
  {Neelakanta}(2014)}]{Armendariz-Picon:2013jej}%
  \BibitemOpen
  \bibfield  {author} {\bibinfo {author} {\bibfnamefont {C.}~\bibnamefont
  {Armendariz-Picon}}\ and\ \bibinfo {author} {\bibfnamefont {J.~T.}\
  \bibnamefont {Neelakanta}},\ }\href {\doibase 10.1088/1475-7516/2014/03/049}
  {\bibfield  {journal} {\bibinfo  {journal} {JCAP}\ }\textbf {\bibinfo
  {volume} {03}},\ \bibinfo {pages} {049} (\bibinfo {year} {2014})},\ \Eprint
  {http://arxiv.org/abs/1309.6971} {arXiv:1309.6971 [astro-ph.CO]} \BibitemShut
  {NoStop}%
\bibitem [{\citenamefont {Kopp}\ \emph {et~al.}(2018)\citenamefont {Kopp},
  \citenamefont {Skordis}, \citenamefont {Thomas},\ and\ \citenamefont
  {Ili\'c}}]{Kopp:2018zxp}%
  \BibitemOpen
  \bibfield  {author} {\bibinfo {author} {\bibfnamefont {M.}~\bibnamefont
  {Kopp}}, \bibinfo {author} {\bibfnamefont {C.}~\bibnamefont {Skordis}},
  \bibinfo {author} {\bibfnamefont {D.~B.}\ \bibnamefont {Thomas}}, \ and\
  \bibinfo {author} {\bibfnamefont {S.}~\bibnamefont {Ili\'c}},\ }\href
  {\doibase 10.1103/PhysRevLett.120.221102} {\bibfield  {journal} {\bibinfo
  {journal} {Phys. Rev. Lett.}\ }\textbf {\bibinfo {volume} {120}},\ \bibinfo
  {pages} {221102} (\bibinfo {year} {2018})},\ \Eprint
  {http://arxiv.org/abs/1802.09541} {arXiv:1802.09541 [astro-ph.CO]}
  \BibitemShut {NoStop}%
\bibitem [{\citenamefont {Ili\'c}\ \emph {et~al.}(2021)\citenamefont {Ili\'c},
  \citenamefont {Kopp}, \citenamefont {Skordis},\ and\ \citenamefont
  {Thomas}}]{Ilic:2020onu}%
  \BibitemOpen
  \bibfield  {author} {\bibinfo {author} {\bibfnamefont {S.}~\bibnamefont
  {Ili\'c}}, \bibinfo {author} {\bibfnamefont {M.}~\bibnamefont {Kopp}},
  \bibinfo {author} {\bibfnamefont {C.}~\bibnamefont {Skordis}}, \ and\
  \bibinfo {author} {\bibfnamefont {D.~B.}\ \bibnamefont {Thomas}},\ }\href
  {\doibase 10.1103/PhysRevD.104.043520} {\bibfield  {journal} {\bibinfo
  {journal} {Phys. Rev. D}\ }\textbf {\bibinfo {volume} {104}},\ \bibinfo
  {pages} {043520} (\bibinfo {year} {2021})},\ \Eprint
  {http://arxiv.org/abs/2004.09572} {arXiv:2004.09572 [astro-ph.CO]}
  \BibitemShut {NoStop}%
\bibitem [{\citenamefont {Naidoo}\ \emph {et~al.}(2022)\citenamefont {Naidoo},
  \citenamefont {Jaber}, \citenamefont {Hellwing},\ and\ \citenamefont
  {Bilicki}}]{Naidoo:2022rda}%
  \BibitemOpen
  \bibfield  {author} {\bibinfo {author} {\bibfnamefont {K.}~\bibnamefont
  {Naidoo}}, \bibinfo {author} {\bibfnamefont {M.}~\bibnamefont {Jaber}},
  \bibinfo {author} {\bibfnamefont {W.~A.}\ \bibnamefont {Hellwing}}, \ and\
  \bibinfo {author} {\bibfnamefont {M.}~\bibnamefont {Bilicki}},\ }\href@noop
  {} {\  (\bibinfo {year} {2022})},\ \Eprint {http://arxiv.org/abs/2209.08102}
  {arXiv:2209.08102 [astro-ph.CO]} \BibitemShut {NoStop}%
\bibitem [{\citenamefont {Amendola}(2000)}]{Amendola:1999er}%
  \BibitemOpen
  \bibfield  {author} {\bibinfo {author} {\bibfnamefont {L.}~\bibnamefont
  {Amendola}},\ }\href {\doibase 10.1103/PhysRevD.62.043511} {\bibfield
  {journal} {\bibinfo  {journal} {Phys. Rev. D}\ }\textbf {\bibinfo {volume}
  {62}},\ \bibinfo {pages} {043511} (\bibinfo {year} {2000})},\ \Eprint
  {http://arxiv.org/abs/astro-ph/9908023} {arXiv:astro-ph/9908023} \BibitemShut
  {NoStop}%
\bibitem [{\citenamefont {Huey}\ and\ \citenamefont
  {Wandelt}(2006)}]{Huey:2004qv}%
  \BibitemOpen
  \bibfield  {author} {\bibinfo {author} {\bibfnamefont {G.}~\bibnamefont
  {Huey}}\ and\ \bibinfo {author} {\bibfnamefont {B.~D.}\ \bibnamefont
  {Wandelt}},\ }\href {\doibase 10.1103/PhysRevD.74.023519} {\bibfield
  {journal} {\bibinfo  {journal} {Phys. Rev. D}\ }\textbf {\bibinfo {volume}
  {74}},\ \bibinfo {pages} {023519} (\bibinfo {year} {2006})},\ \Eprint
  {http://arxiv.org/abs/astro-ph/0407196} {arXiv:astro-ph/0407196} \BibitemShut
  {NoStop}%
\bibitem [{\citenamefont {Cai}\ and\ \citenamefont {Wang}(2005)}]{Cai:2004dk}%
  \BibitemOpen
  \bibfield  {author} {\bibinfo {author} {\bibfnamefont {R.-G.}\ \bibnamefont
  {Cai}}\ and\ \bibinfo {author} {\bibfnamefont {A.}~\bibnamefont {Wang}},\
  }\href {\doibase 10.1088/1475-7516/2005/03/002} {\bibfield  {journal}
  {\bibinfo  {journal} {JCAP}\ }\textbf {\bibinfo {volume} {03}},\ \bibinfo
  {pages} {002} (\bibinfo {year} {2005})},\ \Eprint
  {http://arxiv.org/abs/hep-th/0411025} {arXiv:hep-th/0411025} \BibitemShut
  {NoStop}%
\bibitem [{\citenamefont {Pavon}\ and\ \citenamefont
  {Zimdahl}(2005)}]{Pavon:2005yx}%
  \BibitemOpen
  \bibfield  {author} {\bibinfo {author} {\bibfnamefont {D.}~\bibnamefont
  {Pavon}}\ and\ \bibinfo {author} {\bibfnamefont {W.}~\bibnamefont
  {Zimdahl}},\ }\href {\doibase 10.1016/j.physletb.2005.08.134} {\bibfield
  {journal} {\bibinfo  {journal} {Phys. Lett. B}\ }\textbf {\bibinfo {volume}
  {628}},\ \bibinfo {pages} {206} (\bibinfo {year} {2005})},\ \Eprint
  {http://arxiv.org/abs/gr-qc/0505020} {arXiv:gr-qc/0505020} \BibitemShut
  {NoStop}%
\bibitem [{\citenamefont {del Campo}\ \emph {et~al.}(2006)\citenamefont {del
  Campo}, \citenamefont {Herrera}, \citenamefont {Olivares},\ and\
  \citenamefont {Pavon}}]{delCampo:2006vv}%
  \BibitemOpen
  \bibfield  {author} {\bibinfo {author} {\bibfnamefont {S.}~\bibnamefont {del
  Campo}}, \bibinfo {author} {\bibfnamefont {R.}~\bibnamefont {Herrera}},
  \bibinfo {author} {\bibfnamefont {G.}~\bibnamefont {Olivares}}, \ and\
  \bibinfo {author} {\bibfnamefont {D.}~\bibnamefont {Pavon}},\ }\href
  {\doibase 10.1103/PhysRevD.74.023501} {\bibfield  {journal} {\bibinfo
  {journal} {Phys. Rev. D}\ }\textbf {\bibinfo {volume} {74}},\ \bibinfo
  {pages} {023501} (\bibinfo {year} {2006})},\ \Eprint
  {http://arxiv.org/abs/astro-ph/0606520} {arXiv:astro-ph/0606520} \BibitemShut
  {NoStop}%
\bibitem [{\citenamefont {del Campo}\ \emph {et~al.}(2008)\citenamefont {del
  Campo}, \citenamefont {Herrera},\ and\ \citenamefont
  {Pavon}}]{delCampo:2008sr}%
  \BibitemOpen
  \bibfield  {author} {\bibinfo {author} {\bibfnamefont {S.}~\bibnamefont {del
  Campo}}, \bibinfo {author} {\bibfnamefont {R.}~\bibnamefont {Herrera}}, \
  and\ \bibinfo {author} {\bibfnamefont {D.}~\bibnamefont {Pavon}},\ }\href
  {\doibase 10.1103/PhysRevD.78.021302} {\bibfield  {journal} {\bibinfo
  {journal} {Phys. Rev. D}\ }\textbf {\bibinfo {volume} {78}},\ \bibinfo
  {pages} {021302} (\bibinfo {year} {2008})},\ \Eprint
  {http://arxiv.org/abs/0806.2116} {arXiv:0806.2116 [astro-ph]} \BibitemShut
  {NoStop}%
\bibitem [{\citenamefont {del Campo}\ \emph {et~al.}(2009)\citenamefont {del
  Campo}, \citenamefont {Herrera},\ and\ \citenamefont
  {Pavon}}]{delCampo:2008jx}%
  \BibitemOpen
  \bibfield  {author} {\bibinfo {author} {\bibfnamefont {S.}~\bibnamefont {del
  Campo}}, \bibinfo {author} {\bibfnamefont {R.}~\bibnamefont {Herrera}}, \
  and\ \bibinfo {author} {\bibfnamefont {D.}~\bibnamefont {Pavon}},\ }\href
  {\doibase 10.1088/1475-7516/2009/01/020} {\bibfield  {journal} {\bibinfo
  {journal} {JCAP}\ }\textbf {\bibinfo {volume} {01}},\ \bibinfo {pages} {020}
  (\bibinfo {year} {2009})},\ \Eprint {http://arxiv.org/abs/0812.2210}
  {arXiv:0812.2210 [gr-qc]} \BibitemShut {NoStop}%
\bibitem [{\citenamefont {Wang}\ \emph {et~al.}(2005)\citenamefont {Wang},
  \citenamefont {Gong},\ and\ \citenamefont {Abdalla}}]{Wang:2005jx}%
  \BibitemOpen
  \bibfield  {author} {\bibinfo {author} {\bibfnamefont {B.}~\bibnamefont
  {Wang}}, \bibinfo {author} {\bibfnamefont {Y.-g.}\ \bibnamefont {Gong}}, \
  and\ \bibinfo {author} {\bibfnamefont {E.}~\bibnamefont {Abdalla}},\ }\href
  {\doibase 10.1016/j.physletb.2005.08.008} {\bibfield  {journal} {\bibinfo
  {journal} {Phys. Lett. B}\ }\textbf {\bibinfo {volume} {624}},\ \bibinfo
  {pages} {141} (\bibinfo {year} {2005})},\ \Eprint
  {http://arxiv.org/abs/hep-th/0506069} {arXiv:hep-th/0506069} \BibitemShut
  {NoStop}%
\bibitem [{\citenamefont {Das}\ \emph {et~al.}(2006)\citenamefont {Das},
  \citenamefont {Corasaniti},\ and\ \citenamefont {Khoury}}]{Das:2005yj}%
  \BibitemOpen
  \bibfield  {author} {\bibinfo {author} {\bibfnamefont {S.}~\bibnamefont
  {Das}}, \bibinfo {author} {\bibfnamefont {P.~S.}\ \bibnamefont {Corasaniti}},
  \ and\ \bibinfo {author} {\bibfnamefont {J.}~\bibnamefont {Khoury}},\ }\href
  {\doibase 10.1103/PhysRevD.73.083509} {\bibfield  {journal} {\bibinfo
  {journal} {Phys. Rev. D}\ }\textbf {\bibinfo {volume} {73}},\ \bibinfo
  {pages} {083509} (\bibinfo {year} {2006})},\ \Eprint
  {http://arxiv.org/abs/astro-ph/0510628} {arXiv:astro-ph/0510628} \BibitemShut
  {NoStop}%
\bibitem [{\citenamefont {Sadjadi}\ and\ \citenamefont
  {Honardoost}(2007)}]{Sadjadi:2006qb}%
  \BibitemOpen
  \bibfield  {author} {\bibinfo {author} {\bibfnamefont {H.~M.}\ \bibnamefont
  {Sadjadi}}\ and\ \bibinfo {author} {\bibfnamefont {M.}~\bibnamefont
  {Honardoost}},\ }\href {\doibase 10.1016/j.physletb.2007.02.016} {\bibfield
  {journal} {\bibinfo  {journal} {Phys. Lett. B}\ }\textbf {\bibinfo {volume}
  {647}},\ \bibinfo {pages} {231} (\bibinfo {year} {2007})},\ \Eprint
  {http://arxiv.org/abs/gr-qc/0609076} {arXiv:gr-qc/0609076} \BibitemShut
  {NoStop}%
\bibitem [{\citenamefont {Pan}\ and\ \citenamefont
  {Chakraborty}(2014)}]{Pan:2014afa}%
  \BibitemOpen
  \bibfield  {author} {\bibinfo {author} {\bibfnamefont {S.}~\bibnamefont
  {Pan}}\ and\ \bibinfo {author} {\bibfnamefont {S.}~\bibnamefont
  {Chakraborty}},\ }\href {\doibase 10.1142/S0218271814500928} {\bibfield
  {journal} {\bibinfo  {journal} {Int. J. Mod. Phys. D}\ }\textbf {\bibinfo
  {volume} {23}},\ \bibinfo {pages} {1450092} (\bibinfo {year} {2014})},\
  \Eprint {http://arxiv.org/abs/1410.8281} {arXiv:1410.8281 [gr-qc]}
  \BibitemShut {NoStop}%
\bibitem [{\citenamefont {Di~Valentino}\ \emph {et~al.}(2017)\citenamefont
  {Di~Valentino}, \citenamefont {Melchiorri},\ and\ \citenamefont
  {Mena}}]{DiValentino:2017iww}%
  \BibitemOpen
  \bibfield  {author} {\bibinfo {author} {\bibfnamefont {E.}~\bibnamefont
  {Di~Valentino}}, \bibinfo {author} {\bibfnamefont {A.}~\bibnamefont
  {Melchiorri}}, \ and\ \bibinfo {author} {\bibfnamefont {O.}~\bibnamefont
  {Mena}},\ }\href {\doibase 10.1103/PhysRevD.96.043503} {\bibfield  {journal}
  {\bibinfo  {journal} {Phys. Rev.}\ }\textbf {\bibinfo {volume} {D96}},\
  \bibinfo {pages} {043503} (\bibinfo {year} {2017})},\ \Eprint
  {http://arxiv.org/abs/1704.08342} {arXiv:1704.08342 [astro-ph.CO]}
  \BibitemShut {NoStop}%
\bibitem [{\citenamefont {Kumar}\ and\ \citenamefont
  {Nunes}(2017)}]{Kumar:2017dnp}%
  \BibitemOpen
  \bibfield  {author} {\bibinfo {author} {\bibfnamefont {S.}~\bibnamefont
  {Kumar}}\ and\ \bibinfo {author} {\bibfnamefont {R.~C.}\ \bibnamefont
  {Nunes}},\ }\href {\doibase 10.1103/PhysRevD.96.103511} {\bibfield  {journal}
  {\bibinfo  {journal} {Phys. Rev.}\ }\textbf {\bibinfo {volume} {D96}},\
  \bibinfo {pages} {103511} (\bibinfo {year} {2017})},\ \Eprint
  {http://arxiv.org/abs/1702.02143} {arXiv:1702.02143 [astro-ph.CO]}
  \BibitemShut {NoStop}%
\bibitem [{\citenamefont {Yang}\ \emph {et~al.}(2018)\citenamefont {Yang},
  \citenamefont {Pan}, \citenamefont {Di~Valentino}, \citenamefont {Nunes},
  \citenamefont {Vagnozzi},\ and\ \citenamefont {Mota}}]{Yang:2018euj}%
  \BibitemOpen
  \bibfield  {author} {\bibinfo {author} {\bibfnamefont {W.}~\bibnamefont
  {Yang}}, \bibinfo {author} {\bibfnamefont {S.}~\bibnamefont {Pan}}, \bibinfo
  {author} {\bibfnamefont {E.}~\bibnamefont {Di~Valentino}}, \bibinfo {author}
  {\bibfnamefont {R.~C.}\ \bibnamefont {Nunes}}, \bibinfo {author}
  {\bibfnamefont {S.}~\bibnamefont {Vagnozzi}}, \ and\ \bibinfo {author}
  {\bibfnamefont {D.~F.}\ \bibnamefont {Mota}},\ }\href {\doibase
  10.1088/1475-7516/2018/09/019} {\bibfield  {journal} {\bibinfo  {journal}
  {JCAP}\ }\textbf {\bibinfo {volume} {1809}},\ \bibinfo {pages} {019}
  (\bibinfo {year} {2018})},\ \Eprint {http://arxiv.org/abs/1805.08252}
  {arXiv:1805.08252 [astro-ph.CO]} \BibitemShut {NoStop}%
\bibitem [{\citenamefont {Pan}\ \emph {et~al.}(2019{\natexlab{a}})\citenamefont
  {Pan}, \citenamefont {Yang}, \citenamefont {Singha},\ and\ \citenamefont
  {Saridakis}}]{Pan:2019jqh}%
  \BibitemOpen
  \bibfield  {author} {\bibinfo {author} {\bibfnamefont {S.}~\bibnamefont
  {Pan}}, \bibinfo {author} {\bibfnamefont {W.}~\bibnamefont {Yang}}, \bibinfo
  {author} {\bibfnamefont {C.}~\bibnamefont {Singha}}, \ and\ \bibinfo {author}
  {\bibfnamefont {E.~N.}\ \bibnamefont {Saridakis}},\ }\href {\doibase
  10.1103/PhysRevD.100.083539} {\bibfield  {journal} {\bibinfo  {journal}
  {Phys. Rev.}\ }\textbf {\bibinfo {volume} {D100}},\ \bibinfo {pages} {083539}
  (\bibinfo {year} {2019}{\natexlab{a}})},\ \Eprint
  {http://arxiv.org/abs/1903.10969} {arXiv:1903.10969 [astro-ph.CO]}
  \BibitemShut {NoStop}%
\bibitem [{\citenamefont {Yang}\ \emph {et~al.}(2020)\citenamefont {Yang},
  \citenamefont {Pan}, \citenamefont {Nunes},\ and\ \citenamefont
  {Mota}}]{Yang:2019uog}%
  \BibitemOpen
  \bibfield  {author} {\bibinfo {author} {\bibfnamefont {W.}~\bibnamefont
  {Yang}}, \bibinfo {author} {\bibfnamefont {S.}~\bibnamefont {Pan}}, \bibinfo
  {author} {\bibfnamefont {R.~C.}\ \bibnamefont {Nunes}}, \ and\ \bibinfo
  {author} {\bibfnamefont {D.~F.}\ \bibnamefont {Mota}},\ }\href {\doibase
  10.1088/1475-7516/2020/04/008} {\bibfield  {journal} {\bibinfo  {journal}
  {JCAP}\ }\textbf {\bibinfo {volume} {04}},\ \bibinfo {pages} {008} (\bibinfo
  {year} {2020})},\ \Eprint {http://arxiv.org/abs/1910.08821} {arXiv:1910.08821
  [astro-ph.CO]} \BibitemShut {NoStop}%
\bibitem [{\citenamefont {Pan}\ \emph {et~al.}(2019{\natexlab{b}})\citenamefont
  {Pan}, \citenamefont {Yang}, \citenamefont {Di~Valentino}, \citenamefont
  {Saridakis},\ and\ \citenamefont {Chakraborty}}]{Pan:2019gop}%
  \BibitemOpen
  \bibfield  {author} {\bibinfo {author} {\bibfnamefont {S.}~\bibnamefont
  {Pan}}, \bibinfo {author} {\bibfnamefont {W.}~\bibnamefont {Yang}}, \bibinfo
  {author} {\bibfnamefont {E.}~\bibnamefont {Di~Valentino}}, \bibinfo {author}
  {\bibfnamefont {E.~N.}\ \bibnamefont {Saridakis}}, \ and\ \bibinfo {author}
  {\bibfnamefont {S.}~\bibnamefont {Chakraborty}},\ }\href {\doibase
  10.1103/PhysRevD.100.103520} {\bibfield  {journal} {\bibinfo  {journal}
  {Phys. Rev.}\ }\textbf {\bibinfo {volume} {D100}},\ \bibinfo {pages} {103520}
  (\bibinfo {year} {2019}{\natexlab{b}})},\ \Eprint
  {http://arxiv.org/abs/1907.07540} {arXiv:1907.07540 [astro-ph.CO]}
  \BibitemShut {NoStop}%
\bibitem [{\citenamefont {Pan}\ \emph {et~al.}(2020{\natexlab{a}})\citenamefont
  {Pan}, \citenamefont {Yang},\ and\ \citenamefont
  {Paliathanasis}}]{Pan:2020bur}%
  \BibitemOpen
  \bibfield  {author} {\bibinfo {author} {\bibfnamefont {S.}~\bibnamefont
  {Pan}}, \bibinfo {author} {\bibfnamefont {W.}~\bibnamefont {Yang}}, \ and\
  \bibinfo {author} {\bibfnamefont {A.}~\bibnamefont {Paliathanasis}},\ }\href
  {\doibase 10.1093/mnras/staa213} {\bibfield  {journal} {\bibinfo  {journal}
  {Mon. Not. Roy. Astron. Soc.}\ }\textbf {\bibinfo {volume} {493}},\ \bibinfo
  {pages} {3114} (\bibinfo {year} {2020}{\natexlab{a}})},\ \Eprint
  {http://arxiv.org/abs/2002.03408} {arXiv:2002.03408 [astro-ph.CO]}
  \BibitemShut {NoStop}%
\bibitem [{\citenamefont {Di~Valentino}\ \emph
  {et~al.}(2020{\natexlab{a}})\citenamefont {Di~Valentino}, \citenamefont
  {Melchiorri}, \citenamefont {Mena},\ and\ \citenamefont
  {Vagnozzi}}]{DiValentino:2019ffd}%
  \BibitemOpen
  \bibfield  {author} {\bibinfo {author} {\bibfnamefont {E.}~\bibnamefont
  {Di~Valentino}}, \bibinfo {author} {\bibfnamefont {A.}~\bibnamefont
  {Melchiorri}}, \bibinfo {author} {\bibfnamefont {O.}~\bibnamefont {Mena}}, \
  and\ \bibinfo {author} {\bibfnamefont {S.}~\bibnamefont {Vagnozzi}},\ }\href
  {\doibase 10.1016/j.dark.2020.100666} {\bibfield  {journal} {\bibinfo
  {journal} {Phys. Dark Univ.}\ }\textbf {\bibinfo {volume} {30}},\ \bibinfo
  {pages} {100666} (\bibinfo {year} {2020}{\natexlab{a}})},\ \Eprint
  {http://arxiv.org/abs/1908.04281} {arXiv:1908.04281 [astro-ph.CO]}
  \BibitemShut {NoStop}%
\bibitem [{\citenamefont {Di~Valentino}\ \emph
  {et~al.}(2020{\natexlab{b}})\citenamefont {Di~Valentino}, \citenamefont
  {Melchiorri}, \citenamefont {Mena},\ and\ \citenamefont
  {Vagnozzi}}]{DiValentino:2019jae}%
  \BibitemOpen
  \bibfield  {author} {\bibinfo {author} {\bibfnamefont {E.}~\bibnamefont
  {Di~Valentino}}, \bibinfo {author} {\bibfnamefont {A.}~\bibnamefont
  {Melchiorri}}, \bibinfo {author} {\bibfnamefont {O.}~\bibnamefont {Mena}}, \
  and\ \bibinfo {author} {\bibfnamefont {S.}~\bibnamefont {Vagnozzi}},\ }\href
  {\doibase 10.1103/PhysRevD.101.063502} {\bibfield  {journal} {\bibinfo
  {journal} {Phys. Rev. D}\ }\textbf {\bibinfo {volume} {101}},\ \bibinfo
  {pages} {063502} (\bibinfo {year} {2020}{\natexlab{b}})},\ \Eprint
  {http://arxiv.org/abs/1910.09853} {arXiv:1910.09853 [astro-ph.CO]}
  \BibitemShut {NoStop}%
\bibitem [{\citenamefont {Lucca}\ and\ \citenamefont
  {Hooper}(2020)}]{Lucca:2020zjb}%
  \BibitemOpen
  \bibfield  {author} {\bibinfo {author} {\bibfnamefont {M.}~\bibnamefont
  {Lucca}}\ and\ \bibinfo {author} {\bibfnamefont {D.~C.}\ \bibnamefont
  {Hooper}},\ }\href {\doibase 10.1103/PhysRevD.102.123502} {\bibfield
  {journal} {\bibinfo  {journal} {Phys. Rev. D}\ }\textbf {\bibinfo {volume}
  {102}},\ \bibinfo {pages} {123502} (\bibinfo {year} {2020})},\ \Eprint
  {http://arxiv.org/abs/2002.06127} {arXiv:2002.06127 [astro-ph.CO]}
  \BibitemShut {NoStop}%
\bibitem [{\citenamefont {Anchordoqui}\ \emph {et~al.}(2021)\citenamefont
  {Anchordoqui}, \citenamefont {Di~Valentino}, \citenamefont {Pan},\ and\
  \citenamefont {Yang}}]{Anchordoqui:2021gji}%
  \BibitemOpen
  \bibfield  {author} {\bibinfo {author} {\bibfnamefont {L.~A.}\ \bibnamefont
  {Anchordoqui}}, \bibinfo {author} {\bibfnamefont {E.}~\bibnamefont
  {Di~Valentino}}, \bibinfo {author} {\bibfnamefont {S.}~\bibnamefont {Pan}}, \
  and\ \bibinfo {author} {\bibfnamefont {W.}~\bibnamefont {Yang}},\ }\href
  {\doibase 10.1016/j.jheap.2021.08.001} {\bibfield  {journal} {\bibinfo
  {journal} {JHEAp}\ }\textbf {\bibinfo {volume} {32}},\ \bibinfo {pages} {28}
  (\bibinfo {year} {2021})},\ \Eprint {http://arxiv.org/abs/2107.13932}
  {arXiv:2107.13932 [astro-ph.CO]} \BibitemShut {NoStop}%
\bibitem [{\citenamefont {Gao}\ \emph {et~al.}(2021)\citenamefont {Gao},
  \citenamefont {Zhao}, \citenamefont {Xue},\ and\ \citenamefont
  {Zhang}}]{Gao:2021xnk}%
  \BibitemOpen
  \bibfield  {author} {\bibinfo {author} {\bibfnamefont {L.-Y.}\ \bibnamefont
  {Gao}}, \bibinfo {author} {\bibfnamefont {Z.-W.}\ \bibnamefont {Zhao}},
  \bibinfo {author} {\bibfnamefont {S.-S.}\ \bibnamefont {Xue}}, \ and\
  \bibinfo {author} {\bibfnamefont {X.}~\bibnamefont {Zhang}},\ }\href
  {\doibase 10.1088/1475-7516/2021/07/005} {\bibfield  {journal} {\bibinfo
  {journal} {JCAP}\ }\textbf {\bibinfo {volume} {07}},\ \bibinfo {pages} {005}
  (\bibinfo {year} {2021})},\ \Eprint {http://arxiv.org/abs/2101.10714}
  {arXiv:2101.10714 [astro-ph.CO]} \BibitemShut {NoStop}%
\bibitem [{\citenamefont {Pourtsidou}\ and\ \citenamefont
  {Tram}(2016)}]{Pourtsidou:2016ico}%
  \BibitemOpen
  \bibfield  {author} {\bibinfo {author} {\bibfnamefont {A.}~\bibnamefont
  {Pourtsidou}}\ and\ \bibinfo {author} {\bibfnamefont {T.}~\bibnamefont
  {Tram}},\ }\href {\doibase 10.1103/PhysRevD.94.043518} {\bibfield  {journal}
  {\bibinfo  {journal} {Phys. Rev. D}\ }\textbf {\bibinfo {volume} {94}},\
  \bibinfo {pages} {043518} (\bibinfo {year} {2016})},\ \Eprint
  {http://arxiv.org/abs/1604.04222} {arXiv:1604.04222 [astro-ph.CO]}
  \BibitemShut {NoStop}%
\bibitem [{\citenamefont {An}\ \emph {et~al.}(2018)\citenamefont {An},
  \citenamefont {Feng},\ and\ \citenamefont {Wang}}]{An:2017crg}%
  \BibitemOpen
  \bibfield  {author} {\bibinfo {author} {\bibfnamefont {R.}~\bibnamefont
  {An}}, \bibinfo {author} {\bibfnamefont {C.}~\bibnamefont {Feng}}, \ and\
  \bibinfo {author} {\bibfnamefont {B.}~\bibnamefont {Wang}},\ }\href {\doibase
  10.1088/1475-7516/2018/02/038} {\bibfield  {journal} {\bibinfo  {journal}
  {JCAP}\ }\textbf {\bibinfo {volume} {02}},\ \bibinfo {pages} {038} (\bibinfo
  {year} {2018})},\ \Eprint {http://arxiv.org/abs/1711.06799} {arXiv:1711.06799
  [astro-ph.CO]} \BibitemShut {NoStop}%
\bibitem [{\citenamefont {Lucca}(2021)}]{Lucca:2021dxo}%
  \BibitemOpen
  \bibfield  {author} {\bibinfo {author} {\bibfnamefont {M.}~\bibnamefont
  {Lucca}},\ }\href {\doibase 10.1016/j.dark.2021.100899} {\bibfield  {journal}
  {\bibinfo  {journal} {Phys. Dark Univ.}\ }\textbf {\bibinfo {volume} {34}},\
  \bibinfo {pages} {100899} (\bibinfo {year} {2021})},\ \Eprint
  {http://arxiv.org/abs/2105.09249} {arXiv:2105.09249 [astro-ph.CO]}
  \BibitemShut {NoStop}%
\bibitem [{\citenamefont {Gariazzo}\ \emph {et~al.}(2022)\citenamefont
  {Gariazzo}, \citenamefont {Di~Valentino}, \citenamefont {Mena},\ and\
  \citenamefont {Nunes}}]{Gariazzo:2021qtg}%
  \BibitemOpen
  \bibfield  {author} {\bibinfo {author} {\bibfnamefont {S.}~\bibnamefont
  {Gariazzo}}, \bibinfo {author} {\bibfnamefont {E.}~\bibnamefont
  {Di~Valentino}}, \bibinfo {author} {\bibfnamefont {O.}~\bibnamefont {Mena}},
  \ and\ \bibinfo {author} {\bibfnamefont {R.~C.}\ \bibnamefont {Nunes}},\
  }\href {\doibase 10.1103/PhysRevD.106.023530} {\bibfield  {journal} {\bibinfo
   {journal} {Phys. Rev. D}\ }\textbf {\bibinfo {volume} {106}},\ \bibinfo
  {pages} {023530} (\bibinfo {year} {2022})},\ \Eprint
  {http://arxiv.org/abs/2111.03152} {arXiv:2111.03152 [astro-ph.CO]}
  \BibitemShut {NoStop}%
\bibitem [{\citenamefont {Yang}\ \emph
  {et~al.}(2021{\natexlab{a}})\citenamefont {Yang}, \citenamefont {Pan},
  \citenamefont {Di~Valentino}, \citenamefont {Mena},\ and\ \citenamefont
  {Melchiorri}}]{Yang:2021hxg}%
  \BibitemOpen
  \bibfield  {author} {\bibinfo {author} {\bibfnamefont {W.}~\bibnamefont
  {Yang}}, \bibinfo {author} {\bibfnamefont {S.}~\bibnamefont {Pan}}, \bibinfo
  {author} {\bibfnamefont {E.}~\bibnamefont {Di~Valentino}}, \bibinfo {author}
  {\bibfnamefont {O.}~\bibnamefont {Mena}}, \ and\ \bibinfo {author}
  {\bibfnamefont {A.}~\bibnamefont {Melchiorri}},\ }\href {\doibase
  10.1088/1475-7516/2021/10/008} {\bibfield  {journal} {\bibinfo  {journal}
  {JCAP}\ }\textbf {\bibinfo {volume} {10}},\ \bibinfo {pages} {008} (\bibinfo
  {year} {2021}{\natexlab{a}})},\ \Eprint {http://arxiv.org/abs/2101.03129}
  {arXiv:2101.03129 [astro-ph.CO]} \BibitemShut {NoStop}%
\bibitem [{\citenamefont {Yao}\ and\ \citenamefont {Meng}(2022)}]{Yao:2022kub}%
  \BibitemOpen
  \bibfield  {author} {\bibinfo {author} {\bibfnamefont {Y.-H.}\ \bibnamefont
  {Yao}}\ and\ \bibinfo {author} {\bibfnamefont {X.-H.}\ \bibnamefont {Meng}},\
  }\href@noop {} {\  (\bibinfo {year} {2022})},\ \Eprint
  {http://arxiv.org/abs/2207.05955} {arXiv:2207.05955 [astro-ph.CO]}
  \BibitemShut {NoStop}%
\bibitem [{\citenamefont {Wetterich}(1995)}]{Wetterich:1994bg}%
  \BibitemOpen
  \bibfield  {author} {\bibinfo {author} {\bibfnamefont {C.}~\bibnamefont
  {Wetterich}},\ }\href@noop {} {\bibfield  {journal} {\bibinfo  {journal}
  {Astron. Astrophys.}\ }\textbf {\bibinfo {volume} {301}},\ \bibinfo {pages}
  {321} (\bibinfo {year} {1995})},\ \Eprint
  {http://arxiv.org/abs/hep-th/9408025} {arXiv:hep-th/9408025} \BibitemShut
  {NoStop}%
\bibitem [{\citenamefont {Amendola}\ \emph {et~al.}(2007)\citenamefont
  {Amendola}, \citenamefont {Camargo~Campos},\ and\ \citenamefont
  {Rosenfeld}}]{Amendola:2006dg}%
  \BibitemOpen
  \bibfield  {author} {\bibinfo {author} {\bibfnamefont {L.}~\bibnamefont
  {Amendola}}, \bibinfo {author} {\bibfnamefont {G.}~\bibnamefont
  {Camargo~Campos}}, \ and\ \bibinfo {author} {\bibfnamefont {R.}~\bibnamefont
  {Rosenfeld}},\ }\href {\doibase 10.1103/PhysRevD.75.083506} {\bibfield
  {journal} {\bibinfo  {journal} {Phys. Rev. D}\ }\textbf {\bibinfo {volume}
  {75}},\ \bibinfo {pages} {083506} (\bibinfo {year} {2007})},\ \Eprint
  {http://arxiv.org/abs/astro-ph/0610806} {arXiv:astro-ph/0610806} \BibitemShut
  {NoStop}%
\bibitem [{\citenamefont {Barrow}\ and\ \citenamefont
  {Clifton}(2006)}]{Barrow:2006hia}%
  \BibitemOpen
  \bibfield  {author} {\bibinfo {author} {\bibfnamefont {J.~D.}\ \bibnamefont
  {Barrow}}\ and\ \bibinfo {author} {\bibfnamefont {T.}~\bibnamefont
  {Clifton}},\ }\href {\doibase 10.1103/PhysRevD.73.103520} {\bibfield
  {journal} {\bibinfo  {journal} {Phys. Rev. D}\ }\textbf {\bibinfo {volume}
  {73}},\ \bibinfo {pages} {103520} (\bibinfo {year} {2006})},\ \Eprint
  {http://arxiv.org/abs/gr-qc/0604063} {arXiv:gr-qc/0604063} \BibitemShut
  {NoStop}%
\bibitem [{\citenamefont {Berger}\ and\ \citenamefont
  {Shojaei}(2006)}]{Berger:2006db}%
  \BibitemOpen
  \bibfield  {author} {\bibinfo {author} {\bibfnamefont {M.~S.}\ \bibnamefont
  {Berger}}\ and\ \bibinfo {author} {\bibfnamefont {H.}~\bibnamefont
  {Shojaei}},\ }\href {\doibase 10.1103/PhysRevD.73.083528} {\bibfield
  {journal} {\bibinfo  {journal} {Phys. Rev. D}\ }\textbf {\bibinfo {volume}
  {73}},\ \bibinfo {pages} {083528} (\bibinfo {year} {2006})},\ \Eprint
  {http://arxiv.org/abs/gr-qc/0601086} {arXiv:gr-qc/0601086} \BibitemShut
  {NoStop}%
\bibitem [{\citenamefont {Valiviita}\ \emph {et~al.}(2008)\citenamefont
  {Valiviita}, \citenamefont {Majerotto},\ and\ \citenamefont
  {Maartens}}]{Valiviita:2008iv}%
  \BibitemOpen
  \bibfield  {author} {\bibinfo {author} {\bibfnamefont {J.}~\bibnamefont
  {Valiviita}}, \bibinfo {author} {\bibfnamefont {E.}~\bibnamefont
  {Majerotto}}, \ and\ \bibinfo {author} {\bibfnamefont {R.}~\bibnamefont
  {Maartens}},\ }\href {\doibase 10.1088/1475-7516/2008/07/020} {\bibfield
  {journal} {\bibinfo  {journal} {JCAP}\ }\textbf {\bibinfo {volume} {07}},\
  \bibinfo {pages} {020} (\bibinfo {year} {2008})},\ \Eprint
  {http://arxiv.org/abs/0804.0232} {arXiv:0804.0232 [astro-ph]} \BibitemShut
  {NoStop}%
\bibitem [{\citenamefont {Gavela}\ \emph {et~al.}(2009)\citenamefont {Gavela},
  \citenamefont {Hernandez}, \citenamefont {Lopez~Honorez}, \citenamefont
  {Mena},\ and\ \citenamefont {Rigolin}}]{Gavela:2009cy}%
  \BibitemOpen
  \bibfield  {author} {\bibinfo {author} {\bibfnamefont {M.~B.}\ \bibnamefont
  {Gavela}}, \bibinfo {author} {\bibfnamefont {D.}~\bibnamefont {Hernandez}},
  \bibinfo {author} {\bibfnamefont {L.}~\bibnamefont {Lopez~Honorez}}, \bibinfo
  {author} {\bibfnamefont {O.}~\bibnamefont {Mena}}, \ and\ \bibinfo {author}
  {\bibfnamefont {S.}~\bibnamefont {Rigolin}},\ }\href {\doibase
  10.1088/1475-7516/2009/07/034} {\bibfield  {journal} {\bibinfo  {journal}
  {JCAP}\ }\textbf {\bibinfo {volume} {07}},\ \bibinfo {pages} {034} (\bibinfo
  {year} {2009})},\ \bibinfo {note} {[Erratum: JCAP 05, E01 (2010)]},\ \Eprint
  {http://arxiv.org/abs/0901.1611} {arXiv:0901.1611 [astro-ph.CO]} \BibitemShut
  {NoStop}%
\bibitem [{\citenamefont {Majerotto}\ \emph {et~al.}(2010)\citenamefont
  {Majerotto}, \citenamefont {Valiviita},\ and\ \citenamefont
  {Maartens}}]{Majerotto:2009np}%
  \BibitemOpen
  \bibfield  {author} {\bibinfo {author} {\bibfnamefont {E.}~\bibnamefont
  {Majerotto}}, \bibinfo {author} {\bibfnamefont {J.}~\bibnamefont
  {Valiviita}}, \ and\ \bibinfo {author} {\bibfnamefont {R.}~\bibnamefont
  {Maartens}},\ }\href {\doibase 10.1111/j.1365-2966.2009.16140.x} {\bibfield
  {journal} {\bibinfo  {journal} {Mon. Not. Roy. Astron. Soc.}\ }\textbf
  {\bibinfo {volume} {402}},\ \bibinfo {pages} {2344} (\bibinfo {year}
  {2010})},\ \Eprint {http://arxiv.org/abs/0907.4981} {arXiv:0907.4981
  [astro-ph.CO]} \BibitemShut {NoStop}%
\bibitem [{\citenamefont {Gavela}\ \emph {et~al.}(2010)\citenamefont {Gavela},
  \citenamefont {Lopez~Honorez}, \citenamefont {Mena},\ and\ \citenamefont
  {Rigolin}}]{Gavela:2010tm}%
  \BibitemOpen
  \bibfield  {author} {\bibinfo {author} {\bibfnamefont {M.~B.}\ \bibnamefont
  {Gavela}}, \bibinfo {author} {\bibfnamefont {L.}~\bibnamefont
  {Lopez~Honorez}}, \bibinfo {author} {\bibfnamefont {O.}~\bibnamefont {Mena}},
  \ and\ \bibinfo {author} {\bibfnamefont {S.}~\bibnamefont {Rigolin}},\ }\href
  {\doibase 10.1088/1475-7516/2010/11/044} {\bibfield  {journal} {\bibinfo
  {journal} {JCAP}\ }\textbf {\bibinfo {volume} {11}},\ \bibinfo {pages} {044}
  (\bibinfo {year} {2010})},\ \Eprint {http://arxiv.org/abs/1005.0295}
  {arXiv:1005.0295 [astro-ph.CO]} \BibitemShut {NoStop}%
\bibitem [{\citenamefont {Clemson}\ \emph {et~al.}(2012)\citenamefont
  {Clemson}, \citenamefont {Koyama}, \citenamefont {Zhao}, \citenamefont
  {Maartens},\ and\ \citenamefont {Valiviita}}]{Clemson:2011an}%
  \BibitemOpen
  \bibfield  {author} {\bibinfo {author} {\bibfnamefont {T.}~\bibnamefont
  {Clemson}}, \bibinfo {author} {\bibfnamefont {K.}~\bibnamefont {Koyama}},
  \bibinfo {author} {\bibfnamefont {G.-B.}\ \bibnamefont {Zhao}}, \bibinfo
  {author} {\bibfnamefont {R.}~\bibnamefont {Maartens}}, \ and\ \bibinfo
  {author} {\bibfnamefont {J.}~\bibnamefont {Valiviita}},\ }\href {\doibase
  10.1103/PhysRevD.85.043007} {\bibfield  {journal} {\bibinfo  {journal} {Phys.
  Rev. D}\ }\textbf {\bibinfo {volume} {85}},\ \bibinfo {pages} {043007}
  (\bibinfo {year} {2012})},\ \Eprint {http://arxiv.org/abs/1109.6234}
  {arXiv:1109.6234 [astro-ph.CO]} \BibitemShut {NoStop}%
\bibitem [{\citenamefont {Salvatelli}\ \emph {et~al.}(2014)\citenamefont
  {Salvatelli}, \citenamefont {Said}, \citenamefont {Bruni}, \citenamefont
  {Melchiorri},\ and\ \citenamefont {Wands}}]{Salvatelli:2014zta}%
  \BibitemOpen
  \bibfield  {author} {\bibinfo {author} {\bibfnamefont {V.}~\bibnamefont
  {Salvatelli}}, \bibinfo {author} {\bibfnamefont {N.}~\bibnamefont {Said}},
  \bibinfo {author} {\bibfnamefont {M.}~\bibnamefont {Bruni}}, \bibinfo
  {author} {\bibfnamefont {A.}~\bibnamefont {Melchiorri}}, \ and\ \bibinfo
  {author} {\bibfnamefont {D.}~\bibnamefont {Wands}},\ }\href {\doibase
  10.1103/PhysRevLett.113.181301} {\bibfield  {journal} {\bibinfo  {journal}
  {Phys. Rev. Lett.}\ }\textbf {\bibinfo {volume} {113}},\ \bibinfo {pages}
  {181301} (\bibinfo {year} {2014})},\ \Eprint {http://arxiv.org/abs/1406.7297}
  {arXiv:1406.7297 [astro-ph.CO]} \BibitemShut {NoStop}%
\bibitem [{\citenamefont {Wang}\ \emph {et~al.}(2014)\citenamefont {Wang},
  \citenamefont {Wands}, \citenamefont {Zhao},\ and\ \citenamefont
  {Xu}}]{Wang:2014xca}%
  \BibitemOpen
  \bibfield  {author} {\bibinfo {author} {\bibfnamefont {Y.}~\bibnamefont
  {Wang}}, \bibinfo {author} {\bibfnamefont {D.}~\bibnamefont {Wands}},
  \bibinfo {author} {\bibfnamefont {G.-B.}\ \bibnamefont {Zhao}}, \ and\
  \bibinfo {author} {\bibfnamefont {L.}~\bibnamefont {Xu}},\ }\href {\doibase
  10.1103/PhysRevD.90.023502} {\bibfield  {journal} {\bibinfo  {journal} {Phys.
  Rev. D}\ }\textbf {\bibinfo {volume} {90}},\ \bibinfo {pages} {023502}
  (\bibinfo {year} {2014})},\ \Eprint {http://arxiv.org/abs/1404.5706}
  {arXiv:1404.5706 [astro-ph.CO]} \BibitemShut {NoStop}%
\bibitem [{\citenamefont {Yang}\ and\ \citenamefont
  {Xu}(2014{\natexlab{a}})}]{Yang:2014gza}%
  \BibitemOpen
  \bibfield  {author} {\bibinfo {author} {\bibfnamefont {W.}~\bibnamefont
  {Yang}}\ and\ \bibinfo {author} {\bibfnamefont {L.}~\bibnamefont {Xu}},\
  }\href {\doibase 10.1103/PhysRevD.89.083517} {\bibfield  {journal} {\bibinfo
  {journal} {Phys. Rev. D}\ }\textbf {\bibinfo {volume} {89}},\ \bibinfo
  {pages} {083517} (\bibinfo {year} {2014}{\natexlab{a}})},\ \Eprint
  {http://arxiv.org/abs/1401.1286} {arXiv:1401.1286 [astro-ph.CO]} \BibitemShut
  {NoStop}%
\bibitem [{\citenamefont {Yang}\ and\ \citenamefont
  {Xu}(2014{\natexlab{b}})}]{yang:2014vza}%
  \BibitemOpen
  \bibfield  {author} {\bibinfo {author} {\bibfnamefont {W.}~\bibnamefont
  {Yang}}\ and\ \bibinfo {author} {\bibfnamefont {L.}~\bibnamefont {Xu}},\
  }\href {\doibase 10.1088/1475-7516/2014/08/034} {\bibfield  {journal}
  {\bibinfo  {journal} {JCAP}\ }\textbf {\bibinfo {volume} {08}},\ \bibinfo
  {pages} {034} (\bibinfo {year} {2014}{\natexlab{b}})},\ \Eprint
  {http://arxiv.org/abs/1401.5177} {arXiv:1401.5177 [astro-ph.CO]} \BibitemShut
  {NoStop}%
\bibitem [{\citenamefont {Yang}\ and\ \citenamefont
  {Xu}(2014{\natexlab{c}})}]{Yang:2014hea}%
  \BibitemOpen
  \bibfield  {author} {\bibinfo {author} {\bibfnamefont {W.}~\bibnamefont
  {Yang}}\ and\ \bibinfo {author} {\bibfnamefont {L.}~\bibnamefont {Xu}},\
  }\href {\doibase 10.1103/PhysRevD.90.083532} {\bibfield  {journal} {\bibinfo
  {journal} {Phys. Rev. D}\ }\textbf {\bibinfo {volume} {90}},\ \bibinfo
  {pages} {083532} (\bibinfo {year} {2014}{\natexlab{c}})},\ \Eprint
  {http://arxiv.org/abs/1409.5533} {arXiv:1409.5533 [astro-ph.CO]} \BibitemShut
  {NoStop}%
\bibitem [{\citenamefont {Li}\ \emph {et~al.}(2016)\citenamefont {Li},
  \citenamefont {Zhang},\ and\ \citenamefont {Zhang}}]{Li:2015vla}%
  \BibitemOpen
  \bibfield  {author} {\bibinfo {author} {\bibfnamefont {Y.-H.}\ \bibnamefont
  {Li}}, \bibinfo {author} {\bibfnamefont {J.-F.}\ \bibnamefont {Zhang}}, \
  and\ \bibinfo {author} {\bibfnamefont {X.}~\bibnamefont {Zhang}},\ }\href
  {\doibase 10.1103/PhysRevD.93.023002} {\bibfield  {journal} {\bibinfo
  {journal} {Phys. Rev. D}\ }\textbf {\bibinfo {volume} {93}},\ \bibinfo
  {pages} {023002} (\bibinfo {year} {2016})},\ \Eprint
  {http://arxiv.org/abs/1506.06349} {arXiv:1506.06349 [astro-ph.CO]}
  \BibitemShut {NoStop}%
\bibitem [{\citenamefont {van~de Bruck}\ \emph {et~al.}(2016)\citenamefont
  {van~de Bruck}, \citenamefont {Mifsud}, \citenamefont {Mimoso},\ and\
  \citenamefont {Nunes}}]{vandeBruck:2016jgg}%
  \BibitemOpen
  \bibfield  {author} {\bibinfo {author} {\bibfnamefont {C.}~\bibnamefont
  {van~de Bruck}}, \bibinfo {author} {\bibfnamefont {J.}~\bibnamefont
  {Mifsud}}, \bibinfo {author} {\bibfnamefont {J.~P.}\ \bibnamefont {Mimoso}},
  \ and\ \bibinfo {author} {\bibfnamefont {N.~J.}\ \bibnamefont {Nunes}},\
  }\href {\doibase 10.1088/1475-7516/2016/11/031} {\bibfield  {journal}
  {\bibinfo  {journal} {JCAP}\ }\textbf {\bibinfo {volume} {11}},\ \bibinfo
  {pages} {031} (\bibinfo {year} {2016})},\ \Eprint
  {http://arxiv.org/abs/1605.03834} {arXiv:1605.03834 [gr-qc]} \BibitemShut
  {NoStop}%
\bibitem [{\citenamefont {Feng}\ and\ \citenamefont
  {Zhang}(2016)}]{Feng:2016djj}%
  \BibitemOpen
  \bibfield  {author} {\bibinfo {author} {\bibfnamefont {L.}~\bibnamefont
  {Feng}}\ and\ \bibinfo {author} {\bibfnamefont {X.}~\bibnamefont {Zhang}},\
  }\href {\doibase 10.1088/1475-7516/2016/08/072} {\bibfield  {journal}
  {\bibinfo  {journal} {JCAP}\ }\textbf {\bibinfo {volume} {08}},\ \bibinfo
  {pages} {072} (\bibinfo {year} {2016})},\ \Eprint
  {http://arxiv.org/abs/1607.05567} {arXiv:1607.05567 [astro-ph.CO]}
  \BibitemShut {NoStop}%
\bibitem [{\citenamefont {van~de Bruck}\ \emph {et~al.}(2017)\citenamefont
  {van~de Bruck}, \citenamefont {Mifsud},\ and\ \citenamefont
  {Morrice}}]{vandeBruck:2016hpz}%
  \BibitemOpen
  \bibfield  {author} {\bibinfo {author} {\bibfnamefont {C.}~\bibnamefont
  {van~de Bruck}}, \bibinfo {author} {\bibfnamefont {J.}~\bibnamefont
  {Mifsud}}, \ and\ \bibinfo {author} {\bibfnamefont {J.}~\bibnamefont
  {Morrice}},\ }\href {\doibase 10.1103/PhysRevD.95.043513} {\bibfield
  {journal} {\bibinfo  {journal} {Phys. Rev. D}\ }\textbf {\bibinfo {volume}
  {95}},\ \bibinfo {pages} {043513} (\bibinfo {year} {2017})},\ \Eprint
  {http://arxiv.org/abs/1609.09855} {arXiv:1609.09855 [astro-ph.CO]}
  \BibitemShut {NoStop}%
\bibitem [{\citenamefont {Sharov}\ \emph {et~al.}(2017)\citenamefont {Sharov},
  \citenamefont {Bhattacharya}, \citenamefont {Pan}, \citenamefont {Nunes},\
  and\ \citenamefont {Chakraborty}}]{Sharov:2017iue}%
  \BibitemOpen
  \bibfield  {author} {\bibinfo {author} {\bibfnamefont {G.~S.}\ \bibnamefont
  {Sharov}}, \bibinfo {author} {\bibfnamefont {S.}~\bibnamefont
  {Bhattacharya}}, \bibinfo {author} {\bibfnamefont {S.}~\bibnamefont {Pan}},
  \bibinfo {author} {\bibfnamefont {R.~C.}\ \bibnamefont {Nunes}}, \ and\
  \bibinfo {author} {\bibfnamefont {S.}~\bibnamefont {Chakraborty}},\ }\href
  {\doibase 10.1093/mnras/stw3358} {\bibfield  {journal} {\bibinfo  {journal}
  {Mon. Not. Roy. Astron. Soc.}\ }\textbf {\bibinfo {volume} {466}},\ \bibinfo
  {pages} {3497} (\bibinfo {year} {2017})},\ \Eprint
  {http://arxiv.org/abs/1701.00780} {arXiv:1701.00780 [gr-qc]} \BibitemShut
  {NoStop}%
\bibitem [{\citenamefont {Mifsud}\ and\ \citenamefont {Van
  De~Bruck}(2017)}]{Mifsud:2017fsy}%
  \BibitemOpen
  \bibfield  {author} {\bibinfo {author} {\bibfnamefont {J.}~\bibnamefont
  {Mifsud}}\ and\ \bibinfo {author} {\bibfnamefont {C.}~\bibnamefont {Van
  De~Bruck}},\ }\href {\doibase 10.1088/1475-7516/2017/11/001} {\bibfield
  {journal} {\bibinfo  {journal} {JCAP}\ }\textbf {\bibinfo {volume} {11}},\
  \bibinfo {pages} {001} (\bibinfo {year} {2017})},\ \Eprint
  {http://arxiv.org/abs/1707.07667} {arXiv:1707.07667 [astro-ph.CO]}
  \BibitemShut {NoStop}%
\bibitem [{\citenamefont {Van De~Bruck}\ and\ \citenamefont
  {Mifsud}(2018)}]{VanDeBruck:2017mua}%
  \BibitemOpen
  \bibfield  {author} {\bibinfo {author} {\bibfnamefont {C.}~\bibnamefont {Van
  De~Bruck}}\ and\ \bibinfo {author} {\bibfnamefont {J.}~\bibnamefont
  {Mifsud}},\ }\href {\doibase 10.1103/PhysRevD.97.023506} {\bibfield
  {journal} {\bibinfo  {journal} {Phys. Rev. D}\ }\textbf {\bibinfo {volume}
  {97}},\ \bibinfo {pages} {023506} (\bibinfo {year} {2018})},\ \Eprint
  {http://arxiv.org/abs/1709.04882} {arXiv:1709.04882 [astro-ph.CO]}
  \BibitemShut {NoStop}%
\bibitem [{\citenamefont {Yang}\ \emph
  {et~al.}(2019{\natexlab{a}})\citenamefont {Yang}, \citenamefont {Pan},\ and\
  \citenamefont {Paliathanasis}}]{Yang:2018pej}%
  \BibitemOpen
  \bibfield  {author} {\bibinfo {author} {\bibfnamefont {W.}~\bibnamefont
  {Yang}}, \bibinfo {author} {\bibfnamefont {S.}~\bibnamefont {Pan}}, \ and\
  \bibinfo {author} {\bibfnamefont {A.}~\bibnamefont {Paliathanasis}},\ }\href
  {\doibase 10.1093/mnras/sty2780} {\bibfield  {journal} {\bibinfo  {journal}
  {Mon. Not. Roy. Astron. Soc.}\ }\textbf {\bibinfo {volume} {482}},\ \bibinfo
  {pages} {1007} (\bibinfo {year} {2019}{\natexlab{a}})},\ \Eprint
  {http://arxiv.org/abs/1804.08558} {arXiv:1804.08558 [gr-qc]} \BibitemShut
  {NoStop}%
\bibitem [{\citenamefont {Li}\ \emph {et~al.}(2018)\citenamefont {Li},
  \citenamefont {Yang}, \citenamefont {Wu},\ and\ \citenamefont
  {Jiang}}]{Li:2018jiu}%
  \BibitemOpen
  \bibfield  {author} {\bibinfo {author} {\bibfnamefont {H.}~\bibnamefont
  {Li}}, \bibinfo {author} {\bibfnamefont {W.}~\bibnamefont {Yang}}, \bibinfo
  {author} {\bibfnamefont {Y.}~\bibnamefont {Wu}}, \ and\ \bibinfo {author}
  {\bibfnamefont {Y.}~\bibnamefont {Jiang}},\ }\href {\doibase
  10.1016/j.dark.2018.04.001} {\bibfield  {journal} {\bibinfo  {journal} {Phys.
  Dark Univ.}\ }\textbf {\bibinfo {volume} {20}},\ \bibinfo {pages} {78}
  (\bibinfo {year} {2018})}\BibitemShut {NoStop}%
\bibitem [{\citenamefont {Martinelli}\ \emph {et~al.}(2019)\citenamefont
  {Martinelli}, \citenamefont {Hogg}, \citenamefont {Peirone}, \citenamefont
  {Bruni},\ and\ \citenamefont {Wands}}]{Martinelli:2019dau}%
  \BibitemOpen
  \bibfield  {author} {\bibinfo {author} {\bibfnamefont {M.}~\bibnamefont
  {Martinelli}}, \bibinfo {author} {\bibfnamefont {N.~B.}\ \bibnamefont
  {Hogg}}, \bibinfo {author} {\bibfnamefont {S.}~\bibnamefont {Peirone}},
  \bibinfo {author} {\bibfnamefont {M.}~\bibnamefont {Bruni}}, \ and\ \bibinfo
  {author} {\bibfnamefont {D.}~\bibnamefont {Wands}},\ }\href {\doibase
  10.1093/mnras/stz1915} {\bibfield  {journal} {\bibinfo  {journal} {Mon. Not.
  Roy. Astron. Soc.}\ }\textbf {\bibinfo {volume} {488}},\ \bibinfo {pages}
  {3423} (\bibinfo {year} {2019})},\ \Eprint {http://arxiv.org/abs/1902.10694}
  {arXiv:1902.10694 [astro-ph.CO]} \BibitemShut {NoStop}%
\bibitem [{\citenamefont {Mifsud}\ and\ \citenamefont {van~de
  Bruck}(2019)}]{Mifsud:2019fut}%
  \BibitemOpen
  \bibfield  {author} {\bibinfo {author} {\bibfnamefont {J.}~\bibnamefont
  {Mifsud}}\ and\ \bibinfo {author} {\bibfnamefont {C.}~\bibnamefont {van~de
  Bruck}},\ }\href {\doibase 10.1093/mnras/stz1293} {\bibfield  {journal}
  {\bibinfo  {journal} {Mon. Not. Roy. Astron. Soc.}\ }\textbf {\bibinfo
  {volume} {487}},\ \bibinfo {pages} {900} (\bibinfo {year}
  {2019})}\BibitemShut {NoStop}%
\bibitem [{\citenamefont {Oikonomou}(2019)}]{Oikonomou:2019boy}%
  \BibitemOpen
  \bibfield  {author} {\bibinfo {author} {\bibfnamefont {V.~K.}\ \bibnamefont
  {Oikonomou}},\ }\href {\doibase 10.1103/PhysRevD.99.104042} {\bibfield
  {journal} {\bibinfo  {journal} {Phys. Rev. D}\ }\textbf {\bibinfo {volume}
  {99}},\ \bibinfo {pages} {104042} (\bibinfo {year} {2019})},\ \Eprint
  {http://arxiv.org/abs/1905.00826} {arXiv:1905.00826 [gr-qc]} \BibitemShut
  {NoStop}%
\bibitem [{\citenamefont {Cheng}\ \emph {et~al.}(2020)\citenamefont {Cheng},
  \citenamefont {Ma}, \citenamefont {Wu}, \citenamefont {Zhang},\ and\
  \citenamefont {Chen}}]{Cheng:2019bkh}%
  \BibitemOpen
  \bibfield  {author} {\bibinfo {author} {\bibfnamefont {G.}~\bibnamefont
  {Cheng}}, \bibinfo {author} {\bibfnamefont {Y.-Z.}\ \bibnamefont {Ma}},
  \bibinfo {author} {\bibfnamefont {F.}~\bibnamefont {Wu}}, \bibinfo {author}
  {\bibfnamefont {J.}~\bibnamefont {Zhang}}, \ and\ \bibinfo {author}
  {\bibfnamefont {X.}~\bibnamefont {Chen}},\ }\href {\doibase
  10.1103/PhysRevD.102.043517} {\bibfield  {journal} {\bibinfo  {journal}
  {Phys. Rev. D}\ }\textbf {\bibinfo {volume} {102}},\ \bibinfo {pages}
  {043517} (\bibinfo {year} {2020})},\ \Eprint
  {http://arxiv.org/abs/1911.04520} {arXiv:1911.04520 [astro-ph.CO]}
  \BibitemShut {NoStop}%
\bibitem [{\citenamefont {Yao}\ and\ \citenamefont
  {Meng}(2020{\natexlab{a}})}]{Yao:2020hkw}%
  \BibitemOpen
  \bibfield  {author} {\bibinfo {author} {\bibfnamefont {Y.-H.}\ \bibnamefont
  {Yao}}\ and\ \bibinfo {author} {\bibfnamefont {X.-H.}\ \bibnamefont {Meng}},\
  }\href {\doibase 10.1016/j.dark.2020.100729} {\bibfield  {journal} {\bibinfo
  {journal} {Phys. Dark Univ.}\ }\textbf {\bibinfo {volume} {30}},\ \bibinfo
  {pages} {100729} (\bibinfo {year} {2020}{\natexlab{a}})}\BibitemShut
  {NoStop}%
\bibitem [{\citenamefont {Yao}\ and\ \citenamefont
  {Meng}(2020{\natexlab{b}})}]{Yao:2020pji}%
  \BibitemOpen
  \bibfield  {author} {\bibinfo {author} {\bibfnamefont {Y.}~\bibnamefont
  {Yao}}\ and\ \bibinfo {author} {\bibfnamefont {X.}~\bibnamefont {Meng}},\
  }\href@noop {} {\  (\bibinfo {year} {2020}{\natexlab{b}})},\ \Eprint
  {http://arxiv.org/abs/2011.09160} {arXiv:2011.09160 [astro-ph.CO]}
  \BibitemShut {NoStop}%
\bibitem [{\citenamefont {Jesus}\ \emph {et~al.}(2022)\citenamefont {Jesus},
  \citenamefont {Escobal}, \citenamefont {Benndorf},\ and\ \citenamefont
  {Pereira}}]{Jesus:2020tby}%
  \BibitemOpen
  \bibfield  {author} {\bibinfo {author} {\bibfnamefont {J.~F.}\ \bibnamefont
  {Jesus}}, \bibinfo {author} {\bibfnamefont {A.~A.}\ \bibnamefont {Escobal}},
  \bibinfo {author} {\bibfnamefont {D.}~\bibnamefont {Benndorf}}, \ and\
  \bibinfo {author} {\bibfnamefont {S.~H.}\ \bibnamefont {Pereira}},\ }\href
  {\doibase 10.1140/epjc/s10052-022-10234-y} {\bibfield  {journal} {\bibinfo
  {journal} {Eur. Phys. J. C}\ }\textbf {\bibinfo {volume} {82}},\ \bibinfo
  {pages} {273} (\bibinfo {year} {2022})},\ \Eprint
  {http://arxiv.org/abs/2012.07494} {arXiv:2012.07494 [astro-ph.CO]}
  \BibitemShut {NoStop}%
\bibitem [{\citenamefont {S\'a}(2020)}]{Sa:2020fvn}%
  \BibitemOpen
  \bibfield  {author} {\bibinfo {author} {\bibfnamefont {P.~M.}\ \bibnamefont
  {S\'a}},\ }\href {\doibase 10.1103/PhysRevD.102.103519} {\bibfield  {journal}
  {\bibinfo  {journal} {Phys. Rev. D}\ }\textbf {\bibinfo {volume} {102}},\
  \bibinfo {pages} {103519} (\bibinfo {year} {2020})},\ \Eprint
  {http://arxiv.org/abs/2007.07109} {arXiv:2007.07109 [gr-qc]} \BibitemShut
  {NoStop}%
\bibitem [{\citenamefont {S\'a}(2021)}]{Sa:2021eft}%
  \BibitemOpen
  \bibfield  {author} {\bibinfo {author} {\bibfnamefont {P.~M.}\ \bibnamefont
  {S\'a}},\ }\href {\doibase 10.1103/PhysRevD.103.123517} {\bibfield  {journal}
  {\bibinfo  {journal} {Phys. Rev. D}\ }\textbf {\bibinfo {volume} {103}},\
  \bibinfo {pages} {123517} (\bibinfo {year} {2021})},\ \Eprint
  {http://arxiv.org/abs/2103.01693} {arXiv:2103.01693 [gr-qc]} \BibitemShut
  {NoStop}%
\bibitem [{\citenamefont {Yang}\ \emph
  {et~al.}(2021{\natexlab{b}})\citenamefont {Yang}, \citenamefont {Pan},
  \citenamefont {Arest\'e~Sal\'o},\ and\ \citenamefont
  {de~Haro}}]{Yang:2021oxc}%
  \BibitemOpen
  \bibfield  {author} {\bibinfo {author} {\bibfnamefont {W.}~\bibnamefont
  {Yang}}, \bibinfo {author} {\bibfnamefont {S.}~\bibnamefont {Pan}}, \bibinfo
  {author} {\bibfnamefont {L.}~\bibnamefont {Arest\'e~Sal\'o}}, \ and\ \bibinfo
  {author} {\bibfnamefont {J.}~\bibnamefont {de~Haro}},\ }\href {\doibase
  10.1103/PhysRevD.103.083520} {\bibfield  {journal} {\bibinfo  {journal}
  {Phys. Rev. D}\ }\textbf {\bibinfo {volume} {103}},\ \bibinfo {pages}
  {083520} (\bibinfo {year} {2021}{\natexlab{b}})},\ \Eprint
  {http://arxiv.org/abs/2104.04505} {arXiv:2104.04505 [astro-ph.CO]}
  \BibitemShut {NoStop}%
\bibitem [{\citenamefont {Kang}(2021)}]{Kang:2021osc}%
  \BibitemOpen
  \bibfield  {author} {\bibinfo {author} {\bibfnamefont {J.}~\bibnamefont
  {Kang}},\ }\href {\doibase 10.1016/j.dark.2021.100784} {\bibfield  {journal}
  {\bibinfo  {journal} {Phys. Dark Univ.}\ }\textbf {\bibinfo {volume} {31}},\
  \bibinfo {pages} {100784} (\bibinfo {year} {2021})},\ \Eprint
  {http://arxiv.org/abs/2102.04232} {arXiv:2102.04232 [astro-ph.CO]}
  \BibitemShut {NoStop}%
\bibitem [{\citenamefont {Bonilla}\ \emph {et~al.}(2022)\citenamefont
  {Bonilla}, \citenamefont {Kumar}, \citenamefont {Nunes},\ and\ \citenamefont
  {Pan}}]{Bonilla:2021dql}%
  \BibitemOpen
  \bibfield  {author} {\bibinfo {author} {\bibfnamefont {A.}~\bibnamefont
  {Bonilla}}, \bibinfo {author} {\bibfnamefont {S.}~\bibnamefont {Kumar}},
  \bibinfo {author} {\bibfnamefont {R.~C.}\ \bibnamefont {Nunes}}, \ and\
  \bibinfo {author} {\bibfnamefont {S.}~\bibnamefont {Pan}},\ }\href {\doibase
  10.1093/mnras/stac687} {\bibfield  {journal} {\bibinfo  {journal} {Mon. Not.
  Roy. Astron. Soc.}\ }\textbf {\bibinfo {volume} {512}},\ \bibinfo {pages}
  {4231} (\bibinfo {year} {2022})},\ \Eprint {http://arxiv.org/abs/2102.06149}
  {arXiv:2102.06149 [astro-ph.CO]} \BibitemShut {NoStop}%
\bibitem [{\citenamefont {Harko}\ \emph {et~al.}(2022)\citenamefont {Harko},
  \citenamefont {Asadi}, \citenamefont {Moshafi},\ and\ \citenamefont
  {Sheikhahmadi}}]{Harko:2022unn}%
  \BibitemOpen
  \bibfield  {author} {\bibinfo {author} {\bibfnamefont {T.}~\bibnamefont
  {Harko}}, \bibinfo {author} {\bibfnamefont {K.}~\bibnamefont {Asadi}},
  \bibinfo {author} {\bibfnamefont {H.}~\bibnamefont {Moshafi}}, \ and\
  \bibinfo {author} {\bibfnamefont {H.}~\bibnamefont {Sheikhahmadi}},\ }\href
  {\doibase 10.1016/j.dark.2022.101131} {\bibfield  {journal} {\bibinfo
  {journal} {Phys. Dark Univ.}\ }\textbf {\bibinfo {volume} {38}},\ \bibinfo
  {pages} {101131} (\bibinfo {year} {2022})},\ \Eprint
  {http://arxiv.org/abs/2203.08907} {arXiv:2203.08907 [gr-qc]} \BibitemShut
  {NoStop}%
\bibitem [{\citenamefont {Yengejeh}\ \emph {et~al.}(2023)\citenamefont
  {Yengejeh}, \citenamefont {Fakhry}, \citenamefont {Firouzjaee},\ and\
  \citenamefont {Fathi}}]{Yengejeh:2022tpa}%
  \BibitemOpen
  \bibfield  {author} {\bibinfo {author} {\bibfnamefont {M.~G.}\ \bibnamefont
  {Yengejeh}}, \bibinfo {author} {\bibfnamefont {S.}~\bibnamefont {Fakhry}},
  \bibinfo {author} {\bibfnamefont {J.~T.}\ \bibnamefont {Firouzjaee}}, \ and\
  \bibinfo {author} {\bibfnamefont {H.}~\bibnamefont {Fathi}},\ }\href
  {\doibase 10.1016/j.dark.2022.101144} {\bibfield  {journal} {\bibinfo
  {journal} {Phys. Dark Univ.}\ }\textbf {\bibinfo {volume} {39}},\ \bibinfo
  {pages} {101144} (\bibinfo {year} {2023})},\ \Eprint
  {http://arxiv.org/abs/2206.01030} {arXiv:2206.01030 [astro-ph.CO]}
  \BibitemShut {NoStop}%
\bibitem [{\citenamefont {Chatzidakis}\ \emph {et~al.}(2022)\citenamefont
  {Chatzidakis}, \citenamefont {Giacomini}, \citenamefont {Leach},
  \citenamefont {Leon}, \citenamefont {Paliathanasis},\ and\ \citenamefont
  {Pan}}]{Chatzidakis:2022mpf}%
  \BibitemOpen
  \bibfield  {author} {\bibinfo {author} {\bibfnamefont {S.}~\bibnamefont
  {Chatzidakis}}, \bibinfo {author} {\bibfnamefont {A.}~\bibnamefont
  {Giacomini}}, \bibinfo {author} {\bibfnamefont {P.~G.~L.}\ \bibnamefont
  {Leach}}, \bibinfo {author} {\bibfnamefont {G.}~\bibnamefont {Leon}},
  \bibinfo {author} {\bibfnamefont {A.}~\bibnamefont {Paliathanasis}}, \ and\
  \bibinfo {author} {\bibfnamefont {S.}~\bibnamefont {Pan}},\ }\href {\doibase
  10.1016/j.jheap.2022.10.001} {\bibfield  {journal} {\bibinfo  {journal}
  {JHEAp}\ }\textbf {\bibinfo {volume} {36}},\ \bibinfo {pages} {141} (\bibinfo
  {year} {2022})},\ \Eprint {http://arxiv.org/abs/2206.06639} {arXiv:2206.06639
  [gr-qc]} \BibitemShut {NoStop}%
\bibitem [{\citenamefont {Yang}\ \emph {et~al.}(2022)\citenamefont {Yang},
  \citenamefont {Pan}, \citenamefont {Mena},\ and\ \citenamefont
  {Di~Valentino}}]{Yang:2022csz}%
  \BibitemOpen
  \bibfield  {author} {\bibinfo {author} {\bibfnamefont {W.}~\bibnamefont
  {Yang}}, \bibinfo {author} {\bibfnamefont {S.}~\bibnamefont {Pan}}, \bibinfo
  {author} {\bibfnamefont {O.}~\bibnamefont {Mena}}, \ and\ \bibinfo {author}
  {\bibfnamefont {E.}~\bibnamefont {Di~Valentino}},\ }\href@noop {} {\
  (\bibinfo {year} {2022})},\ \Eprint {http://arxiv.org/abs/2209.14816}
  {arXiv:2209.14816 [astro-ph.CO]} \BibitemShut {NoStop}%
\bibitem [{\citenamefont {Pan}\ \emph {et~al.}(2020{\natexlab{b}})\citenamefont
  {Pan}, \citenamefont {Sharov},\ and\ \citenamefont {Yang}}]{Pan:2020zza}%
  \BibitemOpen
  \bibfield  {author} {\bibinfo {author} {\bibfnamefont {S.}~\bibnamefont
  {Pan}}, \bibinfo {author} {\bibfnamefont {G.~S.}\ \bibnamefont {Sharov}}, \
  and\ \bibinfo {author} {\bibfnamefont {W.}~\bibnamefont {Yang}},\ }\href@noop
  {} {\  (\bibinfo {year} {2020}{\natexlab{b}})},\ \Eprint
  {http://arxiv.org/abs/2001.03120} {arXiv:2001.03120 [astro-ph.CO]}
  \BibitemShut {NoStop}%
\bibitem [{\citenamefont {Mukhanov}\ \emph {et~al.}(1992)\citenamefont
  {Mukhanov}, \citenamefont {Feldman},\ and\ \citenamefont
  {Brandenberger}}]{Mukhanov:1990me}%
  \BibitemOpen
  \bibfield  {author} {\bibinfo {author} {\bibfnamefont {V.~F.}\ \bibnamefont
  {Mukhanov}}, \bibinfo {author} {\bibfnamefont {H.~A.}\ \bibnamefont
  {Feldman}}, \ and\ \bibinfo {author} {\bibfnamefont {R.~H.}\ \bibnamefont
  {Brandenberger}},\ }\href {\doibase 10.1016/0370-1573(92)90044-Z} {\bibfield
  {journal} {\bibinfo  {journal} {Phys. Rept.}\ }\textbf {\bibinfo {volume}
  {215}},\ \bibinfo {pages} {203} (\bibinfo {year} {1992})}\BibitemShut
  {NoStop}%
\bibitem [{\citenamefont {Ma}\ and\ \citenamefont
  {Bertschinger}(1995)}]{Ma:1995ey}%
  \BibitemOpen
  \bibfield  {author} {\bibinfo {author} {\bibfnamefont {C.-P.}\ \bibnamefont
  {Ma}}\ and\ \bibinfo {author} {\bibfnamefont {E.}~\bibnamefont
  {Bertschinger}},\ }\href {\doibase 10.1086/176550} {\bibfield  {journal}
  {\bibinfo  {journal} {Astrophys. J.}\ }\textbf {\bibinfo {volume} {455}},\
  \bibinfo {pages} {7} (\bibinfo {year} {1995})},\ \Eprint
  {http://arxiv.org/abs/astro-ph/9506072} {arXiv:astro-ph/9506072} \BibitemShut
  {NoStop}%
\bibitem [{\citenamefont {Malik}\ and\ \citenamefont
  {Wands}(2009)}]{Malik:2008im}%
  \BibitemOpen
  \bibfield  {author} {\bibinfo {author} {\bibfnamefont {K.~A.}\ \bibnamefont
  {Malik}}\ and\ \bibinfo {author} {\bibfnamefont {D.}~\bibnamefont {Wands}},\
  }\href {\doibase 10.1016/j.physrep.2009.03.001} {\bibfield  {journal}
  {\bibinfo  {journal} {Phys. Rept.}\ }\textbf {\bibinfo {volume} {475}},\
  \bibinfo {pages} {1} (\bibinfo {year} {2009})},\ \Eprint
  {http://arxiv.org/abs/0809.4944} {arXiv:0809.4944 [astro-ph]} \BibitemShut
  {NoStop}%
\bibitem [{\citenamefont {Gordon}\ and\ \citenamefont
  {Hu}(2004)}]{Gordon:2004ez}%
  \BibitemOpen
  \bibfield  {author} {\bibinfo {author} {\bibfnamefont {C.}~\bibnamefont
  {Gordon}}\ and\ \bibinfo {author} {\bibfnamefont {W.}~\bibnamefont {Hu}},\
  }\href {\doibase 10.1103/PhysRevD.70.083003} {\bibfield  {journal} {\bibinfo
  {journal} {Phys. Rev. D}\ }\textbf {\bibinfo {volume} {70}},\ \bibinfo
  {pages} {083003} (\bibinfo {year} {2004})},\ \Eprint
  {http://arxiv.org/abs/astro-ph/0406496} {arXiv:astro-ph/0406496} \BibitemShut
  {NoStop}%
\bibitem [{\citenamefont {Wands}\ \emph {et~al.}(2012)\citenamefont {Wands},
  \citenamefont {De-Santiago},\ and\ \citenamefont {Wang}}]{Wands:2012vg}%
  \BibitemOpen
  \bibfield  {author} {\bibinfo {author} {\bibfnamefont {D.}~\bibnamefont
  {Wands}}, \bibinfo {author} {\bibfnamefont {J.}~\bibnamefont {De-Santiago}},
  \ and\ \bibinfo {author} {\bibfnamefont {Y.}~\bibnamefont {Wang}},\ }\href
  {\doibase 10.1088/0264-9381/29/14/145017} {\bibfield  {journal} {\bibinfo
  {journal} {Class. Quant. Grav.}\ }\textbf {\bibinfo {volume} {29}},\ \bibinfo
  {pages} {145017} (\bibinfo {year} {2012})},\ \Eprint
  {http://arxiv.org/abs/1203.6776} {arXiv:1203.6776 [astro-ph.CO]} \BibitemShut
  {NoStop}%
\bibitem [{\citenamefont {Wang}\ \emph {et~al.}(2013)\citenamefont {Wang},
  \citenamefont {Wands}, \citenamefont {Xu}, \citenamefont {De-Santiago},\ and\
  \citenamefont {Hojjati}}]{Wang:2013qy}%
  \BibitemOpen
  \bibfield  {author} {\bibinfo {author} {\bibfnamefont {Y.}~\bibnamefont
  {Wang}}, \bibinfo {author} {\bibfnamefont {D.}~\bibnamefont {Wands}},
  \bibinfo {author} {\bibfnamefont {L.}~\bibnamefont {Xu}}, \bibinfo {author}
  {\bibfnamefont {J.}~\bibnamefont {De-Santiago}}, \ and\ \bibinfo {author}
  {\bibfnamefont {A.}~\bibnamefont {Hojjati}},\ }\href {\doibase
  10.1103/PhysRevD.87.083503} {\bibfield  {journal} {\bibinfo  {journal} {Phys.
  Rev. D}\ }\textbf {\bibinfo {volume} {87}},\ \bibinfo {pages} {083503}
  (\bibinfo {year} {2013})},\ \Eprint {http://arxiv.org/abs/1301.5315}
  {arXiv:1301.5315 [astro-ph.CO]} \BibitemShut {NoStop}%
\bibitem [{\citenamefont {Aghanim}\ \emph {et~al.}(2019)\citenamefont {Aghanim}
  \emph {et~al.}}]{Aghanim:2019ame}%
  \BibitemOpen
  \bibfield  {author} {\bibinfo {author} {\bibfnamefont {N.}~\bibnamefont
  {Aghanim}} \emph {et~al.} (\bibinfo {collaboration} {Planck}),\ }\href@noop
  {} {\  (\bibinfo {year} {2019})},\ \Eprint {http://arxiv.org/abs/1907.12875}
  {arXiv:1907.12875 [astro-ph.CO]} \BibitemShut {NoStop}%
\bibitem [{\citenamefont {Aghanim}\ \emph {et~al.}(2018)\citenamefont {Aghanim}
  \emph {et~al.}}]{Aghanim:2018oex}%
  \BibitemOpen
  \bibfield  {author} {\bibinfo {author} {\bibfnamefont {N.}~\bibnamefont
  {Aghanim}} \emph {et~al.} (\bibinfo {collaboration} {Planck}),\ }\href@noop
  {} {\  (\bibinfo {year} {2018})},\ \Eprint {http://arxiv.org/abs/1807.06210}
  {arXiv:1807.06210 [astro-ph.CO]} \BibitemShut {NoStop}%
\bibitem [{\citenamefont {Beutler}\ \emph {et~al.}(2011)\citenamefont
  {Beutler}, \citenamefont {Blake}, \citenamefont {Colless}, \citenamefont
  {Jones}, \citenamefont {Staveley-Smith}, \citenamefont {Campbell},
  \citenamefont {Parker}, \citenamefont {Saunders},\ and\ \citenamefont
  {Watson}}]{Beutler:2011hx}%
  \BibitemOpen
  \bibfield  {author} {\bibinfo {author} {\bibfnamefont {F.}~\bibnamefont
  {Beutler}}, \bibinfo {author} {\bibfnamefont {C.}~\bibnamefont {Blake}},
  \bibinfo {author} {\bibfnamefont {M.}~\bibnamefont {Colless}}, \bibinfo
  {author} {\bibfnamefont {D.}~\bibnamefont {Jones}}, \bibinfo {author}
  {\bibfnamefont {L.}~\bibnamefont {Staveley-Smith}}, \bibinfo {author}
  {\bibfnamefont {L.}~\bibnamefont {Campbell}}, \bibinfo {author}
  {\bibfnamefont {Q.}~\bibnamefont {Parker}}, \bibinfo {author} {\bibfnamefont
  {W.}~\bibnamefont {Saunders}}, \ and\ \bibinfo {author} {\bibfnamefont
  {F.}~\bibnamefont {Watson}},\ }\href {\doibase
  10.1111/j.1365-2966.2011.19250.x} {\bibfield  {journal} {\bibinfo  {journal}
  {Mon. Not. Roy. Astron. Soc.}\ }\textbf {\bibinfo {volume} {416}},\ \bibinfo
  {pages} {3017} (\bibinfo {year} {2011})},\ \Eprint
  {http://arxiv.org/abs/1106.3366} {arXiv:1106.3366 [astro-ph.CO]} \BibitemShut
  {NoStop}%
\bibitem [{\citenamefont {Ross}\ \emph {et~al.}(2015)\citenamefont {Ross},
  \citenamefont {Samushia}, \citenamefont {Howlett}, \citenamefont {Percival},
  \citenamefont {Burden},\ and\ \citenamefont {Manera}}]{Ross:2014qpa}%
  \BibitemOpen
  \bibfield  {author} {\bibinfo {author} {\bibfnamefont {A.~J.}\ \bibnamefont
  {Ross}}, \bibinfo {author} {\bibfnamefont {L.}~\bibnamefont {Samushia}},
  \bibinfo {author} {\bibfnamefont {C.}~\bibnamefont {Howlett}}, \bibinfo
  {author} {\bibfnamefont {W.~J.}\ \bibnamefont {Percival}}, \bibinfo {author}
  {\bibfnamefont {A.}~\bibnamefont {Burden}}, \ and\ \bibinfo {author}
  {\bibfnamefont {M.}~\bibnamefont {Manera}},\ }\href {\doibase
  10.1093/mnras/stv154} {\bibfield  {journal} {\bibinfo  {journal} {Mon. Not.
  Roy. Astron. Soc.}\ }\textbf {\bibinfo {volume} {449}},\ \bibinfo {pages}
  {835} (\bibinfo {year} {2015})},\ \Eprint {http://arxiv.org/abs/1409.3242}
  {arXiv:1409.3242 [astro-ph.CO]} \BibitemShut {NoStop}%
\bibitem [{\citenamefont {Alam}\ \emph {et~al.}(2017)\citenamefont {Alam} \emph
  {et~al.}}]{Alam:2016hwk}%
  \BibitemOpen
  \bibfield  {author} {\bibinfo {author} {\bibfnamefont {S.}~\bibnamefont
  {Alam}} \emph {et~al.} (\bibinfo {collaboration} {BOSS}),\ }\href {\doibase
  10.1093/mnras/stx721} {\bibfield  {journal} {\bibinfo  {journal} {Mon. Not.
  Roy. Astron. Soc.}\ }\textbf {\bibinfo {volume} {470}},\ \bibinfo {pages}
  {2617} (\bibinfo {year} {2017})},\ \Eprint {http://arxiv.org/abs/1607.03155}
  {arXiv:1607.03155 [astro-ph.CO]} \BibitemShut {NoStop}%
\bibitem [{\citenamefont {Scolnic}\ \emph {et~al.}(2018)\citenamefont {Scolnic}
  \emph {et~al.}}]{Scolnic:2017caz}%
  \BibitemOpen
  \bibfield  {author} {\bibinfo {author} {\bibfnamefont {D.}~\bibnamefont
  {Scolnic}} \emph {et~al.},\ }\href {\doibase 10.3847/1538-4357/aab9bb}
  {\bibfield  {journal} {\bibinfo  {journal} {Astrophys. J.}\ }\textbf
  {\bibinfo {volume} {859}},\ \bibinfo {pages} {101} (\bibinfo {year}
  {2018})},\ \Eprint {http://arxiv.org/abs/1710.00845} {arXiv:1710.00845
  [astro-ph.CO]} \BibitemShut {NoStop}%
\bibitem [{\citenamefont {Lewis}\ and\ \citenamefont
  {Bridle}(2002)}]{Lewis:2002ah}%
  \BibitemOpen
  \bibfield  {author} {\bibinfo {author} {\bibfnamefont {A.}~\bibnamefont
  {Lewis}}\ and\ \bibinfo {author} {\bibfnamefont {S.}~\bibnamefont {Bridle}},\
  }\href {\doibase 10.1103/PhysRevD.66.103511} {\bibfield  {journal} {\bibinfo
  {journal} {Phys. Rev. D}\ }\textbf {\bibinfo {volume} {66}},\ \bibinfo
  {pages} {103511} (\bibinfo {year} {2002})},\ \Eprint
  {http://arxiv.org/abs/astro-ph/0205436} {arXiv:astro-ph/0205436} \BibitemShut
  {NoStop}%
\bibitem [{\citenamefont {Gelman}\ and\ \citenamefont
  {Rubin}(1992)}]{Gelman:1992zz}%
  \BibitemOpen
  \bibfield  {author} {\bibinfo {author} {\bibfnamefont {A.}~\bibnamefont
  {Gelman}}\ and\ \bibinfo {author} {\bibfnamefont {D.~B.}\ \bibnamefont
  {Rubin}},\ }\href {\doibase 10.1214/ss/1177011136} {\bibfield  {journal}
  {\bibinfo  {journal} {Statist. Sci.}\ }\textbf {\bibinfo {volume} {7}},\
  \bibinfo {pages} {457} (\bibinfo {year} {1992})}\BibitemShut {NoStop}%
\bibitem [{\citenamefont {Riess}\ \emph {et~al.}(2022)\citenamefont {Riess}
  \emph {et~al.}}]{Riess:2021jrx}%
  \BibitemOpen
  \bibfield  {author} {\bibinfo {author} {\bibfnamefont {A.~G.}\ \bibnamefont
  {Riess}} \emph {et~al.},\ }\href {\doibase 10.3847/2041-8213/ac5c5b}
  {\bibfield  {journal} {\bibinfo  {journal} {Astrophys. J. Lett.}\ }\textbf
  {\bibinfo {volume} {934}},\ \bibinfo {pages} {L7} (\bibinfo {year} {2022})},\
  \Eprint {http://arxiv.org/abs/2112.04510} {arXiv:2112.04510 [astro-ph.CO]}
  \BibitemShut {NoStop}%
\bibitem [{\citenamefont {Yang}\ \emph
  {et~al.}(2021{\natexlab{c}})\citenamefont {Yang}, \citenamefont
  {Di~Valentino}, \citenamefont {Pan}, \citenamefont {Wu},\ and\ \citenamefont
  {Lu}}]{Yang:2021flj}%
  \BibitemOpen
  \bibfield  {author} {\bibinfo {author} {\bibfnamefont {W.}~\bibnamefont
  {Yang}}, \bibinfo {author} {\bibfnamefont {E.}~\bibnamefont {Di~Valentino}},
  \bibinfo {author} {\bibfnamefont {S.}~\bibnamefont {Pan}}, \bibinfo {author}
  {\bibfnamefont {Y.}~\bibnamefont {Wu}}, \ and\ \bibinfo {author}
  {\bibfnamefont {J.}~\bibnamefont {Lu}},\ }\href {\doibase
  10.1093/mnras/staa3914} {\bibfield  {journal} {\bibinfo  {journal} {Mon. Not.
  Roy. Astron. Soc.}\ }\textbf {\bibinfo {volume} {501}},\ \bibinfo {pages}
  {5845} (\bibinfo {year} {2021}{\natexlab{c}})},\ \Eprint
  {http://arxiv.org/abs/2101.02168} {arXiv:2101.02168 [astro-ph.CO]}
  \BibitemShut {NoStop}%
\bibitem [{\citenamefont {Yang}\ \emph
  {et~al.}(2019{\natexlab{b}})\citenamefont {Yang}, \citenamefont {Pan},
  \citenamefont {Di~Valentino}, \citenamefont {Saridakis},\ and\ \citenamefont
  {Chakraborty}}]{Yang:2018qmz}%
  \BibitemOpen
  \bibfield  {author} {\bibinfo {author} {\bibfnamefont {W.}~\bibnamefont
  {Yang}}, \bibinfo {author} {\bibfnamefont {S.}~\bibnamefont {Pan}}, \bibinfo
  {author} {\bibfnamefont {E.}~\bibnamefont {Di~Valentino}}, \bibinfo {author}
  {\bibfnamefont {E.~N.}\ \bibnamefont {Saridakis}}, \ and\ \bibinfo {author}
  {\bibfnamefont {S.}~\bibnamefont {Chakraborty}},\ }\href {\doibase
  10.1103/PhysRevD.99.043543} {\bibfield  {journal} {\bibinfo  {journal} {Phys.
  Rev.}\ }\textbf {\bibinfo {volume} {D99}},\ \bibinfo {pages} {043543}
  (\bibinfo {year} {2019}{\natexlab{b}})},\ \Eprint
  {http://arxiv.org/abs/1810.05141} {arXiv:1810.05141 [astro-ph.CO]}
  \BibitemShut {NoStop}%
\bibitem [{\citenamefont {Heavens}\ \emph
  {et~al.}(2017{\natexlab{a}})\citenamefont {Heavens}, \citenamefont {Fantaye},
  \citenamefont {Sellentin}, \citenamefont {Eggers}, \citenamefont {Hosenie},
  \citenamefont {Kroon},\ and\ \citenamefont {Mootoovaloo}}]{Heavens:2017hkr}%
  \BibitemOpen
  \bibfield  {author} {\bibinfo {author} {\bibfnamefont {A.}~\bibnamefont
  {Heavens}}, \bibinfo {author} {\bibfnamefont {Y.}~\bibnamefont {Fantaye}},
  \bibinfo {author} {\bibfnamefont {E.}~\bibnamefont {Sellentin}}, \bibinfo
  {author} {\bibfnamefont {H.}~\bibnamefont {Eggers}}, \bibinfo {author}
  {\bibfnamefont {Z.}~\bibnamefont {Hosenie}}, \bibinfo {author} {\bibfnamefont
  {S.}~\bibnamefont {Kroon}}, \ and\ \bibinfo {author} {\bibfnamefont
  {A.}~\bibnamefont {Mootoovaloo}},\ }\href {\doibase
  10.1103/PhysRevLett.119.101301} {\bibfield  {journal} {\bibinfo  {journal}
  {Phys. Rev. Lett.}\ }\textbf {\bibinfo {volume} {119}},\ \bibinfo {pages}
  {101301} (\bibinfo {year} {2017}{\natexlab{a}})},\ \Eprint
  {http://arxiv.org/abs/1704.03467} {arXiv:1704.03467 [astro-ph.CO]}
  \BibitemShut {NoStop}%
\bibitem [{\citenamefont {Heavens}\ \emph
  {et~al.}(2017{\natexlab{b}})\citenamefont {Heavens}, \citenamefont {Fantaye},
  \citenamefont {Mootoovaloo}, \citenamefont {Eggers}, \citenamefont {Hosenie},
  \citenamefont {Kroon},\ and\ \citenamefont {Sellentin}}]{Heavens:2017afc}%
  \BibitemOpen
  \bibfield  {author} {\bibinfo {author} {\bibfnamefont {A.}~\bibnamefont
  {Heavens}}, \bibinfo {author} {\bibfnamefont {Y.}~\bibnamefont {Fantaye}},
  \bibinfo {author} {\bibfnamefont {A.}~\bibnamefont {Mootoovaloo}}, \bibinfo
  {author} {\bibfnamefont {H.}~\bibnamefont {Eggers}}, \bibinfo {author}
  {\bibfnamefont {Z.}~\bibnamefont {Hosenie}}, \bibinfo {author} {\bibfnamefont
  {S.}~\bibnamefont {Kroon}}, \ and\ \bibinfo {author} {\bibfnamefont
  {E.}~\bibnamefont {Sellentin}},\ }\href@noop {} {\  (\bibinfo {year}
  {2017}{\natexlab{b}})},\ \Eprint {http://arxiv.org/abs/1704.03472}
  {arXiv:1704.03472 [stat.CO]} \BibitemShut {NoStop}%
\end{thebibliography}%

\end{document}